# Medical Image Registration Using Deep Neural Networks: A Comprehensive Review

Hamid Reza Boveiri [a,*], Raouf Khayami [a], Reza Javidan [a], Ali Reza MehdiZadeh [b]

[a] Department of Computer Engineering and IT, Shiraz University of Technology, Shiraz, Iran
[b] Department of Medical Physics and Engineering, Shiraz University of Medical Science, Shiraz, Iran
* Corresponding Author Email: hr.boveiri@sutech.ac.ir

***Abstract*****—**Image-guided interventions are saving the lives of a large number of patients where the image registration problem should indeed be considered as the most complex and complicated issue to be tackled. On the other hand, the recently huge progress in the field of machine learning made by the possibility of implementing deep neural networks on the contemporary many-core GPUs opened up a promising window to challenge with many medical applications, where the registration is not an exception. In this paper, a comprehensive review on the state-of-the-art literature known as medical image registration using deep neural networks is presented. The review is systematic and encompasses all the related works previously published in the field. Key concepts, statistical analysis from different points of view, confiding challenges, novelties and main contributions, key-enabling techniques, future directions and prospective trends all are discussed and surveyed in details in this comprehensive review. This review allows a deep understanding and insight for the readers active in the field who are investigating the state-of-the-art and seeking to contribute the future literature.

***Keywords*****—**Convolutional Neural Network (CNN); Deep Learning; Deep Reinforcement Learning; Deformable Registration; Generative Adversarial Network (GAN); Image-guided Intervention; Medical Image Registration; One-shot Registration; Staked Auto-Encoders (SAEs).

♦

## 1. Introduction

In most medical interventions, there are a number of cases in which some images need to be captured for diagnosis, prognosis, treatment and follow-up purposes. These images can be vary in terms of temporal, spatial, dimensional or modular. Image fusion causing information synergy can have a significant contribution to guide and support physicians in the process of decision making, mostly in online and real-time fasion. Lack of alignment is unavoidable for these images taken in different times and conditions; hence, can challenge the quality and accuracy of the subsequent analyses. Image registration is the process of aligning two (or more) given images based on an identical geometrical coordination system. The aim is at finding an optimum spatial transformation that registers the structures-of-interest in the best way. This problem is important in numerous ways in the field of machine vision e.g. for remote sensing, object tracing, satellite imaging and so on (Goshtasby 2017).

Image registration is also fundamental to the image-guided intervention where e.g. telesurgery, image-guided radiotherapy (IGRT), and precision medicine cannot be operational without using image registration techniques (Peters & Cleary 2008). To exemplify, in IGRT, a pre-interventional image (typically high-quality 3D image), on which the treatment planning is conducted, needs to be registered on an operational image (typically low-quality and noisy 2D) so that the linear accelerator (linac) machine can be calibrated, and the radiation fragment can be conducted with maximal precision and minimal risk of radiation to the adjacent healthy organs referred to as minimally invasive procedure. In this process, the challenges like different modalities of inputted images, low-quality and noise of interventional images, deformation of abdominal cavity's organs (because of the spontaneous contraction/inflation), movement of thorax cavity's organs (because of the respiration and heartbeats), changing the size of organs and regions-of-interest (RoIs) due to the weight loss/gain during the treatment process can compromises the quality of solving the problem. In practice, special considerations should be taken into account, and other image processing techniques need to collaborate, which makes the issue very challenging and complicated (Hajnal 2001).

Basically, conventional image registration is an iterative optimization process that requires extracting proper features, selecting a similarity measure (to evaluate the registration quality), choosing the model of transformation, and finally, a mechanism to investigate the search space (Oliveira & Tavares 2014). As illustrated in Fig. 1, a couple of images are inputted to the system, of which one is considered as fixed and the other as moving image. The optimal alignment can be achieved via iteratively sliding the moving image over the fixed image. At first, the considered similarity measure identifies the amount of correspondence between the inputted images. An optimization algorithm, using an updating mechanism, calculates the new transformation's parameters. Appling these parameters on the moving image leads a new supposedly better-aligned image. If the termination criteria are satisfied, the algorithm is terminated, else a new iteration should be started. In each iteration, the moving image get a better correspondence with the fixed image, and the iterations continue until no further registration can be achieved, or some predefined criteria are satisfied. The system output can be either the transformation parameters or final interpolated fused image.

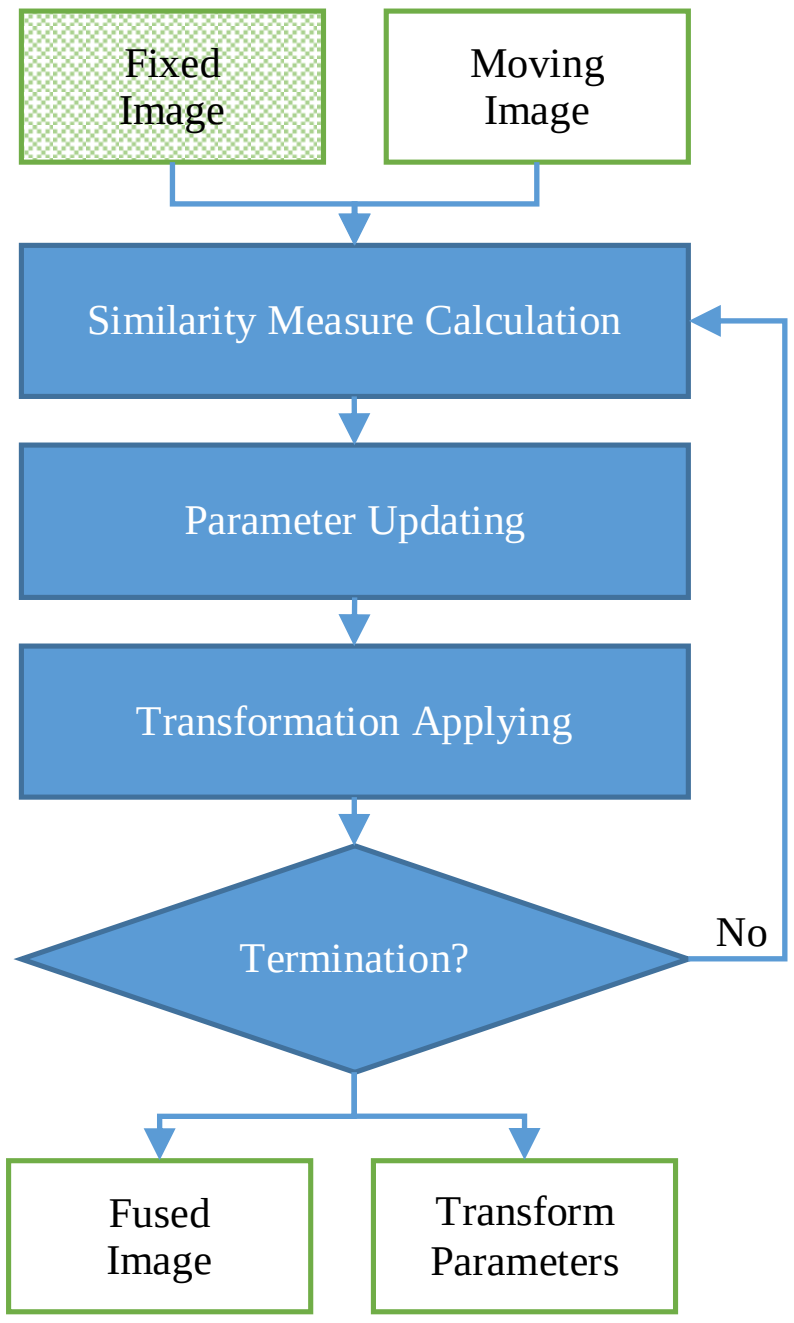

**Figure 1:** The workflow of conventional image registration techniques based on optimization procedures

There are two main drawbacks to this strategy as follows:

- This iterative manner is very slow where runtimes in the tens of minutes are norm for common deformable image registration techniques even with an efficient implementation on the contemporary GPUs (like a NVIDIA TitanX); while the practical use in clinical operations is real-time, and such a prolonged wasting time is not appreciated.
- Most similarity measures have a lots of local optima around the global one, specially where dealing with images from different modalities (referred to as multimodal image registration); they lose their efficiency causing premature convergence or stagnation which are two prevalent confining dilemmas in the optimization field.

Accordingly, to circumvent these two confining problems, learning-based registration approaches have gained increasing popularity in the recent years; meanwhile, deep neural networks (DNNs), as one of the most powerful techniques ever seen by the community of machine intelligence, have been applied to the various image processing applications. Of course, medical image registration is not an exception, and a number of deep learning based approaches were proposed in the literature; however, the number of works and the used techniques is very limited, and there is a promising potential for investigation (Litjens et al. 2017).

In this paper, a comprehensive systematic review on the medical image registration using deep neural networks is presented. We have gathered all the relevant state-of-the-arts, from the first one in 2013 up the last one in 2019. The works are analyses based on the different statistical perspectives and measures of interest e.g. years, publication titles, publication types, publishers, authors (total number of publications and citations), keywords, techniques, comparison metrics, datasets, organs of interest, and modalities. The approaches are categorized to the three major generations based on the breakthroughs affecting the contributions. Also, the seminal works in each category and generation are introduced and analyzed in depth, and key contributions and philosophies behind them are presented in details. Finally, there is a discussion introducing confining challenges, open problems and prospective directions. This review allows a deep understanding and insight for the readers active in the field who are investigating the state-of-the-art and seeking to contribute the future literature.

Rest of the paper is organized as follows. In the next Section, reference gathering methodology and the challenged faced are described. Section 3 and 4 are devoted to the taxonomy of medical image registration, and the problem formulation, respectively. The architectures of deep neural networks used in the literature are investigated in Section 5. Section 6 is an in-depth literature review on seminal works. Statistical analysis on the state-of-the-art is presented in Section 7. Section 8 is devoted to the discussion on the confining challenges and open problems in the field. And finally, the conclusion and future trends are presented in the last Section.

## 2. Reference Gathering Methodology

To investigate the literature, the systematic search was conducted on two major scientific databases i.e. "SCOPUS" and "PubMed" (since the topic is multidisciplinary between engineering and medicine). "Medical Image Registration" was the main keyword searched accompanied with one of the followings as the underlying technique i.e. "Deep Learning," "Deep Neural Network," or "Convolutional." Moreover, another search was conducted using the aforementioned keywords in major scientific datasets i.e. "Google Scholar" and "CrossRef" to validate the comprehensiveness of the results. The searches were restricted to the Title, Abstract and Keywords whenever it was feasible. A total number of 25 references were detected, among them, 6 cases was irrelevant (e.g. just the results were compared to the deep learning approaches) and 21 cases was completely matched with our criteria.

At this point, we detected that some references were overlooked, and that these keywords are insufficient to conduct a comprehensive search; for example, there was a large number of conference papers with no "Deep Learning," "Deep Neural Network," or "Convolutional" in the Title, Abstract or Keywords, but e.g. start with CNN, SAEs or GAN without defining these abbreviations. Additionally, we found that there are a number of references that use e.g. "Image Registration," "Image Correspondence," "Pose Estimation" etc. instead of our main keyword "Medical Image Registration," but the paper is completely within the scope of the review. Such case was true for our backup search in "Google Scholar" and "CrossRef." On this basis, with a comprehensive list of keywords extracted from the already detected references, we conducted a comprehensive systematic search in the "SCOPUS" and "PubMed" as well as an exhaustive ad-hoc search in "Google Scholar" and "CrossRef." We also reviewed all the citations and references of all the selected papers, and searched within the journals/conferences in which the selected papers were published. Totally, we reviewed more than 500 potential papers, and among them we could extract a complete list of 80 papers that constitute the state-of-the-art literature. These references along with their underlying techniques, datasets, evaluation metrics and other related information are listed in the Table 1.

In this regards, we faced four determinative challenges to constitute this systematic review as follows:

- Routinely, we were highly concerned not to overlook any work in order to draw justifiable conclusions, where this topic is in its infancy, and the number of related works is limited.
- Secondly, there was not any application/software to analyze such data to support the authors of systematic reviews, and we were forced to develop a simple desktop application, and use embedded-SQL to draw our statistical analysis.
- Thirdly, we are not able to conduct any experimental benchmark review on this topic because, as we have already contacted most of the authors, this topic is related to the medical science where most of the utilized datasets are private, and the implementations have copyright, and cannot be acquired by us.
- Finally and most importantly, we were highly concerned about the soundness and usefulness of our asserts and conclusions specially from the medical/clinical point of view. Fortunately, our co-Author Dr. MehdiZadeh, the associate Professor and Dean of the Department of Medical Physics and Engineering, SUMS, has been with us in all the phases, and meticulously read, revised, and certified all the discussions and conclusions to be sound and valuable from the medical/clinical perspective.

**Table 1:** Founded references along with their used techniques, datasets and related evaluation metrics

| Author & Year | Reference | Technique | Dataset (Organ) | Metrics | Description |
|---|---|---|---|---|---|
| Wu et al. (2013) | G. Wu, M. Kim, Q. Wang, Y. Gao, S. Liao, and D. Shen, "Unsupervised Deep Feature Learning for Deformable Registration of MR Brain Images," Lecture Notes in Computer Science, pp. 649–656, 2013. | CNN | IXI - T1 and T2-weighted MRI (Adult Brain)<br>ANDI - MRI (Adult Brain) | Dice | Using hybrid ISA and CNN to automatically extract features and feed them to HAMMER for final deformable registration |
| Zhao and Jia (2015) | L. Zhao and K. Jia, "Deep Adaptive Log-Demons: Diffeomorphic Image Registration with Very Large Deformations," Computational and Mathematical Methods in Medicine, vol. 2015, pp. 1–16, 2015. | CNN | BrainWeb – MRI (Brain)<br>Empire 10 – 3D CT (Lung) | Dice | Using CNN to automatically extract features and feed them to Demons for final registration |
| Cheng et al. (2016) | X. Cheng, L. Zhang, and Y. Zheng, "Deep similarity learning for multimodal medical images," Computer Methods in Biomechanics and Biomedical Engineering: Imaging & Visualization, vol. 6, no. 3, pp. 248–252, Apr. 2018. (Accepted & Online from 2016) | SAEs | CT and MRI (Head) - Multimodal | CSPE | Using SAEs as a multimodal similarity for rigid registration. |
| Yang et al. (2016) | X. Yang, X. Han, E. Park, S. Aylward, R. Kwitt, and M. Niethammer, "Registration of Pathological Images," Lecture Notes in Computer Science, pp. 97–107, 2016. | SAEs | OASIS – MRI (Brain)<br>BRATS – MRI (Brain) | TRE and Deformation Error | Directly regressing the unimodal deformable registration parameters via a convolutional SAE. |
| Yang et al. (2016) | X. Yang, R. Kwitt, and M. Niethammer, "Fast Predictive Image Registration," Lecture Notes in Computer Science, pp. 48–57, 2016. | CNN | OASIS – MRI (Brain) | Deformation Error | Directly regressing the deformable registration parameters via hybrid CNN and LDDMM |
| Wu et al. (2016) | G. Wu, M. Kim, Q. Wang, B. C. Munsell, and D. Shen, "Scalable High-Performance Image Registration Framework by Unsupervised Deep Feature Representations Learning," IEEE Transactions on Biomedical Engineering, vol. 63, no. 7, pp. 1505–1516, Jul. 2016. | SAEs | LONI - MRI (Adult Brain)<br>ANDI - MRI (Adult Brain) | Dice | Using SAEs to extract features and feed them to a traditional approaches for final deformable registration |
| Simonovsky et al. (2016) | M. Simonovsky, B. Gutiérrez-Becker, D. Mateus, N. Navab, and N. Komodakis, "A Deep Metric for Multimodal Registration," Medical Image Computing and Computer-Assisted Intervention -- MICCAI 2016, pp. 10–18, 2016. | CNN | IXI - T1 and T2-weighted MRI (Adults' Brain)<br>ALBERTs - T1 and T2-weighted MRI (Neonatals' Brain) | Dice and Jaccard | Using CNN as a multimodal similarity measure to guide conventional iterative approaches |
| Miao et al. (2016) | S. Miao, Z. J. Wang, Y. Zheng, and R. Liao, "Real-time 2D/3D registration via CNN regression," 2016 IEEE 13th International Symposium on Biomedical Imaging (ISBI), Apr. 2016. | CNN | XEF (X-ray Echo Fusion)<br>VIPS (Visual Implant Planning System)<br>TKA (Total Knee Arthroplasty) | mTREproj | Directly regressing the rigid registration parameters using CNN based on the implanted particles |
| Miao et al. (2016) | S. Miao, Z. J. Wang, and R. Liao, "A CNN Regression Approach for Real-Time 2D/3D Registration," IEEE Transactions on Medical Imaging, vol. 35, no. 5, pp. 1352–1363, May 2016. | CNN | XEF (X-ray Echo Fusion)<br>VIPS (Visual Implant Planning System)<br>TKA (Total Knee Arthroplasty) | mTREproj | Directly regressing the rigid registration parameters using CNN based on the implanted particles |
| Liao et al. (2017) | Rui Liao, Shun Miao, Pierre de Tournemire, Sasa Grbic, Ali Kamen, Tommaso Mansi, and Dorin Comaniciu. "An Artificial Agent for Robust Image Registration," in *AAAI*, pp. 4168-4175, 2017. | CNN | E1 – CT and CBCT (Spine)<br>E2 – CT and CBCT (Cardiac) | TRE and MME | Using reinforcement learning to train CNN to be used as agents for approximating Affine registration's parameters in an iterative manner |
| Krebs et al. (2017) | J. Krebs, T. Mansi, H. Delingette, L. Zhang, F. C. Ghesu, S. Miao, A. K. Maier, N. Ayache, R. Liao, and A. Kamen, "Robust Non-rigid Registration Through Agent-Based Action Learning," Lecture Notes in Computer Science, pp. 344–352, 2017. | CNN | PROMISE12 – MRI – Prostate<br>Prostate-3T – MRI – (Prostate) | Dice and Hausdorff | Using reinforcement learning to train CNN to be used as agents for approximating Affine registration's parameters in an iterative manner |
| Yang et al. (2017) | X. Yang, R. Kwitt, M. Styner, and M. Niethammer, "Fast predictive multimodal image registration," 2017 IEEE 14th International Symposium on Biomedical Imaging (ISBI 2017), Apr. 2017. | CNN | IBIS 3D Autism – TI and T2 MRI (multimodal) (Brain) | SSD | Directly regressing the deformable registration parameters via hybrid CNN and LDDMM |
| Wang et al. (2017) | S. Wang, M. Kim, G. Wu, and D. Shen, "Scalable High Performance Image Registration Framework by Unsupervised Deep Feature Representations Learning," Deep Learning for Medical Image Analysis, pp. 245–269, 2017. | SAEs | LONI - MRI (Adult Brain)<br>ANDI - MRI (Adult Brain) | Dice | Using SAEs to extract features and feed them to conventional approaches for final deformable registration |
| Miao et al. (2017) | S. Miao, J. Z. Wang, and R. Liao, "Convolutional Neural Networks for Robust and Real-Time 2-D/3-D Registration," Deep Learning for Medical Image Analysis, pp. 271–296, 2017. | CNN | XEF (X-ray Echo Fusion)<br>VIPS (Visual Implant Planning System)<br>TKA (Total Knee Arthroplasty) | mTREproj | Directly regressing the rigid registration parameters using CNN based on the implanted particles |
| Sokooti et al. (2017) | H. Sokooti, B. de Vos, F. Berendsen, B. P. F. Lelieveldt, I. Išgum, and M. Staring, "Nonrigid Image Registration Using Multi-scale 3D Convolutional Neural Networks," Lecture Notes in Computer Science, pp. 232–239, 2017. | CNN | SPREAD - 3D CT (Chest) | MAE and TRE | Directly regressing the defprmable registration parameters using CNN |
| de Vos et al. (2017) | B. D. de Vos, F. F. Berendsen, M. A. Viergever, M. Staring, and I. Išgum, "End-to-End Unsupervised Deformable Image | CNN | Sunnybrook Cardiac Data – 3D MRI (Cardiac) | Dice and MAD | Directly regressing the deformable registration parameters using CNN |
| | Registration with a Convolutional Neural Network," Lecture Notes in Computer Science, pp. 204–212, 2017. | | | | |
| Bhatia et al. (2017) | P. S. Bhatia, F. Reda, M. Harder, Y. Zhan, and X. S. Zhou, "Real time coarse orientation detection in MR scans using multi-planar deep convolutional neural networks," Medical Imaging 2017: Image Processing, Feb. 2017. | CNN | Private – MRI (Elbow) | Accuracy | Directly regressing the unimodal rigid registration parameters using CNN |
| Zheng et al. (2017) | J. Zheng, S. Miao, and R. Liao, "Learning CNNs with Pairwise Domain Adaption for Real-Time 6DoF Ultrasound Transducer Detection and Tracking from X-Ray Images," Medical Image Computing and Computer-Assisted Intervention − MICCAI 2017, pp. 646–654, 2017. | CNN | Private – X-Ray and TEE (Transducer) | Projected TRE | Directly regressing the multimodal rigid registration parameters (pose) of TEE Transducer using CNN |
| Pei et al. (2017) | Y. Pei, Y. Zhang, H. Qin, G. Ma, Y. Guo, T. Xu, and H. Zha, "Non-rigid Craniofacial 2D-3D Registration Using CNN-Based Regression," Lecture Notes in Computer Science, pp. 117–125, 2017. | CNN | NewTom – CBCT (Craniofacial) | MCD and MID | Directly regressing the unimodal deformable CT and X-Ray registration parameters using CNN |
| Eppenhof and Pluim (2017) | K. A. J. Eppenhof and J. P. W. Pluim, "Supervised local error estimation for nonlinear image registration using convolutional neural networks," Medical Imaging 2017: Image Processing, Feb. 2017. | CNN | Private – 2D DSA (Brain)<br>DIRLAB – 3D CT (Lung) | RMSD, NRMSD and PCC | Proposing a supervised method for the estimation of the unimodal registration error map for deformable image registration using CNN |
| Ghosal and Ray (2017) | S. Ghosal and N. Ray, "Deep deformable registration: Enhancing accuracy by fully convolutional neural net," Pattern Recognition Letters, vol. 94, pp. 81–86, Jul. 2017. | CNN (VGG-net) (Simonyan and Zisserman 2015) | IXI - T1 and T2-weighted MRI (Adults' Brain)<br>ANDI - MRI (Adult Brain) | SSIM, PSNR, and SSD | Learn a CNN to work as SSD as a new unimodal similarity metrics to work with any conventional deformable registration method. |
| Ma et al. (2017) | K. Ma, J. Wang, V. Singh, B. Tamersoy, Y.-J. Chang, A. Wimmer, and T. Chen, "Multimodal Image Registration with Deep Context Reinforcement Learning," Lecture Notes in Computer Science, pp. 240–248, 2017. | DRL (DuelingNet) (Wang et al. 2016) | ABD – depth and CT (Chest and Abdomen) | Hausdorff | Directly regressing the rigid multimodal registration parameters using Deep Reinforcement Learning (DRL) |
| Salehi et al. (2017) | M. Salehi, R. Prevost, J.-L. Moctezuma, N. Navab, and W. Wein, "Precise Ultrasound Bone Registration with Learning-Based Segmentation and Speed of Sound Calibration," Medical Image Computing and Computer-Assisted Intervention − MICCAI 2017, pp. 682–690, 2017. | CNN | Private – CT and US (Bone) | Precision, Recall and Dice | Directly regressing the deformable multimodal (CT-US) registration parameters using a weakly-supervised trained CNN |
| Rohe et al. (2017) | M.-M. Rohé, M. Datar, T. Heimann, M. Sermesant, and X. Pennec, "SVF-Net: Learning Deformable Image Registration Using Shape Matching," Lecture Notes in Computer Science, pp. 266–274, 2017. | CNN | Private - 3D MRI (Cardiac) | Dice, Hausdorff, LCC and RVLJ | Directly regressing the 3D unimodal deformable registration parameters using CNN |
| Yoo et al. (2017) | I. Yoo, D. G. C. Hildebrand, W. F. Tobin, W.-C. A. Lee, and W.-K. Jeong, "ssEMnet: Serial-Section Electron Microscopy Image Registration Using a Spatial Transformer Network with Learned Features," Lecture Notes in Computer Science, pp. 249–257, 2017. | SAEs | CREMI TEM – EM (Brain) | Dice | Using SAEs to extract unimodal structural features and feed to a STN for final deformable registration |
| Cao et al (2017) | X. Cao, J. Yang, J. Zhang, D. Nie, M. Kim, Q. Wang, and D. Shen, "Deformable Image Registration Based on Similarity-Steered CNN Regression," Lecture Notes in Computer Science, pp. 300–308, 2017. | CNN | LONI LPBA40 (Brain MRI)<br>ANDI (Brain MRI) | Dice and ASSD | Directly regressing the unimodal deformable registration parameters using CNN (Transfer Learning) |
| Uzunova et al. (2017) | H. Uzunova, M. Wilms, H. Handels, and J. Ehrhardt, "Training CNNs for Image Registration from Few Samples with Model-based Data Augmentation," Lecture Notes in Computer Science, pp. 223–231, 2017. | CNN (FlowNet) (Dosovitskiy et al. 2015) | LONI LBPA40 (Brain MRI)<br>Private - cine Cardiac MRI | Jaccard and ASCD | Directly regressing the 2D unimodal deformable registration parameters using a weakly-supervised CNN |
| Yang et al. (2017) | X. Yang, R. Kwitt, M. Styner, and M. Niethammer, "Quicksilver: Fast predictive image registration – A deep learning approach," NeuroImage, vol. 158, pp. 378–396, Sep. 2017. | CNN | OASIS – MRI (Brain)<br>IBIS 3D Autism – TI and T2 MRI (multimodal) (Brain) | SSD and Deformation Error | Directly regressing the deformable registration parameters via hybrid CNN and LDDMM |
| Zheng et al. (2018) | J. Zheng, S. Miao, Z. Jane Wang, and R. Liao, "Pairwise domain adaptation module for CNN-based 2-D/3-D registration," Journal of Medical Imaging, vol. 5, no. 02, p. 1, Jan. 2018. | CNN | TEE – X-ray (Spine)<br>Spine – X-ray and CT (Spine) | TRE | Directly regressing the rigid unimodal registration parameters using CNN based on the implanted particles |
| Hu et al. (2018) | Y. Hu, E. Gibson, N. Ghavami, E. Bonmati, C. M. Moore, M. Emberton, T. Vercauteren, J. A. Noble, and D. C. Barratt, "Adversarial Deformation Regularization for Training Image Registration Neural Networks," Lecture Notes in Computer Science, pp. 774–782, 2018. | GAN | SmartTarget – MRI-T2 and TRUS (Prostate) - Multimodal | TRE and Dice | Directly regressing the multimodal deformable registration via a weakly-supervised anatomical-label-driven GAN |
| Hu et al. (2018) | Y. Hu, M. Modat, E. Gibson, N. Ghavami, E. Bonmati, C. M. Moore, M. Emberton, J. A. Noble, D. C. Barratt, and T. Vercauteren, "Label-driven weakly-supervised learning for multimodal deformarle image registration," 2018 IEEE 15th International Symposium on Biomedical Imaging (ISBI 2018), Apr. 2018. | CNN | SmartTarget – MRI-T2 and TRUS (Prostate) - Multimodal | TRE and Dice | Directly regressing the deformable multimodal registration parameters using a weakly-supervisedly trained CNN |
| Hu et al. (2018) | Y. Hu, M. Modat, E. Gibson, W. Li, N. Ghavami, E. Bonmati, G. Wang, S. Bandula, C. M. Moore, M. Emberton, S. Ourselin, J. A. Noble, D. C. Barratt, and T. Vercauteren, "Weakly-supervised convolutional neural networks for multimodal image registration," Medical Image Analysis, vol. 49, pp. 1–13, Oct. 2018. | CNN | SmartTarget – MRI-T2 and TRUS (Prostate) - Multimodal | TRE and Dice | Directly regressing the deformable multimodal registration parameters using a weakly-supervised trained CNN |
| Dalca et al. (2018) | A. V. Dalca, G. Balakrishnan, J. Guttag, and M. R. Sabuncu, "Unsupervised Learning for Fast Probabilistic Diffeomorphic | CNN | ADNI, OASIS<br>ABIDE, ADHD200 | Dice | Directly regressing the unimodal deformable |
| | Registration,” Lecture Notes in Computer Science, pp. 729–738, 2018. | | MCIC PPMI<br>HABS Harvard GSP<br>MRI (Brain) | | diffeomorphic registration parameters using CNN with unsupervised learning |
| Balakrishnan et al. (2018) | G. Balakrishnan, A. Zhao, M. R. Sabuncu, A. V. Dalca, and J. Guttag, “An Unsupervised Learning Model for Deformable Medical Image Registration,” 2018 IEEE/CVF Conference on Computer Vision and Pattern Recognition, Jun. 2018. | CNN | ADNI OASIS<br>ABIDE ADHD200<br>MCIC PPMI<br>HABS Harvard GSP<br>MRI (Brain) | Dice | Directly regressing the unimodal deformable registration parameters using CNN with unsupervised learning |
| Shu et al. (2018) | C. Shu, X. Chen, Q. Xie, and H. Han, “An unsupervised network for fast microscopic image registration,” Medical Imaging 2018: Digital Pathology, Mar. 2018. | CNN | Private – EM (Brain) | Dice | Directly regressing the unimodal deformable registration parameters using CNN |
| Awan and Rajpoot (2018) | R. Awan and N. Rajpoot, “Deep Autoencoder Features for Registration of Histology Images,” Medical Image Understanding and Analysis, pp. 371–378, 2018. | SAEs | Bioimaging Challenge 2015 – EM (Breast) | RMSE | Using a convolutional SAEs as a multimodal similarity for rigid registration. |
| Stergios et al (2018) | C. Stergios, S. Mihir, V. Maria, C. Guillaume, R. Marie-Pierre, M. Stavroula, and P. Nikos, “Linear and Deformable Image Registration with 3D Convolutional Neural Networks,” Lecture Notes in Computer Science, pp. 13–22, 2018. | CNN | Private – MRI (Lung) | Dice | Directly regressing the unimodal deformable registration parameters using CNN in unsupervised manner |
| Onieva et al. (2018) | J. Onieva Onieva, B. Marti-Fuster, M. Pedrero de la Puente, and R. San José Estépar, “Diffeomorphic Lung Registration Using Deep CNNs and Reinforced Learning,” Lecture Notes in Computer Science, pp. 284–294, 2018. | CNN (RegNet) (Sokooti et al. 2017) | COPDGene – CT (Lung) | Deformation Error | Directly regressing the unimodal diffeomorphic deformable registration parameters using CNN |
| Mahapatra et al. (2018) | D. Mahapatra, Z. Ge, S. Sedai, and R. Chakravorty, “Joint Registration And Segmentation Of Xray Images Using Generative Adversarial Networks,” Lecture Notes in Computer Science, pp. 73–80, 2018. | GAN | NIH ChestXray14 – X-Ray (Chest) | TRE, Dice and Hausdorff | Directly regressing the unimodal deformable registration parameters using GAN |
| Mahapatra et al. (2018) | D. Mahapatra, B. Antony, S. Sedai, and R. Garnavi, “Deformable medical image registration using generative adversarial networks,” 2018 IEEE 15th International Symposium on Biomedical Imaging (ISBI 2018), Apr. 2018. | GAN | Private – CFI and FA (Retina)<br>Sunnybrook – 3D MRI (Cardiac) | Dice, Hausdorff, MAD, MSE, Deformation Error | Directly regressing the multimodal deformable registration parameters using GAN |
| Sentker et al. (2018) | T. Sentker, F. Madesta, and R. Werner, “GDL-FIRE 4D: Deep Learning-Based Fast 4D CT Image Registration,” Lecture Notes in Computer Science, pp. 765–773, 2018. | CNN | DIRLAB – 4D CT (Lung and Liver)<br>CREATIS– 4D CT (Lung and Liver) | TRE | Directly regressing the unimodal deformable 4D CT registration parameters using CNN |
| Ito and Ino (2018) | M. Ito and F. Ino, “An Automated Method for Generating Training Sets for Deep Learning based Image Registration,” Proceedings of the 11th International Joint Conference on Biomedical Engineering Systems and Technologies, 2018. | CNN (GoogLeNet) (Szegedy et al. 2015) | ANDI - MRI (Brain) | Precision and Recall | Automated method for generating training set for image registration, aiming at realizing non-rigid registration with deep learning. |
| Yan et al. (2018) | P. Yan, S. Xu, A. R. Rastinehad, and B. J. Wood, “Adversarial Image Registration with Application for MR and TRUS Image Fusion,” Lecture Notes in Computer Science, pp. 197–204, 2018. | GAN | NIH MSH – T2 MRI and TRUS (Prostate) | TRE and D-Score | Directly regressing the multimodal rigid registration parameters using GAN |
| Abanovie et al. (2018) | E. Abanovie, G. Stankevieius, and D. Matuzevieius, “Deep Neural Network-based Feature Descriptor for Retinal Image Registration,” 2018 IEEE 6th Workshop on Advances in Information, Electronic and Electrical Engineering (AIEEE), Nov. 2018. | CNN | Chase DB Diaret DB<br>DTSET1 DTSET2<br>HRF-base Messidor1<br>RODREP FIRE<br>CFI and OCT (Retina) | Matching Performance | Using CNN as a unimodal similarity measure to guide conventional rigid iterative approaches |
| Miao et al. (2018) | S. Miao, S. Piat, P. Fischer, A. Tuysuzoglu, P. Mewes, T. Mansi, R. Liao, “Dilated FCN for multi-agent 2D/3D medical image registration,” in Thirty-Second AAAI Conference on Artificial Intelligence, Apr. 2018. | CNN | Private – CBCT and X-Ray (Spine) | TRE | Directly regressing the multimodal rigid registration parameters using CNN |
| Toth et al. (2018) | D. Toth, S. Miao, T. Kurzendorfer, C. A. Rinaldi, R. Liao, T. Mansi, K. Rhode, and P. Mountney, “3D/2D model-to-image registration by imitation learning for cardiac procedures,” International Journal of Computer Assisted Radiology and Surgery, vol. 13, no. 8, pp. 1141–1149, May 2018. | CNN | LIDC-IDRI – CT (Cardiac)<br>Private - MRI (Cardiac) | TRE | Directly regressing the multimodal rigid model to image registration parameters using CNN |
| Blendowski and Heinrich (2018) | M. Blendowski and M. P. Heinrich, “3D-CNNs for Deep Binary Descriptor Learning in Medical Volume Data,” Informatik aktuell, pp. 23–28, 2018. | CNN | DIRLAB – 3D CT (Lung) | Retrieval Rate | Using CNN as a unimodal similarity measure to guide conventional iterative approaches |
| Liu et al. (2018) | X. Liu, D. Jiang, M. Wang, and Z. Song, “Image synthesis-based multi-modal image registration framework by using deep fully convolutional networks,” Medical & Biological Engineering & Computing, vol. 57, no. 5, pp. 1037–1048, Dec. 2018. | CNN | BrainWeb – T1, T2 and PD MRI (Brain)<br>IXI - T1 and T2-weighted MRI (Adult Brain) | Mean TRE | Using CNN as a multimodal to unimodal similarity measure convertor to guide conventional deformable iterative approaches or deep learning. |
| Eppenhof et al. (2018) | K. A. J. Eppenhof, M. W. Lafarge, P. Moeskops, M. Veta, and J. P. W. Pluim, “Deformable image registration using convolutional | CNN | DIRLAB – 3D CT (Lung)<br>CREATIS– 3D CT (Lung) | TRE | Directly regressing the unimodal deformable |
| | neural networks," Medical Imaging 2018: Image Processing, Mar. 2018. | | | | registration parameters using a combination of TPS and CNN |
| Krebs et al. (2018) | J. Krebs, T. Mansi, B. Mailhé, N. Ayache, and H. Delingette, "Unsupervised Probabilistic Deformation Modeling for Robust Diffeomorphic Registration," Lecture Notes in Computer Science, pp. 101–109, 2018. | SAE | ACDC – MRI (Cardiac) | Dice, RMSE, MDM, MDG and Hausdorff | Directly regressing the unimodal deformable diffeomorphic registration parameters using SAEs |
| Kearney et al. (2018) | V. Kearney, S. Haaf, A. Sudhyadhom, G. Valdes, and T. D. Solberg, "An unsupervised convolutional neural network-based algorithm for deformable image registration," Physics in Medicine & Biology, vol. 63, no. 18, p. 185017, Sep. 2018. | CNN | Private – CBCT and CT (Head and Neck) | NMI, FSIM, and RMSEc | Directly regressing the unimodal deformable registration parameters using CNN |
| Sheikhjafari et al. (2018) | A. Sheikhjafari, M. Noga, K. Punithakumar, N. Ray, "Unsupervised deformable image registration with fully connected generative neural network," in proc. the 1st Conference on Medical Imaging with Deep Learning (MIDL 2018), The Netherlands, 2018. | SAE | ACDC – MRI (Cardiac) | Dice | Using SAEs to produce a low-dimensional vector from image and feed them to a optimizer and fully-connected network for final registration |
| Sun et al. (2018) | Y. Sun, A. Moelker, W. J. Niessen, and T. van Walsum, "Towards Robust CT-Ultrasound Registration Using Deep Learning Methods," Lecture Notes in Computer Science, pp. 43–51, 2018. | CNN | Private – CT and US (Liver) | MAE | Directly regressing the multimodal deformable registration parameters using CNN |
| Sun and Zhang (2018) | L. Sun and S. Zhang, "Deformable MRI-Ultrasound Registration Using 3D Convolutional Neural Network," Lecture Notes in Computer Science, pp. 152–158, 2018. | CNN | RESECT – T1 and T2 MRI and US (Brain) | TRE | Directly regressing the multimodal deformable registration parameters using CNN |
| Cao et al. (2018) | X. Cao, J. Yang, L. Wang, Z. Xue, Q. Wang, and D. Shen, "Deep Learning Based Inter-modality Image Registration Supervised by Intra-modality Similarity," Lecture Notes in Computer Science, pp. 55–63, 2018. | CNN (U-Net) (Ronneberger et al. 2015) | Private – CT and MRI (Prostate) | Dice and ASD | Directly regressing the multimodal deformable registration parameters using CNN |
| Cao et al. (2018) | X. Cao, J. Yang, J. Zhang, Q. Wang, P.-T. Yap, and D. Shen, "Deformable Image Registration Using a Cue-Aware Deep Regression Network," IEEE Transactions on Biomedical Engineering, vol. 65, no. 9, pp. 1900–1911, Sep. 2018. | CNN | ANDI - MRI (Brain)<br>LONI - MRI (Brain)<br>IXI - MRI (Brain) | Dice and ASD | Directly regressing the unimodal deformable registration parameters using CNN |
| Ferrante et al. (2018) | E. Ferrante, O. Oktay, B. Glocker, and D. H. Milone, "On the Adaptability of Unsupervised CNN-Based Deformable Image Registration to Unseen Image Domains," Lecture Notes in Computer Science, pp. 294–302, 2018. | CNN | Sunnybrook Cardiac Data – 3D MRI (Cardiac)<br>JSRT – X-Ray (Chest) | Dice, MAD, and MCD | Using transfer-learning for zero-shot multimodal deformable image registration |
| Li and Fan (2018) | H. Li and Y. Fan, "Non-rigid image registration using self-supervised fully convolutional networks without training data," 2018 IEEE 15th International Symposium on Biomedical Imaging (ISBI 2018), Apr. 2018. | CNN | LONI LPBA40 - MRI (Brain)<br>ANDI – MRI (Brain) | Dice | Deformable unimodal image registration using a self-supervised CNN (without training data) |
| Fan et al. (2018) | J. Fan, X. Cao, Z. Xue, P.-T. Yap, and D. Shen, "Adversarial Similarity Network for Evaluating Image Alignment in Deep Learning Based Registration," Lecture Notes in Computer Science, pp. 739–746, 2018. | GAN | LPBA40 | IBSR18<br>CUMC12 | MGH10<br>MRI (Brain) | Dice | Directly regressing the unimodal deformable registration parameters using GAN |
| Zhu et al. (2018) | X. Zhu, M. Ding, T. Huang, X. Jin, and X. Zhang, "PCANet-Based Structural Representation for Nonrigid Multimodal Medical Image Registration," Sensors, vol. 18, no. 5, p. 1477, May 2018. | CNN (PCANet) (Chan et al. 2015) | BrainWeb – T1, T2 and PD MRI (Brain)<br>AANLib – MRI (Brain)<br>RIRE – CT and MRI (Brain) | TRE | Using PCANet to automatically extract structural features and feed them to L-BFGS for final deformable registration |
| Sloan et al. (2018) | J. M. Sloan, K. A. Goatman, and J. P. Siebert, "Learning Rigid Image Registration - Utilizing Convolutional Neural Networks for Medical Image Registration," Proceedings of the 11th International Joint Conference on Biomedical Engineering Systems and Technologies, 2018. | CNN | OASIS – MRI (Brain)<br>IXI - T1 and T2-weighted MRI (Adult Brain)<br>ISLES2015 – MRI (Brain) | MSE | Directly regressing the multimodal rigid registration parameters using CNN |
| Schaffert et al. (2019) | R. Schaffert, J. Wang, P. Fischer, A. Borsdorf, and A. Maier, "Metric-Driven Learning of Correspondence Weighting for 2-D/3-D Image Registration," Lecture Notes in Computer Science, pp. 140–152, 2019. | CNN (PointNet) (Qi et al. 2017) | C-arm – CT (Spine) | Mean TRE, Mean RPD, Success Rate and Capture Range | Directly regressing the unimodal rigid registration parameters using CNN |
| de Vos et al. (2019) | B. D. de Vos, F. F. Berendsen, M. A. Viergever, H. Sokooti, M. Staring, and I. Išgum, "A deep learning framework for unsupervised affine and deformable image registration," Medical Image Analysis, vol. 52, pp. 128–143, Feb. 2019. | CNN | NLST - 3D CT (Chest)<br>Sunnybrook Cardiac Data – 3D MRI (Cardiac) | Dice, Hausdorff, and ASSD | Directly regressing the Affine and deformable registration parameters using CNN |
| Zhu et al (2019) | N. Zhu, M. Najafi, B. Han, S. Hancock, and D. Hristov, "Feasibility of Image Registration for Ultrasound-Guided Prostate Radiotherapy Based on Similarity Measurement by a Convolutional Neural Network," Technology in Cancer Research & Treatment, vol. 18, pp. 153303381882196, Jan. 2019. | CNN | Private – 3D US (Prostate) | Registration Error | Using CNN as a unimodal similarity measure to guide conventional rigid patch-based approaches |
| Blendowski and Heinrich (2019) | M. Blendowski and M. P. Heinrich, "Combining MRF-based deformable registration and deep binary 3D-CNN descriptors for large lung motion estimation in COPD patients," International Journal of Computer Assisted Radiology and Surgery, vol. 14, no. 1, pp. 43–52, 2019. | CNN | DIRLAB – 3D CT (Lung) | TRE | Using CNN as a unimodal similarity measure to guide conventional deformable iterative approaches |
| Haskins et al. (2019) | G. Haskins, J. Kruecker, U. Kruger, S. Xu, P. A. Pinto, B. J. Wood, and P. Yan, "Learning deep similarity metric for 3D MR–TRUS image registration," International Journal of Computer Assisted Radiology and Surgery, vol. 14, no. 3, pp. 417–425, 2019. | CNN | NIH – MRI and TRUS (Prostate) | TRE | Using CNN as a multimodal similarity measure to guide conventional iterative approaches |
| Sun et al. (2019) | S. Sun, J. Hu, M. Yao, J. Hu, X. Yang, Q. Song, and X. Wu, “Robust Multimodal Image Registration Using Deep Recurrent Reinforcement Learning,” Lecture Notes in Computer Science, pp. 511–526, 2019. | RNN and CNN | Private – MRI-CT (Nasopharynx) | TRE | Directly regressing the rigid multimodal registration parameters using the combination of CNN and RNN (LSTM) |
| Salehi et al. (2019) | S. S. Mohseni Salehi, S. Khan, D. Erdogmus, and A. Gholipour, “Real-Time Deep Pose Estimation With Geodesic Loss for Image-to-Template Rigid Registration,” IEEE Transactions on Medical Imaging, vol. 38, no. 2, pp. 470–481, Feb. 2019. | CNN | Private – T1 and T2 MRI (Newborn Brain)<br>Private – T1 and T2 MRI (Fetus Brain) | Registration Error (Degree) | Directly regressing the multimodal deformable registration parameters (pose) using CNN |
| Fan et al. (2019) | J. Fan, X. Cao, P.-T. Yap, and D. Shen, “BIRNet: Brain image registration using dual-supervised fully convolutional networks,” Medical Image Analysis, vol. 54, pp. 193–206, May 2019. | CNN | LONI LPBA40 - MRI (Brain)<br>IBSR18- MRI (Brain)<br>CUMC12- MRI (Brain)<br>MGH10- MRI (Brain)<br>IXI - MRI (Brain) | Dice | Directly regressing the unimodal deformable diffeomorphic registration parameters using a dual-supervised CNN |
| Krebs et al. (2019) | J. Krebs, H. e Delingette, B. Mailhe, N. Ayache, and T. Mansi, “Learning a Probabilistic Model for Diffeomorphic Registration,” IEEE Transactions on Medical Imaging, Early Access, pp. 1–12, 2019. | SAEs | ACDC – MRI (Cardiac) | Dice, RMSE, MDM, MDG, Hausdorff and Grad Det-Jac | Directly regressing the unimodal deformable diffeomorphic registration parameters using SAEs |
| Balakrishnan et al. (2019) | G. Balakrishnan, A. Zhao, M. R. Sabuncu, J. Guttag, and A. V. Dalca, “VoxelMorph: A Learning Framework for Deformable Medical Image Registration,” IEEE Transactions on Medical Imaging, Early Access, pp. 1–13, 2019. | CNN | Buckner40, OASIS<br>ABIDE, ADHD200<br>MCIC, PPMI<br>HABS, Harvard GSP<br>MRI (Brain) | Dice | Directly regressing the unimodal deformable registration parameters using CNN with unsupervised learning |
| Elmahdy et al. (2019) | M. S. Elmahdy, T. Jagt, R. T. Zinkstok, Y. Qiao, R. Shahzad, H. Sokooti, S. Yousefi, L. Incrocci, C. A. M. Marijnen, M. Hoogeman, and M. Staring, “Robust contour propagation using deep learning and image registration for online adaptive proton therapy of prostate cancer,” Medical Physics, May 2019. | CNN and GAN | LUMC – CT (Prostate)<br>EMC– CT (Prostate)<br>HMC– CT (Prostate) | Dice, MSD, and Hausdorff | Directly regressing the unimodal deformable diffeomorphic registration parameters using a combination of CNN and GAN |
| Yu et al. (2019) | H. Yu, X. Zhou, H. Jiang, H. Kang, Z. Wang, T. Hara, and H. Fujita, “Learning 3D non-rigid deformation based on an unsupervised deep learning for PET/CT image registration,” in Medical Imaging 2019: Biomedical Applications in Molecular, Structural, and Functional Imaging, Mar. 2019. | CNN | Private – PET and CT (Body) | NCC and MI | Directly regressing the multimodal deformable registration parameters using CNN |
| Che et al. (2019) | T. Che, Y. Zheng, X. Sui, Y. Jiang, J. Cong, W. Jiao, and B. Zhao, “DGR-Net: Deep Groupwise Registration of Multispectral Images,” Information Processing in Medical Imaging, pp. 706–717, 2019. | CNN (U-Net) (Ronneberger et al. 2015) | Annidis RHA – MSI (Retina) | Dice, Ratio of Registration, and TRE | Directly regressing the multimodal deformable registration parameters using CNN with unsupervised learning |
| Van Kranen et al. (2019) | S. R. Van Kranen, T. Kanehira, R. Rozendaal, and J. Sonke, “Unsupervised deep learning for fast and accurate CBCT to CT deformable image registration,” Radiotherapy and Oncology, vol. 133, pp. S267–S268, Apr. 2019. | CNN | Private – CBCT and CT (Head and Neck) | Accuracy | Directly regressing the unimodal deformable registration parameters using CNN |
| Che et al. (2019) | T. Che, Y. Zheng, J. Cong, Y. Jiang, Y. Niu, W. Jiao, B. Zhao, and Y. Ding, “Deep Group-Wise Registration for Multi-Spectral Images From Fundus Images,” IEEE Access, vol. 7, pp. 27650–27661, 2019. | CNN (U-Net) (Ronneberger et al. 2015) | Annidis RHA – MSI (Retina) | Dice, Ratio of Registration, and CPD | Directly regressing the multimodal deformable registration parameters using CNN |
| Hering and Heldmann (2019) | A. Hering and S. Heldmann, “Unsupervised learning for large motion thoracic CT follow-up registration,” Medical Imaging 2019: Image Processing, Mar. 2019. | CNN | Private – CT (Lung) | Dice | Directly regressing the unimodal deformable registration parameters using CNN |
| Hering et al. (2019) | A. Hering, S. Kuckertz, S. Heldmann, and M. P. Heinrich, “Enhancing Label-Driven Deep Deformable Image Registration with Local Distance Metrics for State-of-the-Art Cardiac Motion Tracking,” Bildverarbeitung für die Medizin 2019, pp. 309–314, 2019 | CNN (U-Net) (Ronneberger et al. 2015) | ACDC – MRI (Cardiac) | Dice | Directly regressing the unimodal deformable registration parameters using CNN |
| Liu et al. (2019) | C. Liu, L. Ma, Z. Lu, X. Jin, and J. Xu, “Multimodal medical image registration via common representations learning and differentiable geometric constraints,” Electronics Letters, vol. 55, no. 6, pp. 316–318, Mar. 2019. | CNN (Xception) (Chollet 2017) | APCH - DRR and DR (Body) | Success Rate | Directly regressing the multimodal deformable registration parameters using differentiable geometric constraints and CNN (incorporating background knowledge to CNN) |
| Foote et al. (2019) | M. D. Foote, B. E. Zimmerman, A. Sawant, and S. C. Joshi, “Real-Time 2D-3D Deformable Registration with Deep Learning and Application to Lung Radiotherapy Targeting,” Information Processing in Medical Imaging, pp. 265–276, 2019. | CNN (DenseNet) (Gao et al. 2017) | RCCT – 4D CT (Lung) | Distance Error | Directly regressing the unimodal deformable registration parameters using CNN |
| | | CNN | ANDI, LPBA40 | | |

| | | | | | | |
|---|---|---|---|---|---|---|
| Duan et al. (2019) | L. Duan, G. Yuan, L. Gong, T. Fu, X. Yang, X. Chen, and J. Zheng, “Adversarial learning for deformable registration of brain MR image using a multi-scale fully convolutional network,” Biomedical Signal Processing and Control, vol. 53, p. 101562, Aug. 2019. | (U-Net) | IBSR18<br>MGH10<br>MRI (Brain) | CUMC12 | Dice and Distance Error | Directly regressing the unimodal deformable registration parameters using CNN |

## 3. Taxonomy of Image Registration

Typically, a good image registration needs selecting proper features, a similarity metric (to assess the quality), a transformation model, and a search strategy in the state space. So far, a large number of conventional medical image registration methods have been proposed in the literature which can be classified based on different metrics; A popular yet still-alive taxonomy is represented in Fig. 2, where the classification is based on the image dimension (2D, 3D, 4D, etc.), modality, source of the features (intrinsic vs. extrinsic), transformation domain, transformation model (rigid, affine, and deformable), kind of the fusion (interpolation vs. approximation), user interaction (manual, semiautomatic, and automatic), and parameter investigation method (iterative vs. direct) (Maints & Viergever, 1998).

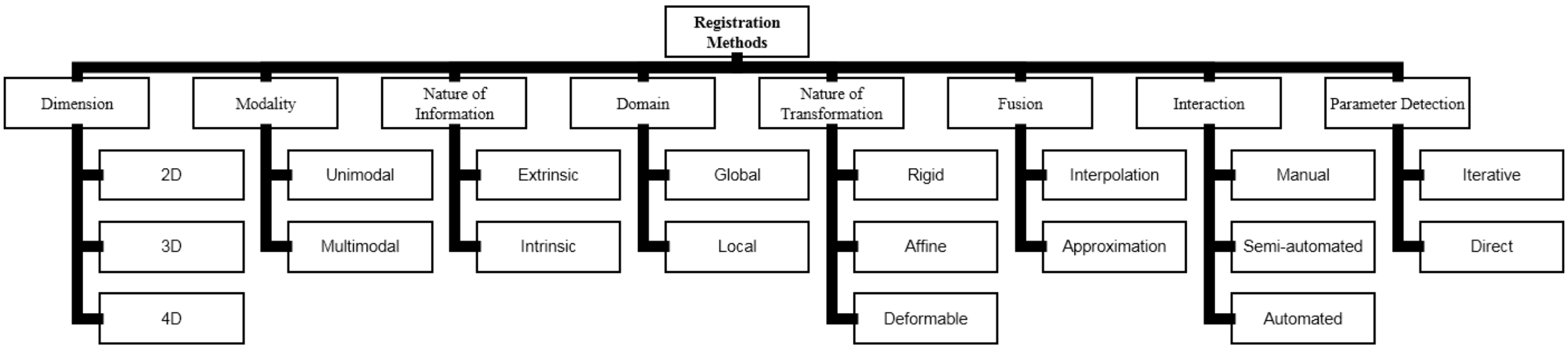

**Figure 2:** A taxonomy on image registration methods (Maints & Viergever, 1998)

At the beginning in machine vision field, image registration was referred to align 2D images; however, medical imaging as illustrated in Fig. 3 is 3D in nature so that most of the medical images captured by current imaging devices like MRI and CT are 3-dimentional. Conventional 2D imaging modalities like mammography of X-ray can also be represented as 3D without losing the generality. Besides, there are plenty of procedures in which a number of discrete (Fluoroscopy) or continuous (Sonography) images are taken by the physicians so that the transformation needs to be considered as 4D, adding time as the fourth dimension. These interventional 4D images are often in low-quality with much amount of noise due to the restrictive nature of the operational devices, which also challenges the registration process.

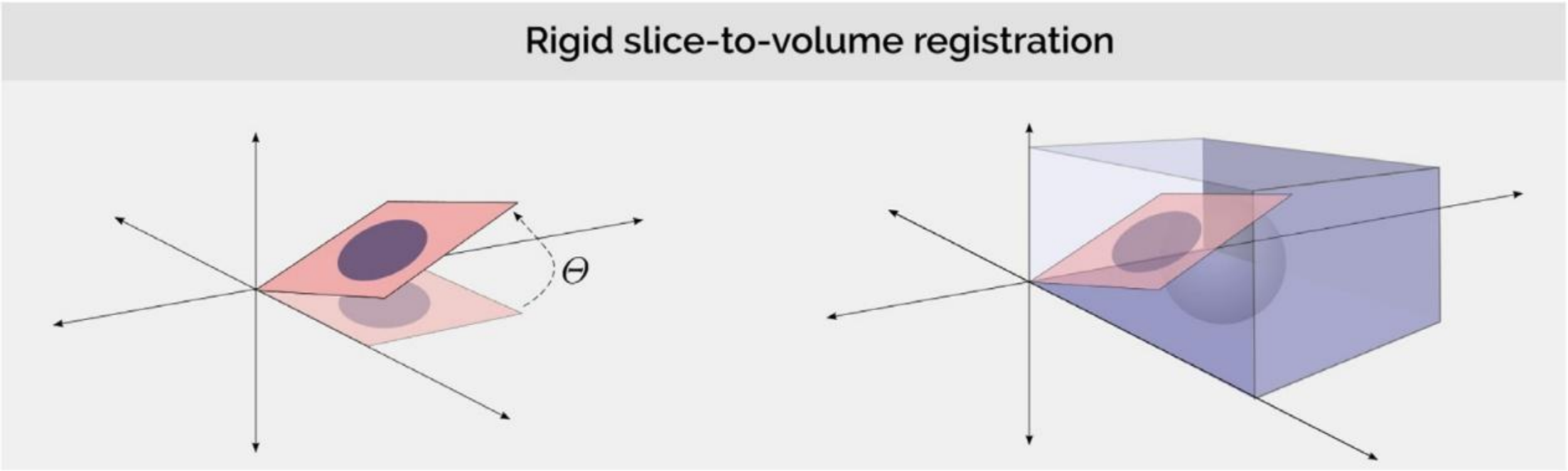

**Figure 3:** An example of registering a 2D plane in a 3D volume (Ferrante & Paragious, 2017)

The pre-interventional image on which the treatment plan is developed is routinely a 3D high-quality MRI or CT image since it is taken by modern diagnosis devices with no operational limitation, while intra-operational images taken for treatment and therapy purposes can be captured with the same (unimodal) or different modalities (multimodal). A common case is to register a pre-interventional 3D T1-weighted MRI with operational Transrectal Ultrasound (TRUS) images of the prostate gland. Often, same modalities with different imaging parameters e.g. T1 and T2 weighted MRI, as illustrated in Fig. 4., are considered as multimodal. Of course, considering different modalities, multimodal image registration is far much complicated since each modality is sensitive to some specific parameters varying in the different body's tissues to generate contrast while may not be detectible in other modalities. Moreover, some similarity metrics e.g. Sum of Squared Differences (SSD) cannot be applied to multimodal registration regarding different colors or intensities' range of points for different modalities.

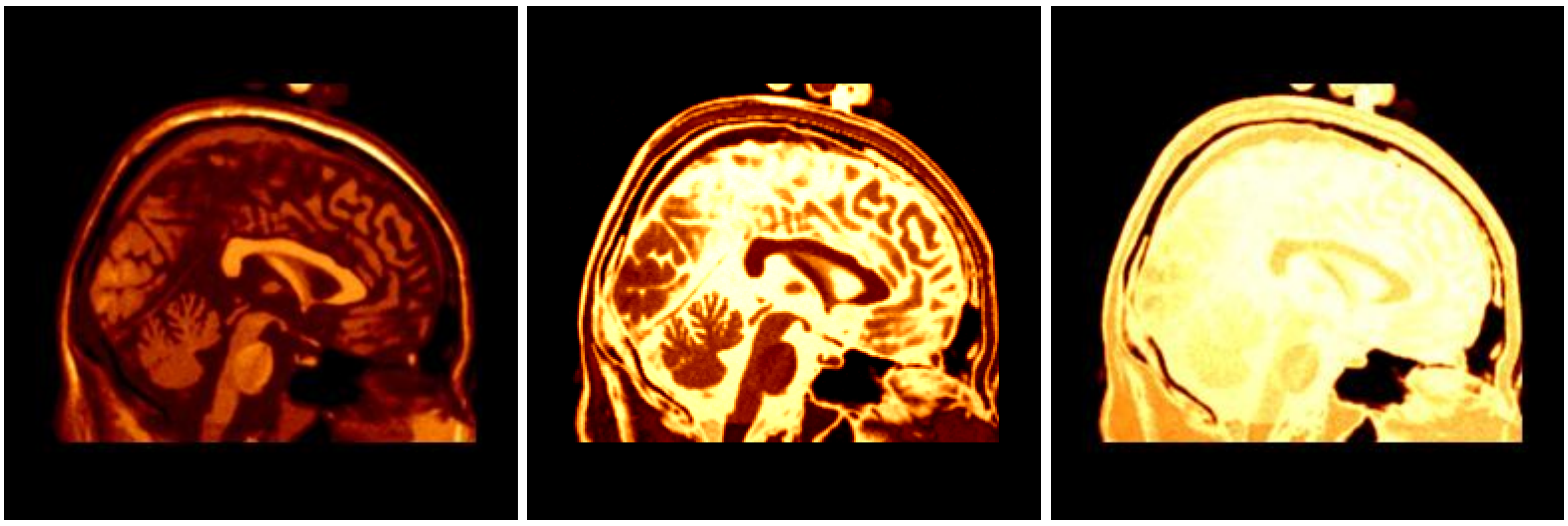

**Figure 4:** Sagittal images of human brain: T1-weighted, T2-weighted, and Proton-Density MRI from left to right, respectively (Images from BrainWeb project)

Medical image registration can be intrinsic or extrinsic based on the nature of extracted features from the images. Although intrinsic registration that is based on the anatomical structures in the body is by far much common, extrinsic registration based on the external objects implanted in the body is still alive with numerous fans specially for skin registration. These objects, like what can be seen in the Fig. 5, can be implanted in the body for different proposes, yet are good indicators to verify the quality of registration. Besides, they may be reflective tiny objects or sensors to facilitate and accurate the registration process. Extrinsic registration methods simply use the locations of these objects and can identify the transformation parameters very simple, accurate and fast.

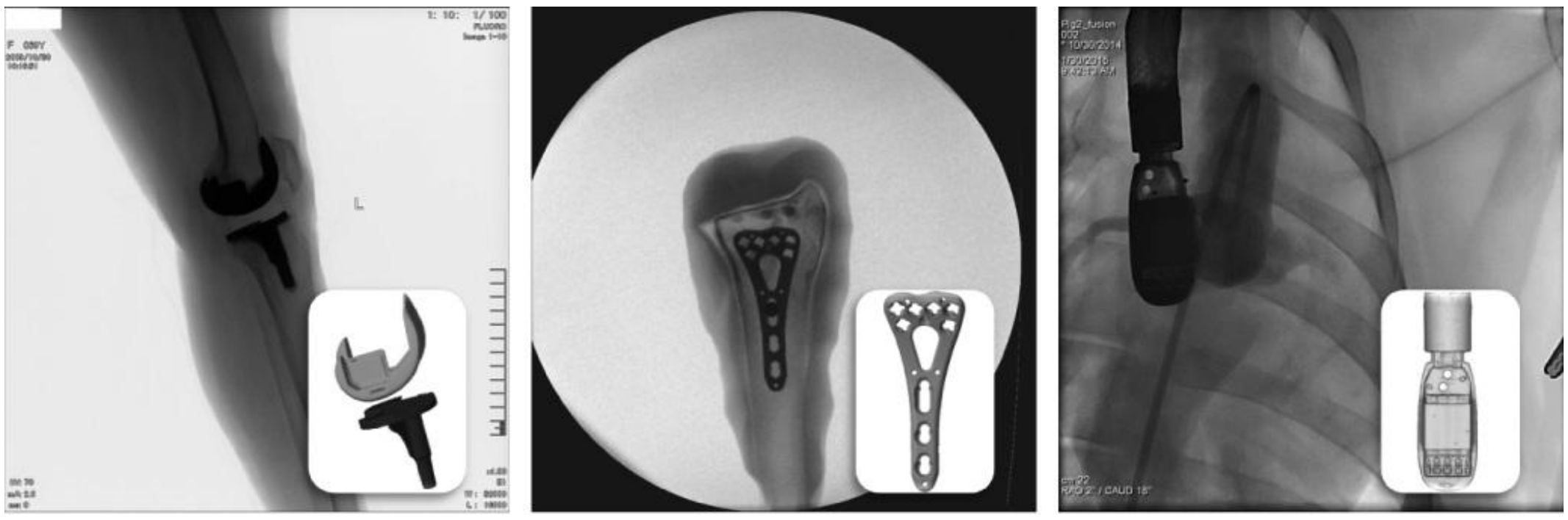

**Figure 5:** An example of implanted external objects used for extrinsic registration (Miao et al. 2016)

Another parameter to classify registration methods is the part of image evolved in the registration process. A method is considered as global where all the image's points are used in the process, and can be subjected to the prospective manipulation, while other approaches use only a part/some parts of the image (at each iteration) are referred to as local approaches. Intensity-based methods are usually global, while feature-based ones are typically local. Also, numerous hybrid approaches reported in the literature have adopted multistage strategy; that is, a global pre-registration is made at the beginning followed by much locally registrations at each stage.

Capability to consider deformation is another decisive factor to categorize image registration methods specially in the field of medicine. From this point of view, the underlying transformation of image registration methods are divided into the three kinds of rigid, Affine and deformable. The rigid transformation model only considers translation and rotation of image along and around the coordinate axes, respectively, so that the transformation can be modeled using 6 parameters (also called degrees-of-freedom). Affine transformation also considers scaling and sheering so that the number of parameters is doubled to the 12 ones. Affine transformation is nonlinear transformation that preserves points, straight lines and planes i.e. sets of parallel lines remain parallel after the Affine transformation; however, angles between lines or distances between points may not be preserved. Although Affine transformation is nonlinear, like rigid transformation, it is not able to capture the deformation complexity inherit in the flexible members and organs inside the human body. Basically, rigid and Affine transformations have no more than the two following applications:

- These approaches can be used for tough structures of the body like bones and skull registration, where there are a lots of fans among the experts and physicians because of the simplicity and speed.
- They can be exploited as a global pre-registration for much complex multistage deformable approaches in order to avoid stalking in local minima, and increase the convergence speed.

Practically, increasing in the number of transformation parameters is inevitable to model the deformable organs inside the body so that a large number of efforts have been made to introduce deformable models matched with different organs. Deformable models are almost always local, and target part of the image in each step. Usually, at the beginning, a mesh of control points is considered forming in this way a deformation network, as illustrated in Fig. 6. The number of control points (and the space among them) identifies the amount of deformation the model can capture. In each iteration, some of these control points are subjected to displacement, which cause a change in the place and intensity of other related points in the image (usually using an interpolation mechanism). In a common deformable approach named Thin-Plate Spline (TPS) (Duchon 1977), a change in a control point broadcasts to all other points in the image; this is a global approach causing high overload on the system. In contrast, B-Spline based approaches (Rueckert et al. 1999) have been introduced in which displacing a control point only affect the adjacent points, for not only increasing the locality of transformation in order to better capture the underlying deformation, but also decreasing the computational overload.

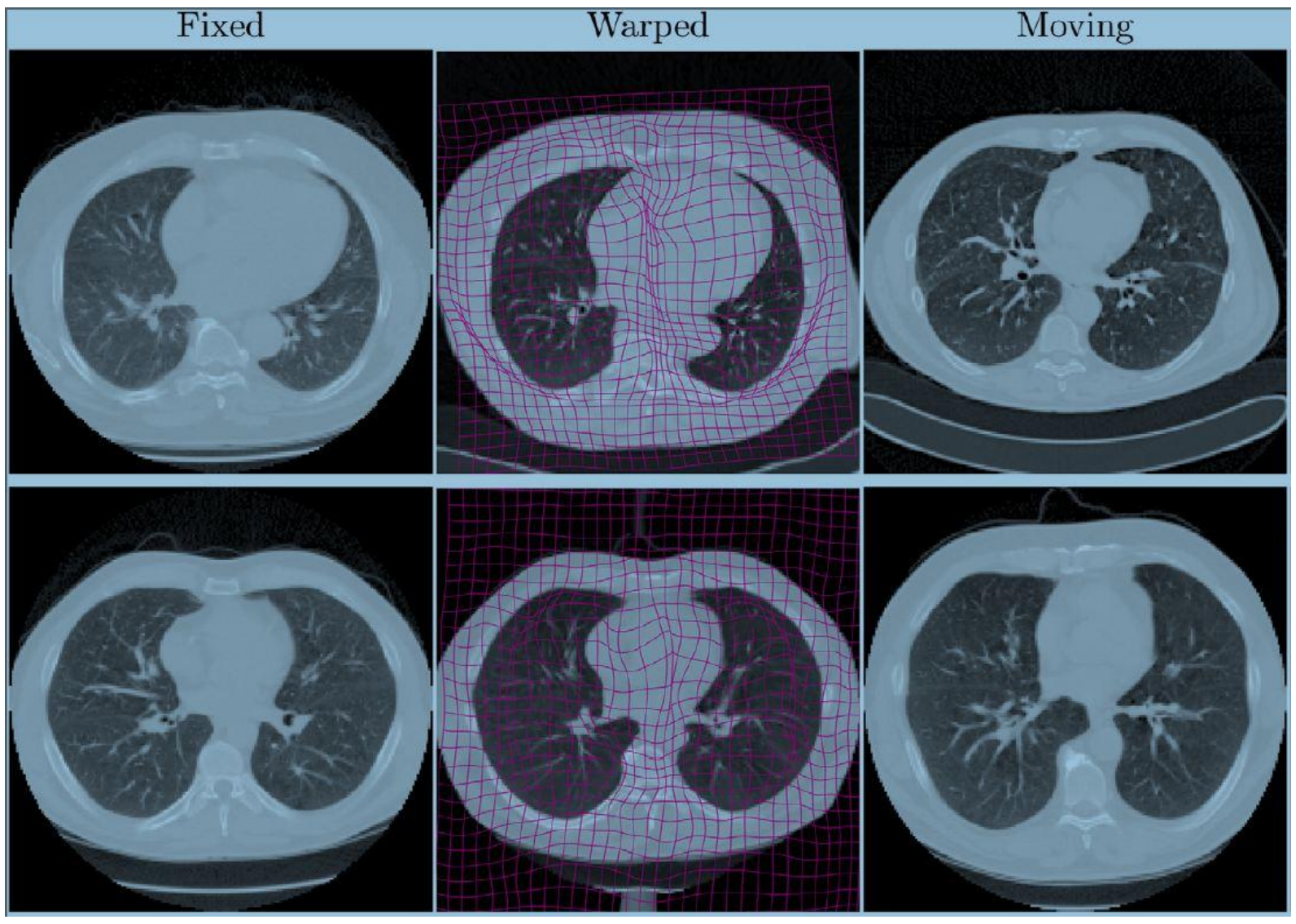

**Figure 6:** A mesh of control points as a deformation network used in deformable image registration

Generally, three levels of interaction can be considered for image registration techniques. In manual approaches, user can completely interact with the system, and may identify some parameters or provide an approximation of the transformation for the method. In contrast, there is no interaction with the machine for fully-automated approaches, and whole the process is done by the approach itself. Semi-automated approaches try to exploit the knowledge of expert user, though minimally. This minimal exploitation can be actually of high value, causing a tremendous decrease in the search space to decrease the risk of failed registration, and increase the overall speed; however, human intervention can challenge whole the process since the different levels of interactions cannot be validated, controlled or measured quantitatively.

## 4. Problem Formulation

The purpose of image registration is to find a geometrical transformation $T : I_F \rightarrow I_M$ to align a moving image $I_M$ to the fixed image $I_F$ in the best way. This alignment can be achieved conventionally by optimizing a similarity measure $L$ so that

$$\hat{\mu} = \min L(T_\mu; I_F, I_M) \tag{1}$$

where $T_\mu$ is the transformation model recognized by the parameters $\mu$. There are a number of similarity measures divided into the two categories of intensity-based and feature-based ones. Generally, intensity-based measures, e.g. Mean Square Difference (MSD), consider a complete mechanical correspondence between the given images, which may not be required from the perspective of human experts. On the other hand, feature-based measures, e.g. Mutual Information which is based on the information theory, look for a satisfying structural correspondence between the Organs-of-Interest (OoIs) in the inputted images; practically, the approaches based on the feature-based measures try to detect peers of structural features like lines, corners, landmarks, contours etc. in both the fixed and moving images, and to align them. The detection and selecting these structural features can be manually (referred to as handcrafted features), or completely automated like what is seen using deep Convolutional Neural Networks (CNNs).

Whatever the similarity measure is, the Eq. (1) needs to be optimized. The optimization process can be conducted in two ways; conventionally, the aforementioned Eq. is optimized using iterative approaches like hill-climbing and gradient decent; while in machine learning approaches, a transformation model is created based on some previously learned samples, and the parameters are regressed directly in one shot using (2).

$$\mu = f_\theta(I_F, I_M) \tag{2}$$

where $\theta$ is the set of the machine-learning model's parameters (the network's weights) identified based on the train on the learning data. In other words, the $L$ acts as a loss-function for the learning model to be optimized using some conventional training algorithms, like back-propagation, to learn $\theta$ so that the similarity between the couple of the inputted images can be maximized.

## 5. Deep Neural Networks

The theory of deep neural networks backs to the late-70s (Fukushima 1980), and was first applied to medical image processing by Lo in 1992 (Lo et al. 1992). Nevertheless, since the proper infrastructure for such a huge computation was not available those days, the first operational implementation dates to 1998 where in (Lecun et al. 1998) a Convolutional Neural Network (CNN) was utilized to recognize hand-written numerical characters for post office application. Numerical character recognition was such a simple machine vision task for which the computational power provided by the hardware of that time allows to effectively use deep learning. Unfortunately, the proposed approach lost its way to be applied to other much complex and complicated problems till 2012 where Krizhevsky et al. (2012) was prospered to train a deep CNN on graphic card equipped with a many-core GPU, and won the grand world image processing championship named ImageNet (Krizhevsky et al. 2012). Since then, other champions were in the same family of deep learning techniques, each of which with a contributing novelty, till it was announced the recognition power of the deep learning proposed approaches outperform the human expert, and the championship was practically closed (Russakovsky et al. 2015).

Accordingly, deep neural networks of various kinds penetrated to all areas of machine vision, and turned to be the dominant technique used by a large number of experts active in the field. Some active domains in the field of medical image analysis can be enumerated as organ detection, landmark localization, lesion detection and classification, treatment planning and follow-up (Litjens et al. 2017). Of course, image registration as one of the most important and challenging dilemma in image-guided intervention was not an exception, and a number of approaches as listed in Table 1 have been introduced so far. Deep neural networks can have different architectures and topologies made each of which suitable for some specific applications. As shown in Fig. 7, five kinds of deep neural networks have already been applied to medical image registration namely CNN (66 times, 80.5%), Staked Auto-Encoders (SAEs) (8 times, 9.8%), Generative Adversarial Network (GAN) (6 times. 7.3%), Recurrent Neural Network (RNN) (1 time, 1.2%) and Deep Reinforcement Learning (DRL) (1 time, 1.2%).

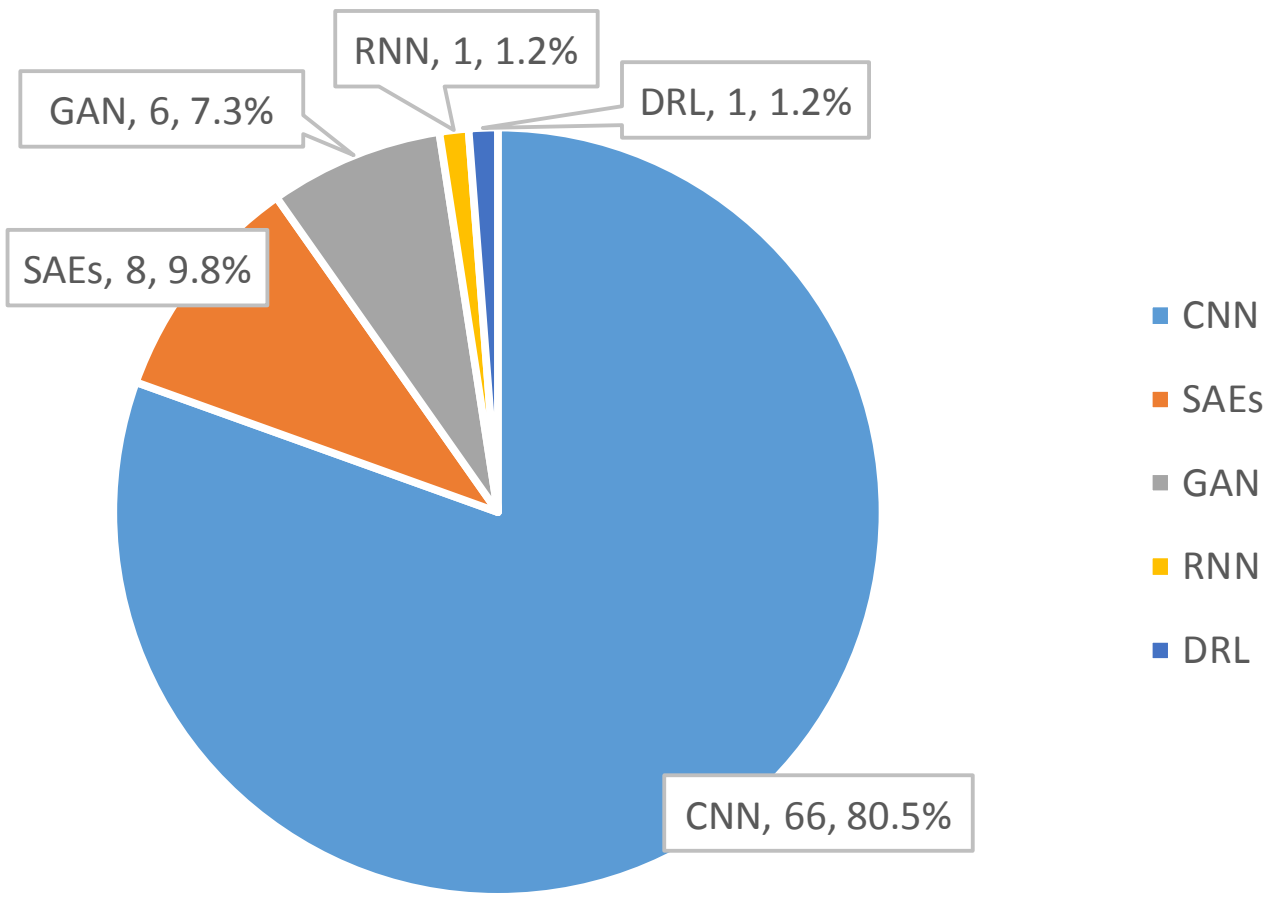

**Figure 7:** Deep learning techniques used in the literature with their frequencies

An Auto-Encoders (AEs), as illustrated in Fig. 8-(a), is a very simple network which tries to reconstruct the input pattern as output using a single hidden layer. Of course, the hidden layer should be smaller than the inputted pattern so that it maps to a compacter space of the hidden layer with the most discriminating capability. Denoising AEs (DAEs) is a resemble network trying to reconstruct the inputted patterns with some noise applied. Applying some noise to the input elevates the generalization capability of the model. Deep architecture of AEs called Staked Auto-Encoders (SAEs), as presented in Fig. 8-(b), has more hidden layers staked on the top of one other. Generally, the computational burden of training such a network is not affordable; hence, to make it practical, usually each layer is trained separately, and a final low-cost integrated training fine tunes whole the network. In the literature of medical image registration, this network has only been used to provide the most significant and discriminating features from the images to feed to an alternative registration method, instead of using handcrafted features.

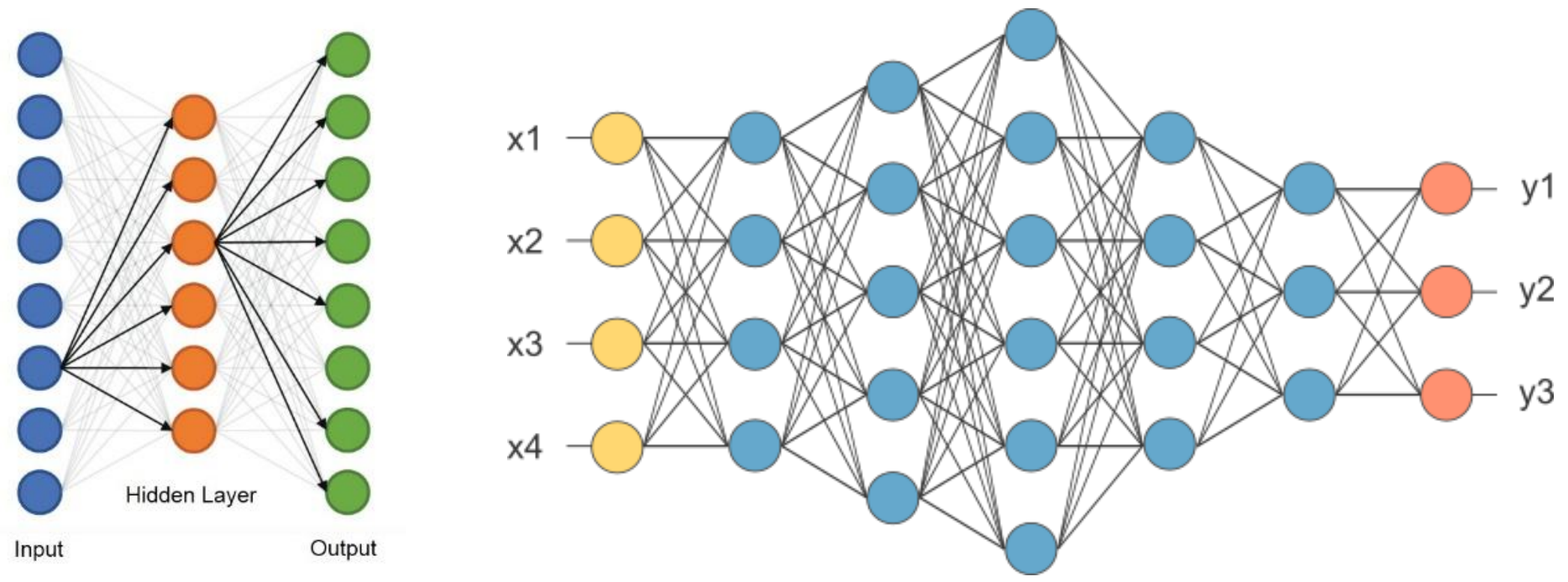

**Figure 8:** (a) Left: Auto-Encoders (AEs) network (b) Right: Staked Auto-Encoders (SAEs) network

CNNs should be considered as one of the most successful and powerful deep learning techniques in which whole the given image (or some extracted patches) is feeded directly to the network. And, this is versus the traditional neural network based image processing approaches in which some handcrafted features were extracted at first, and provided to the network. As represented in Fig. 9, a typical CNN has some interleaving kernel and pooling layers to be ended with a typical fully-connected two or three-layer network. Kernels are trained to extract the most significant features via convolving with the input, while pooling layers decrease the curse of dimensionality, and make the results invariant to the different geometrical transformations. The output of each layer so-called a feature-map is inputted to the next layer, and where the number layers is high, a hierarchical feature-set can be achieved, and the network can be regarded as deep CNN. The feature-maps of the last layer is concatenated and vectorized to feed a fully-connected two or three-layer network for the final classification. In a large number of cases, e.g. in deformable image registration, we are witnessed the so-called U-Net (Ronneberger et al. 2015) where the final fully-connected layer can be dropped out, like Fig. 10, so that a direct end-to-end registration field can be achieved.

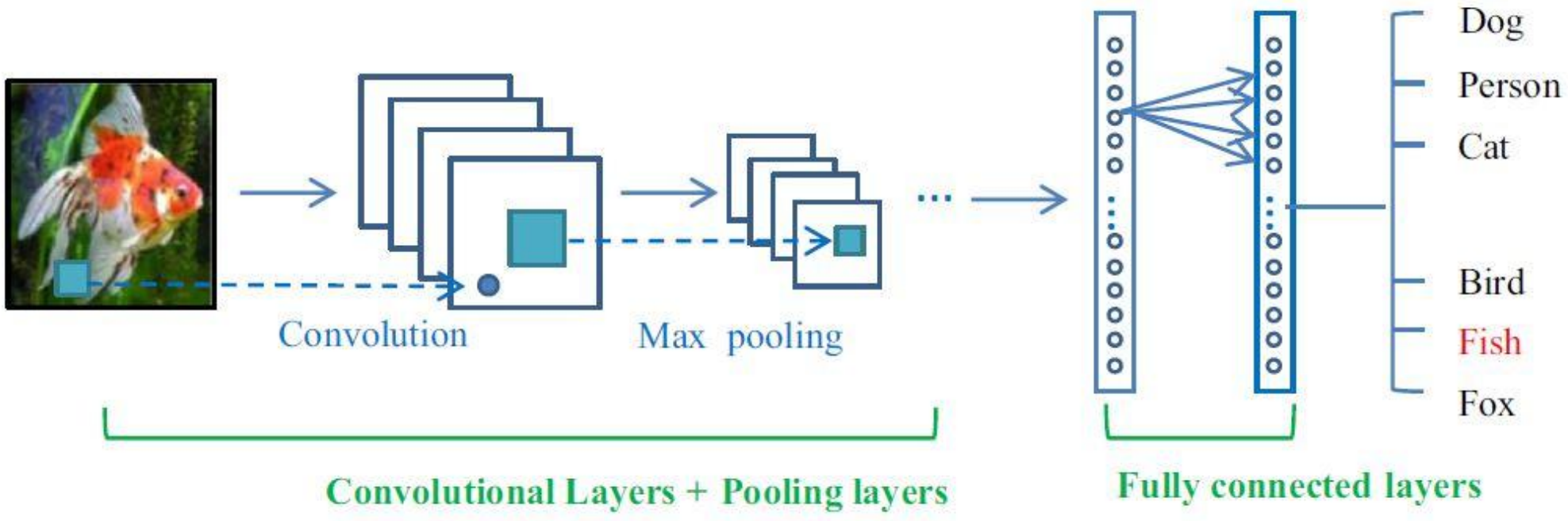

**Figure 9:** Convolutional Neural Network (CNN) architecture

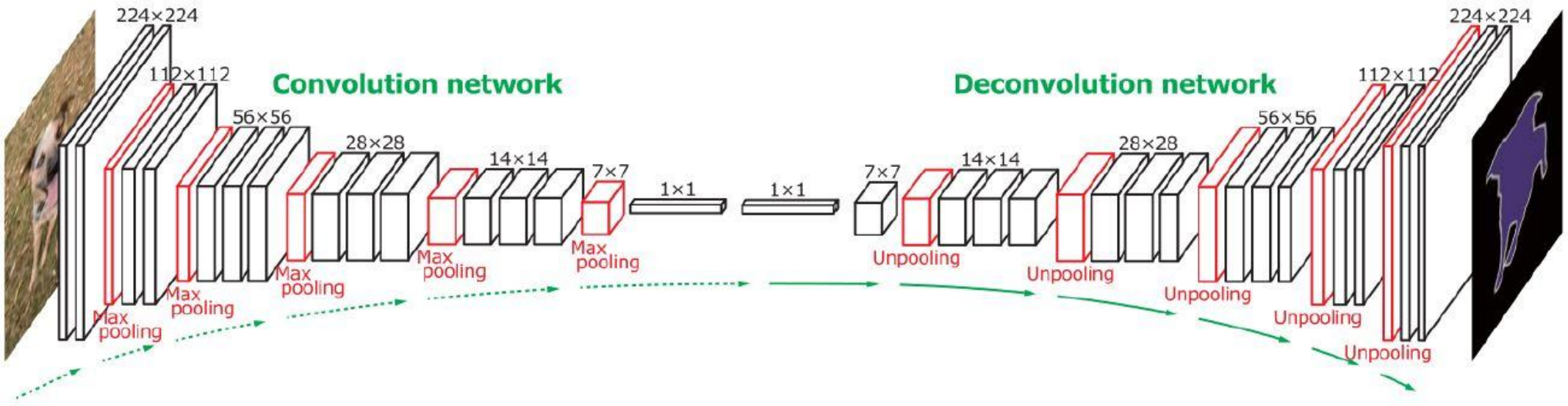

**Figure 10:** Fully Convolutional Neural Network (fCNN or U-Net) architecture

CNN has also the capability of getting heterogeneous patterns with different representations as input. Each representation is regarded as a channel, and the network having this capability is called multi-channel. To exemplify, let's consider small-sized patches extracted from images and inputted to the network for the classification issue. Should the context be informative, one can consider some larger patches around the selected patch, compact them, and feed them to the network as a separate channel. Of course, these compacted larger-sized patches cannot be processed by the network via the same channel with the original-sized patches. Another example is where we are facing color images so that are able to use three channels of RGB instead of one channel of intensity for each image's point. The network can fuse the channel in early layers, or postpone it to the last layers regarded as multi-steam network. Most of the CNNs applied to medical image registration are of this type where a couple of patches extracted from the given fixed and moving images (almost with different modalities and representations) are feeded separately as two channels, processed separately in two pathways (pipelines) where the information fusion is usually postponed to the late layers (Fig. 11).

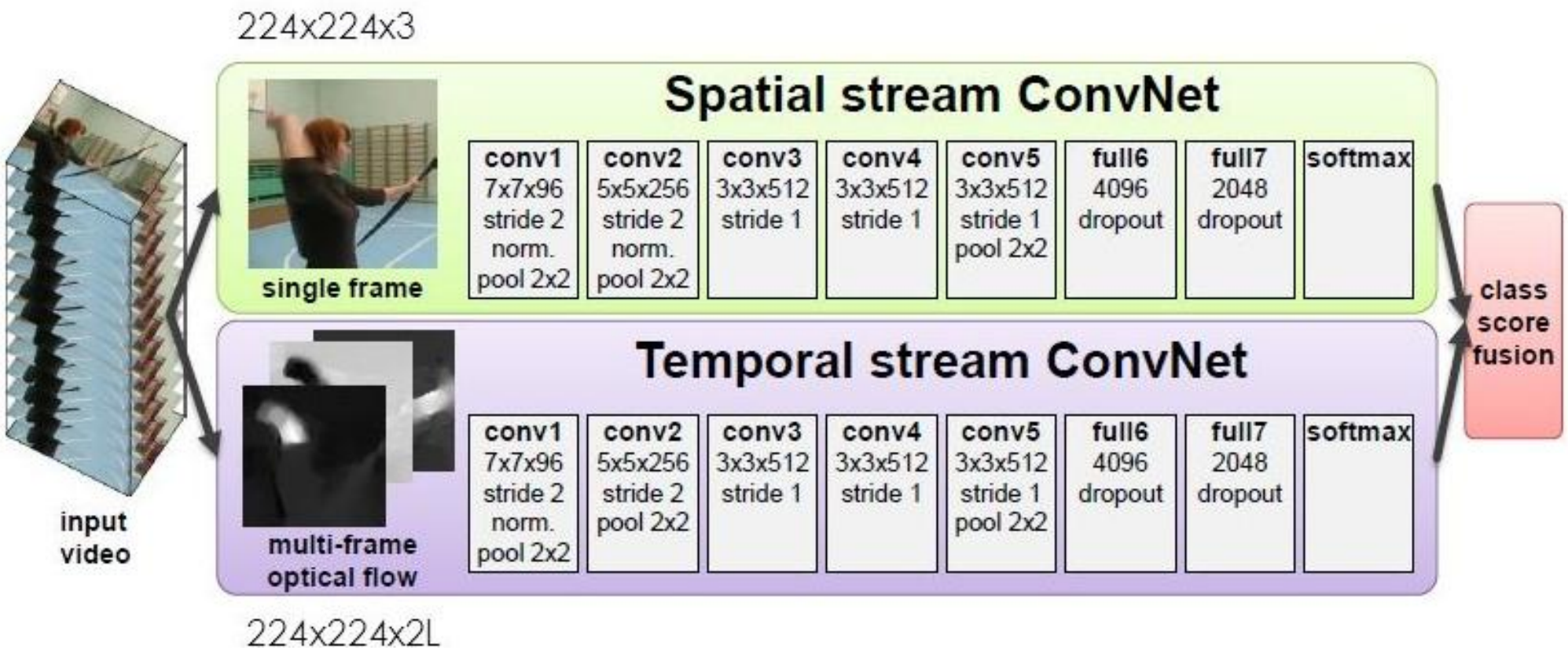

**Figure 11:** Multi-stream CNN architecture

Proposed by Goodfellow *et al.* in 2014 (Goodfellow et al. 2014), a Generative Adversarial Network (GAN) is composed of two competing subnetworks, the generator and the discriminator, as demonstrated in the Fig. 12. The generator is trained on ground-truth dataset to synthesize fake samples, while the discriminator should discriminate between fake (synthesized) data and the real one, as a binary output. Based on the survival competition between the generator and the discriminator, just like the game theory, the network can be trained on a small set of data so that the generated samples cannot be discriminated, and the network goes for equilibrium. As the generator is trained adversarially based on the discriminator's feedback, the network takes its name. While the original GAN was applied for the image noise removal, it has gained increasing popularity in recent years, and applied to almost all the problems in medical imaging (Yi et al. 2018). In the context of image registration, the generator takes the inputted fixed and moving images, and try to produce such transformation parameters so that the transformed moving image called warped image cannot be discriminated from the ground-truth by the discriminator, the situation expected from an expert registration agent.

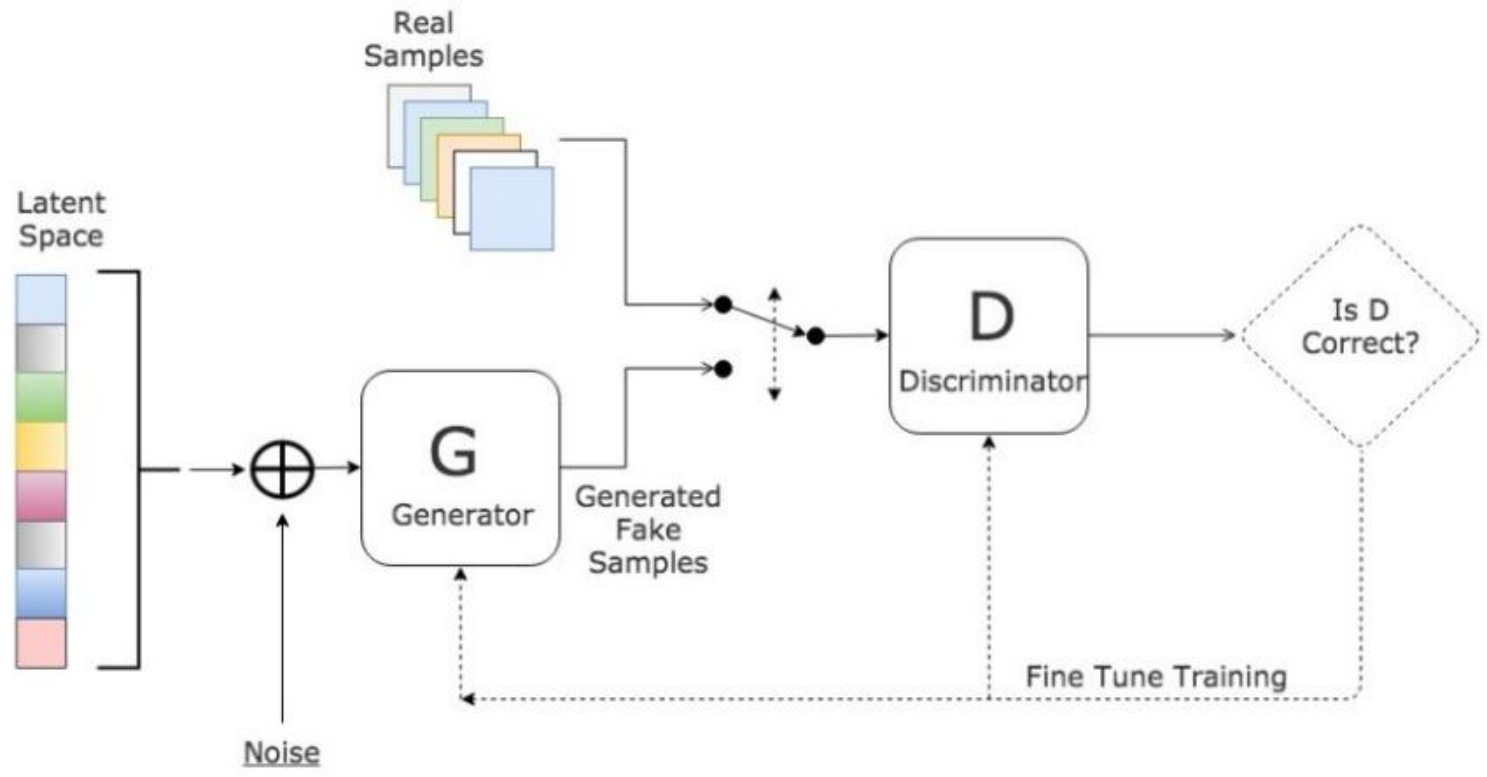

**Figure 12:** Generative Adversarial Network (GAN) architecture

Recurrent Neural Network (RNN) is just like the traditional Multi-Layer Perceptron (MLP) or AEs with an extra data loop (feedback) from the hidden-layer nodes to themselves, as illustrated in Fig. 13. While the other networks like CNNs or AEs are suitable for spatial analysis, these feedback loops make the RNNs the most powerful for temporal analysis. Routinely, the previous states are held across the hidden layers, and accordingly, the next state can be estimated using the current state inputted to the network, and the previous stored ones. Where the number of hidden layers goes more to store farer states, the network is considered as deep. Again, just like SAEs, the computational burden of training such heavily connected network is unaffordable in traditional manner; hence, research was constantly followed by the community, and fortunately simplified memory models such as Long Short Term Memory (LSTM) (Hochreiter and Schmidhuber 1997) and Gated Recurrent Units (GRD) (Cho et al. 2014) were introduced and popularly applied in the recent years. Worse mentioning that in the context of image registration, the RNNs are mostly used for optical-flow, and where one of the modalities are associated with temporal dimensionality e.g. TRUS or X-ray fluoroscopy.

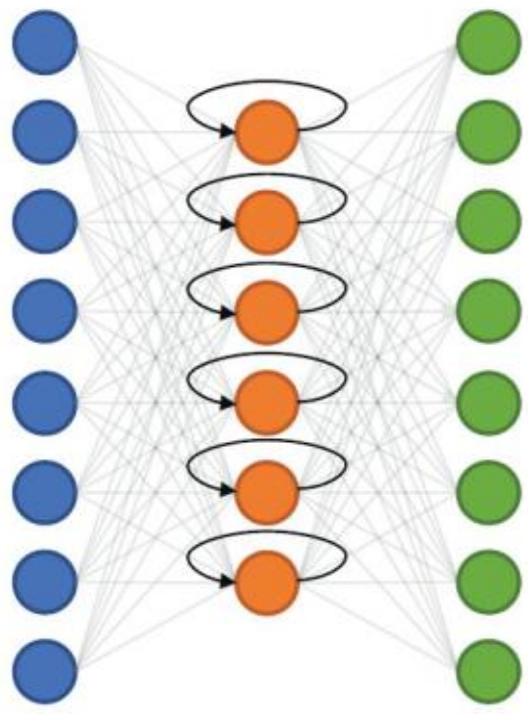

**Figure 13:** Recurrent Neural Network (RNN) architecture

Last but not least is the Deep Reinforcement Learning (DRL). It is based on the theory of stochastic processes and Markov chain. There is an agent with some internal states, transition probabilities, and a reward/penalty rate. It learns iteratively to interact with the environment. At each iteration, the DRL machine choose an action from its action-list based on the environment's feedback, its current internal states and transition probabilities via a probabilistic decision making process. The selected action is applied to the environment, and based on the desirability of its feedback, the machine gains reward or penalty. In this manner, the DRL machine learns to select the best action in each situation, where the best action is the one with the most probability to get reward from the environment. In the context of image registration, such DRL agents were applied to specifically rigid or affine transformations where the number of states are restricted, and affordable for the agent to be converged. For example, the agent can select the actions of 1 degree clockwise/counterclockwise rotation or 1 mm (millimeter) translation in all the directions. These selected actions are applied to the moving image, and based on the desirability of actions e.g. a similarity metric, the agent gets reward/penalty. It updates its internal transition probabilities based on a learning algorithms such as Q-Learning to maximize its performance. Fig. 14 is a clear illustration of this concept.

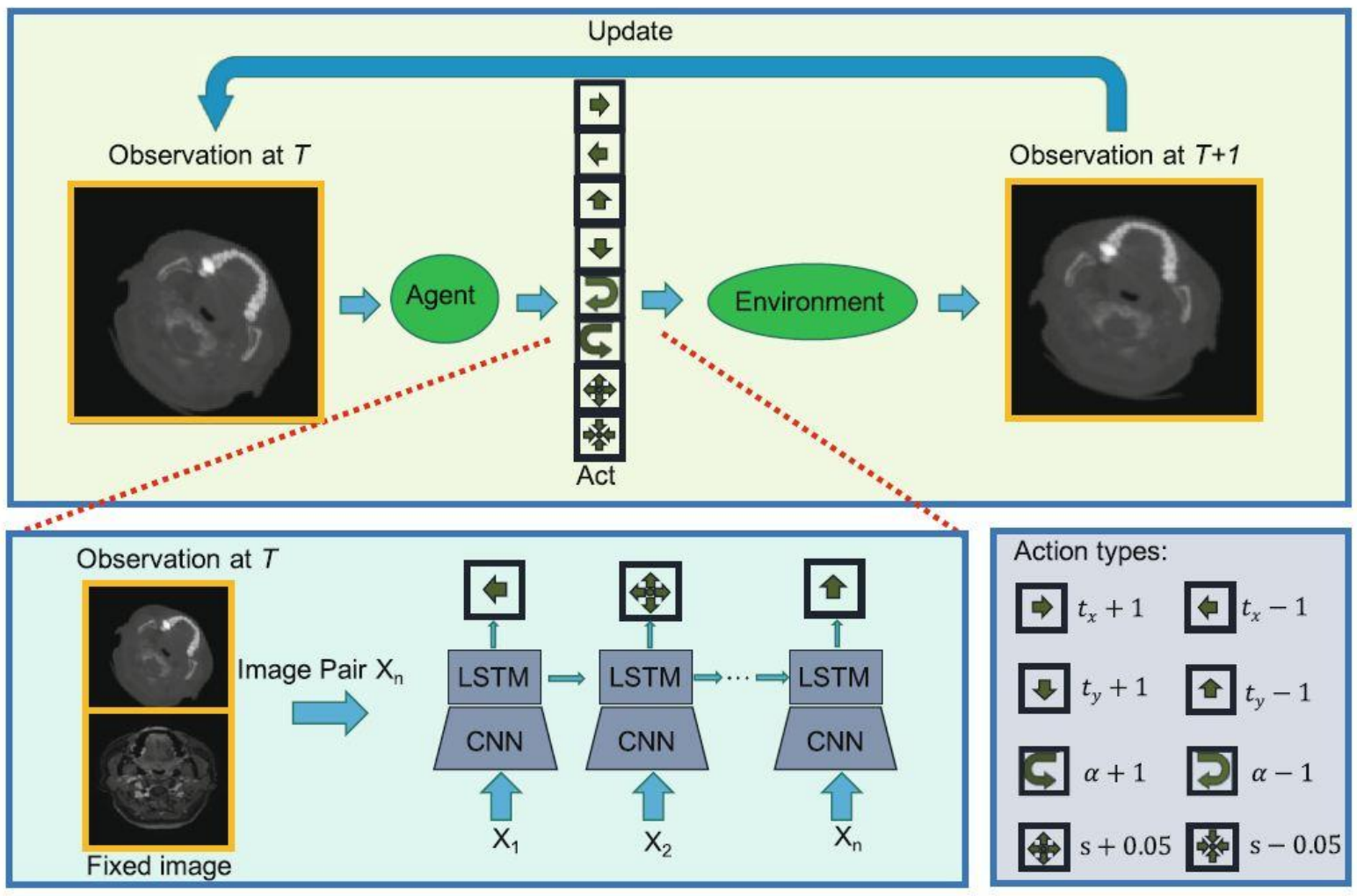

**Figure 14:** Deep Reinforcement Learning (DRL) architecture applied to medical image registration (Sun et al. 2019)

## 6. Literature Review

Conventional image registration is conducted using iterative optimization algorithms. In each iteration, a better alignment is supposed to be achieved based on a predefined similarity measure. The operations continue till no better registration can be achieved, or some predefined criteria are satisfied. Prolonged running time, and defective nature of the introduced similarity measures specially for multimodal registration, which causes getting trapped in local minima, are of most challenging issues to be tackled for exploiting this paradigm. To resolve the aforementioned problems, deep learning based approaches have gained increasing popularity in recent years. The underlying philosophy behind these approaches are divided into the two following categories:

- Deep neural network acts as an approximator of the similarity between inputted images as a complete and no-faulty similarity metric in order to help other registration methods.
- Deep neural network acts as a regressor to directly estimate the transformation parameters in one-shot in order to maximize the runtime speed.

Based on the literature breakthroughs, a taxonomy with five generations can be concluded named Deep Similarity Metrics (DSM), Supervised End-to-End Registration (SE2ER), Deep Reinforcement Learning (or Agent-Based Registration) (DRL), Unsupervised End-to-End Registration (UE2ER), Weakly/Semi-Supervised End-to-End Registration (WSE2ER) (Fig. 15). Inspired by e.g. (Nowak and Jurie 2007) and (Huang et al. 2012), the first generation of works were based on the utilization of different kinds of DNNs to learn visual similarity metrics from a large set of paired annotated ground-truths. We called them deep similarity measures/metrics. As illustrated in Fig. 16, the learned model after the train is supposed to be able to precisely and meaningfully model the structural differences between the inputted pairs of images/patches. (Wu et al. 2013) and (Cheng et al. 2016) are the most important representatives for this primary generation. Deep similarity measures are basically provided to the conventional iterative deformable registration algorithms in order to produce final transformation parameters. As a large number of similar approaches have been conducted so far, nowadays, we can definitely argue that this paradigm, in its basic form, can be a potent rival to the conventional multimodal similarity measures e.g. Mutual Information (MI) if and only if the adequate number of clearly annotated ground-truths are available, which is a severe confining factor to develop such approaches. Moreover, it has been revealed that, for unimodal registration, if the similarity measure can be properly selected based on the context and modality, the utilization of deep similarity measures has no strong justification. A list of works belong to this generation are presented in Table 2.

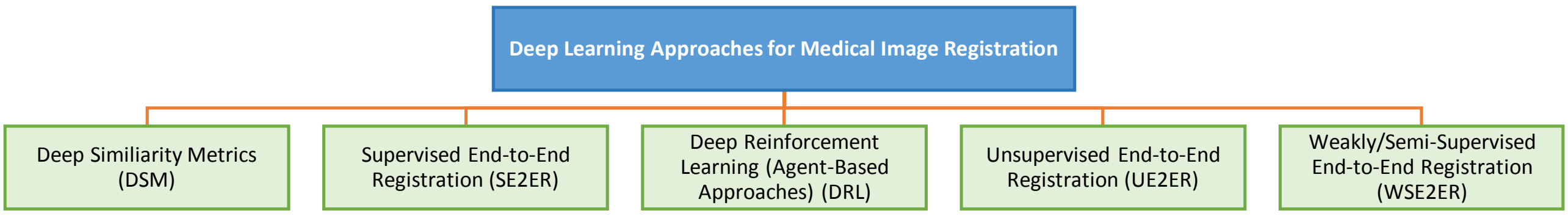

**Figure 15:** The taxonomy on deep learning approaches for medical image registration

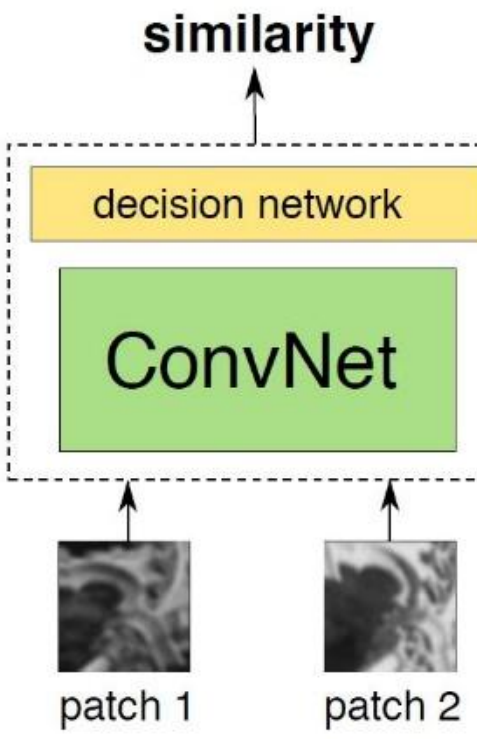

**Figure 16:** A deep similarity metric based on the CNN (Zagoruyko and Komodakis 2015)

**Table 2:** The First Generation of Deep Learning Approaches for Medical Image Registration (Deep Similarity Metrics)

| Reference | Title | Technique | Modality | Transformation |
|---|---|---|---|---|
| Wu et al. (2013) | Unsupervised deep feature learning for deformable registration of MR brain images | CNN | Multimodal | Deformable |
| Zhao et al. (2015) | Deep adaptive log-demons: diffeomorphic image registration with very large deformations | CNN | Unimodal | Deformable |
| Simonovsky et al. (2016) | A deep metric for multimodal registration | CNN | Multimodal | Deformable |
| Cheng et al. (2016) | Deep similarity learning for multimodal medical images | SAEs | Multimodal | Rigid |
| Wu et al. (2016) | Scalable high-performance image registration framework by unsupervised deep feature representations learning | SAEs | Unimodal | Deformable |
| Yang et al. (2016) | Registration of pathological images | SAEs | Unimodal | Deformable |
| Wang et al. (2017) | Scalable high performance image registration framework by unsupervised deep feature representations learning | SAEs | Unimodal | Deformable |
| Ghosal et al. (2017) | Deep deformable registration: Enhancing accuracy by fully convolutional neural net | CNN | Unimodal | Deformable |
| Awan et al. (2018) | Deep Autoencoder Features for Registration of Histology Images | CNN | Unimodal | Rigid |
| Blendowski et al. (2018) | 3D-CNNs for deep binary descriptor learning in medical volume data | CNN | Unimodal | Deformable |
| Abanovie et al. (2018) | Deep Neural Network-based Feature Descriptor for Retinal Image Registration | CNN | Unimodal | Rigid |
| Zhu et al. (2018) | PCANet-Based Structural Representation for Nonrigid Multimodal Medical Image Registration | CNN | Multimodal | Deformable |
| Liu et al. (2019) | Image synthesis-based multi-modal image registration framework by using deep fully convolutional networks | CNN | Unimodal | Deformable |
| Zhu et al. (2019) | Feasibility of Image Registration for Ultrasound-Guided Prostate Radiotherapy Based on Similarity Measurement by a Convolutional Neural Network | CNN | Unimodal | Rigid |
| Haskins et al. (2019) | Learning deep similarity metric for 3D MR-TRUS image registration | CNN | Multimodal | Deformable |
| Blendowski et al. (2019) | Combining MRF-based deformable registration and deep binary 3D-CNN descriptors for large lung motion estimation in COPD patients | CNN | Unimodal | Deformable |

The second generation belongs to the end-to-end supervised registration, where the different kinds of DNNs are trained on ground-truth to construct regression models to produce the transformation parameters in one-shot. For affine and deformable transformation models, CNN and U-Net (i.e. fully CNN) are predominant techniques, respectively (Miao et al. 2016) and (Sokooti et al. 2017). The main deformable framework is illustrated in Fig. 17. First of all, a grid of control points is considered as a Dense Displacement Field (DDR). Each control point can be freely translated in horizontal and vertical directions in order to capture the underlying deformation. The number of control points and the space among them govern the accuracy of model to capture the deformation. In the so-called Thin-Plate Spline (TPS) each movement in a control point is broadcasted to the all ones (global transformation), while so-called B-Spline approaches only considered adjacent control points (local transformation) to tract the computational overhead. Deformations are controlled by the regularizer that penalizes inacceptable transformations. Just like conventional approaches there is a heavy disputation and disagreement on the regularization term to be dictated to the DNN, which turned to be the source of many innovations in the field (Sotiras et al. 2013).

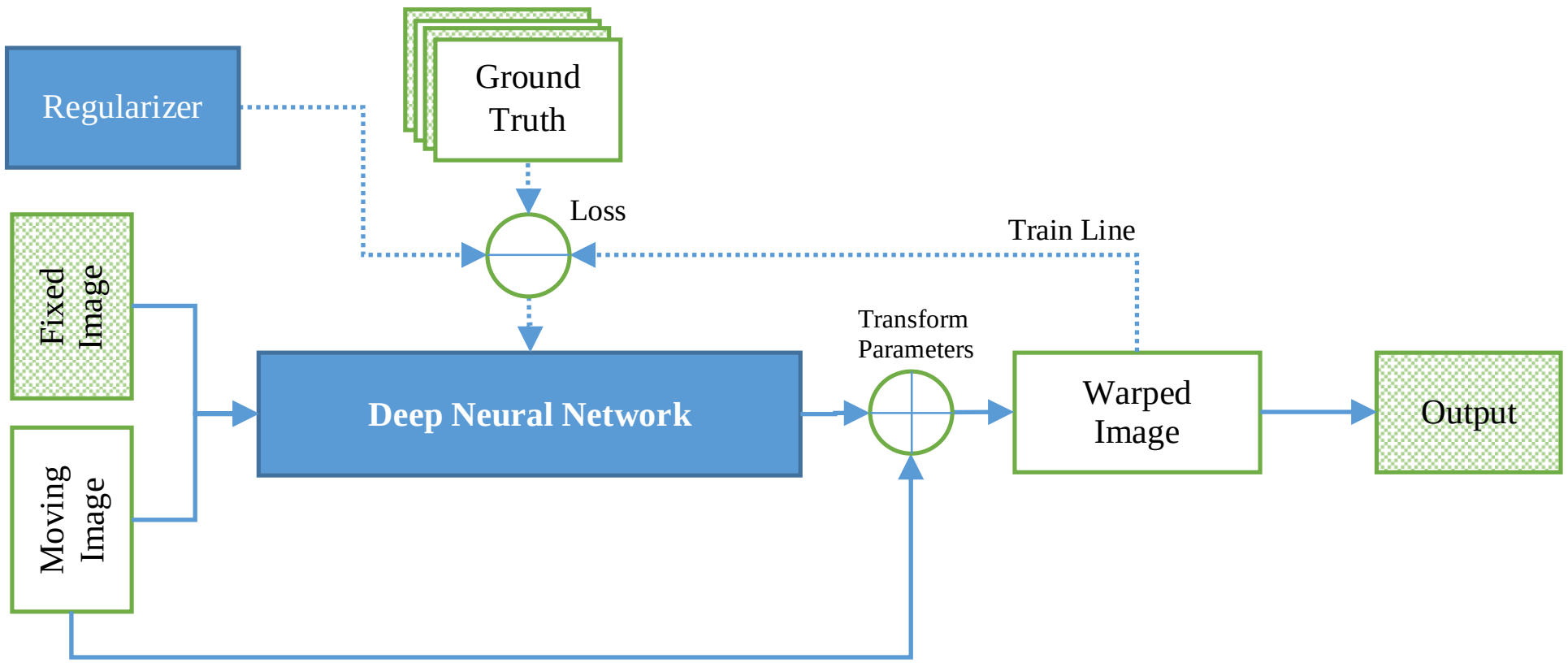

**Figure 17:** The main framework for supervised end-to-end medical image registration

Another source of innovations in this generation is belonged to the introduction of Spatial Transformer Network (STN) by Jaderberag *et al.* in 2015 (Jaderberag et al. 2015). It is an explicit module that can be injected in the different kinds of DNNs to make the flow of data across the hidden layers being transformation invariant; when situated next to and collaborated with pooling layers that are implicitly translation and scaling invariance, they can be complementary to introduce a complete set of spatial invariance whose synergic impact can drastically enhance the performance of CNNs applied to many different image processing applications, where the medical image registration cannot be an exception. The STN, as illustrated in Fig. 18, is composed of three sequential components. First, a localization network, with very flexible structure e.g. a regular MLP, which learns to regress the transformation parameters for the inputted feature-map based on a predefined similarity measure as the loss function. Second, a grid generator whose aim is at applying the estimated transformation parameters by the localization network to the inputted feature-map. Finally, a sampler that works as an interpolator to construct the final outputted warped image. Since the STN is fully differentiable, it can be inserted anywhere in the network, and its location is context-specific, and source of disagreement in the community. STN is not flaw-free; large transformations can cause severe distortion in the output, as it is not tolerable by the sampler, and boundary interpolating is also very hard for the sampler since some of the output should be bring from outside of the input, which is not exist. Recently, Lin and Lucey in 2017 (Lin and Lucey 2017) introduced Inverse Compositional STN (IC-STN), and argued that we can postpone the reconstruction by the sampler, and sending transformation parameters accompanied with the output where the CNN itself decides how to treat with the transformation. Indeed, the issue is in its infancy, and the problem is still open. A list of works belong to this generation are presented in Table 3.

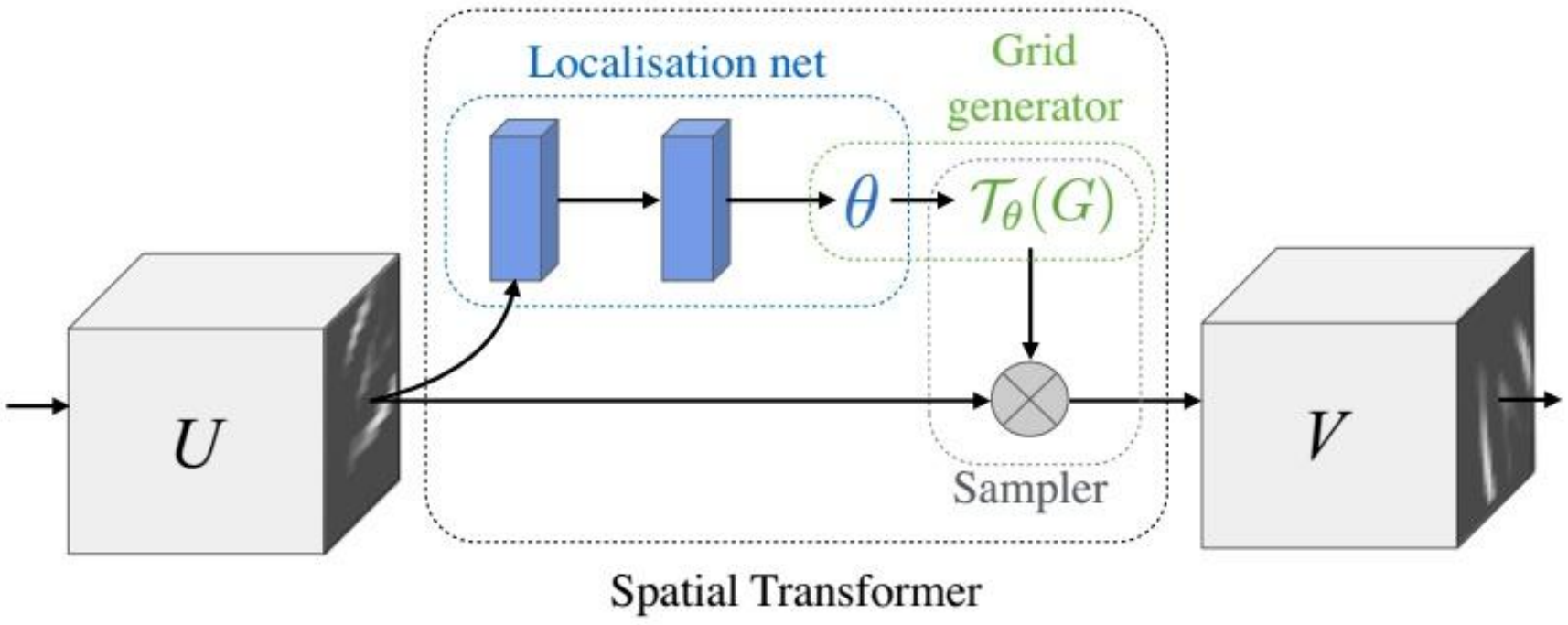

**Figure 18:** Spatial Transformer Network (STN) (Jaderberg et al. 2015)

**Table 3:** The Second Generation of Deep Learning Approaches for Medical Image Registration (Supervised End-to-End Registration)

| Reference | Title | Technique | Modality | Transformation |
|---|---|---|---|---|
| Miao et al. (2016) | A CNN regression approach for real-time 2D/3D registration | CNN | Unimodal | Rigid |
| Yang et al. (2016) | Fast predictive image registration | CNN | Unimodal | Deformable |
| Miao et al. (2016) | Real-time 2D/3D registration via CNN regression | CNN | Unimodal | Rigid |
| Miao et al. (2017) | Convolutional Neural Networks for Robust and Real-Time 2-D/3-D Registration | CNN | Unimodal | Rigid |
| Yang et al. (2017) | Fast predictive multimodal image registration | CNN | Multimodal | Deformable |
| Yang et al. (2017) | Quicksilver: Fast predictive image registration - A deep learning approach | CNN | Multimodal | Deformable |
| Yoo et al. (2017) | ssEMnet: Serial-section electron microscopy image registration using a spatial transformer network with learned features | SAEs | Unimodal | Deformable |
| Sokooti et al. (2017) | Nonrigid image registration using multi-scale 3D convolutional neural networks | CNN | Unimodal | Deformable |
| Cao et al. (2017) | Deformable image registration based on similarity-steered CNN regression | CNN | Unimodal | Deformable |
| Rohe et al. (2017) | SVF-Net: learning deformable image registration using shape matching | CNN | Unimodal | Deformable |
| Bhatia et al. (2017) | Real time coarse orientation detection in MR scans using multi-planar deep convolutional neural networks | CNN | Unimodal | Rigid |
| Zheng et al. (2017) | Learning CNNS with pairwise domain adaption for real-time 6dof ultrasound transducer detection and tracking from x-ray images | CNN | Multimodal | Rigid |
| Pei et al. (2017) | Non-rigid craniofacial 2D-3D registration using CNN-based regression | CNN | Unimodal | Deformable |
| Eppenhof et al. (2017) | Supervised local error estimation for nonlinear image registration using convolutional neural networks | CNN | Unimodal | Deformable |
| Eppenhof et al. (2018) | Deformable image registration using convolutional neural networks | CNN | Unimodal | Deformable |
| Zheng et al. (2018) | Pairwise domain adaptation module for CNN-based 2-D/3-D registration | CNN | Multimodal | Rigid |
| Sloan et al. (2018) | Learning Rigid Image Registration-Utilizing Convolutional Neural Networks for Medical Image Registration | CNN | Multimodal | Rigid |
| Sun et al. (2018) | Deformable mri-ultrasound registration using 3d convolutional neural network | CNN | Multimodal | Deformable |
| Yan et al. (2018) | Adversarial image registration with application for mr and trus image fusion | GAN | Multimodal | Rigid |
| Cao et al. (2018) | Deep learning based inter-modality image registration supervised by intra-modality similarity | CNN | Multimodal | Deformable |
| Cao et al. (2018) | Deformable image registration using a cue-aware deep regression network | CNN | Unimodal | Deformable |
| Onieva et al. (2018) | Diffeomorphic Lung Registration Using Deep CNNs and Reinforced Learning | CNN | Unimodal | Deformable |
| Mahapatra et al. (2018) | Joint registration and segmentation of xray images using generative adversarial networks | GAN | Unimodal | Deformable |
| Mahapatra et al. (2018) | Deformable medical image registration using generative adversarial networks | GAN | Multimodal | Deformable |
| Sentker et al. (2018) | GDL-FIRE 4D: Deep Learning-Based Fast 4D CT Image Registration | CNN | Unimodal | Deformable |
| Sun et al. (2018) | Towards Robust CT-Ultrasound Registration Using Deep Learning Methods | CNN | Multimodal | Deformable |
| Salehi et al. (2019) | Real-time deep pose estimation with geodesic loss for image-to-template rigid registration | CNN | Multimodal | Deformable |
| Elmahdy et al. (2019) | Robust contour propagation using deep learning and image registration for online adaptive proton therapy of prostate cancer | CNN GAN | Unimodal | Deformable |
| Liu et al. (2019) | Multimodal medical image registration via common representations learning and differentiable geometric constraints | CNN | Multimodal | Deformable |
| Foote et al. (2019) | Real-Time 2D-3D Deformable Registration with Deep Learning and Application to Lung Radiotherapy Targeting | CNN | Unimodal | Deformable |

The third generation belongs to the Deep Reinforcement Learning (DRL), where just like Fig. 14, the deep agent (or multiple agents) learns to produce the final transformation step-by-step so that the positive feedback from the environment (here from the similarity measure) can be maximized. Instead of the first deep similarity measure paradigm, the similarity measures are routinely provided in a conventional way e.g. Normalized MI (NMI) or Local Cross Correlation (LCC). The most confining factor to extinct the generation of this paradigm is the inability of the agents to interact with the huge state space introduced by deformable registration field. Without the ability to capture deformation essential for prosperous registration of elastic organs as well as relatively prolonged registration time, the paradigm is doomed to construction. A list of works belong to this generation are presented in Table 4.

**Table 4:** The Third Generation of Deep Learning Approaches for Medical Image Registration (Deep Reinforcement Learning)

| Reference | Title | Technique | Modality | Transformation |
|---|---|---|---|---|
| Ma et al. (2017) | Multimodal image registration with deep context reinforcement learning | DRL | Multimodal | Rigid |
| Krebs et al. (2017) | Robust non-rigid registration through agent-based action learning | CNN | Unimodal | Rigid |
| Liao et al. (2017) | An Artificial Agent for Robust Image Registration | CNN | Unimodal | Rigid |
| Toth et al. (2018) | 3D/2D model-to-image registration by imitation learning for cardiac procedures | CNN | Multimodal | Rigid |
| Miao et al. (2018) | Dilated FCN for multi-agent 2D/3D medical image registration | CNN | Multimodal | Rigid |
| Sun et al. (2019) | Robust Multimodal Image Registration Using Deep Recurrent Reinforcement Learning | CNN and RNN | Multimodal | Rigid |

As the previous generations were based on the ground-truth to construct the model, and generally the annotated datasets in medicine, and specifically for image registration, are small-sized, not suitable for the exhaustive deep learning, the fourth generation belongs to the unsupervised end-to-end registration, where different kinds of DNNs are trained without any ground-truth to construct regression models to produce the transformation parameters in one-shot. Instead of using enormous grand-truth set, they use data augmentation techniques on a few number of inputted samples as seeds where a traditional similarity measure (or a combination of them) is used as loss function to guide the learning process, as illustrated in Fig. 19. Most of the approaches in this generation have been successful on unimodal registration, while multimodal registration is far more complicated as multimodal similarity measures are still inefficient, and a network trained on them inherits this inefficiency, accordingly. Wu *et al.* (2016) can be a good representative where a SAEs is trained to extract the features, and a CNN makes the final transformation estimation. Utilization of SAEs instead of conventional multimodal similarity measures like MI is an evidence to our aforementioned argument. While synthesizing fake samples by data augmentation techniques, the rule of regularization term is critical to control the applied deformations in order to be realistic. Yet, practitioners and experts are dubious on this regard, and we expect a winding way ahead with a significant research focus in the near future. A list of works belong to this generation are presented in Table 5.

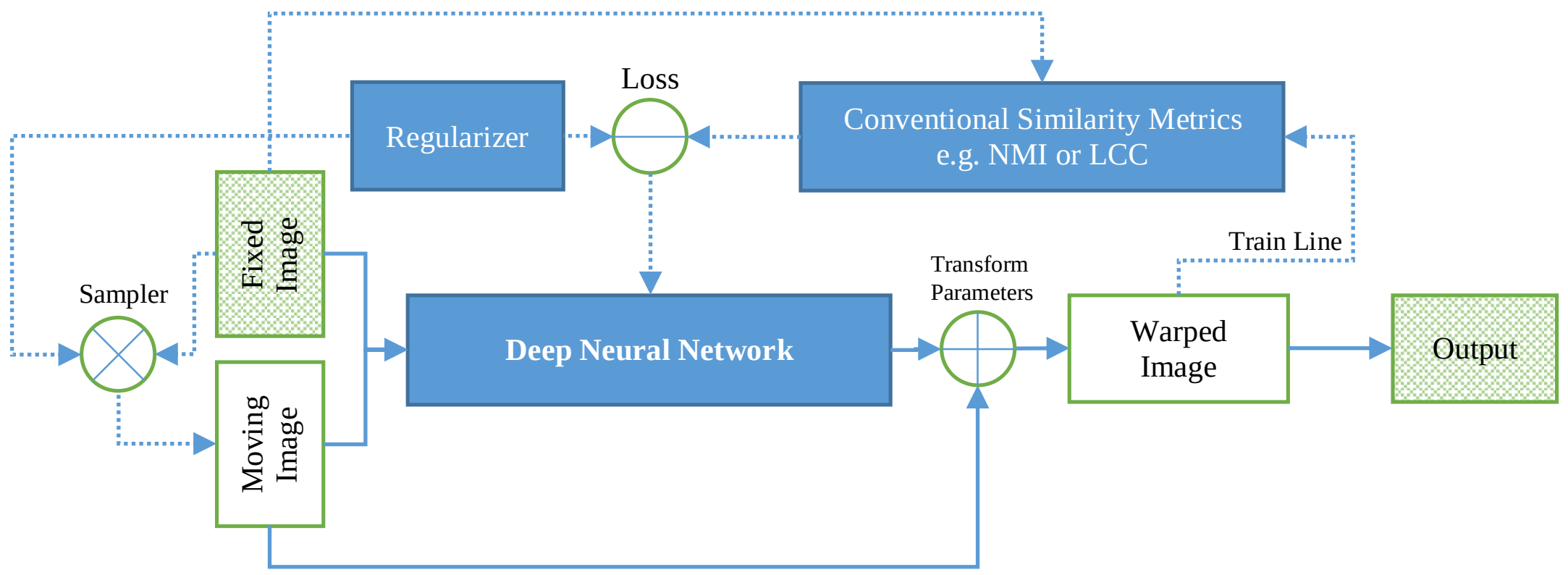

**Figure 19:** The main framework for unsupervised end-to-end medical image registration

**Table 5:** The Fourth Generation of Deep Learning Approaches for Medical Image Registration (Unsupervised End-to-End Registration)

| Reference | Title | Technique | Modality | Transformation |
|---|---|---|---|---|
| de Vos et al. (2017) | End-to-end unsupervised deformable image registration with a convolutional neural network | CNN | Unimodal | Deformable |
| Li et al. (2018) | Non-rigid image registration using self-supervised fully convolutional networks without training data | CNN | Unimodal | Deformable |
| Dalca et al. (2018) | Unsupervised learning for fast probabilistic diffeomorphic registration | CNN | Unimodal | Deformable |
| Balakrishnan et al. (2018) | An unsupervised learning model for deformable medical image registration | CNN | Unimodal | Deformable |
| Shu et al. (2018) | An unsupervised network for fast microscopic image registration | CNN | Unimodal | Deformable |
| Stergios et al. (2018) | Linear and Deformable Image Registration with 3D Convolutional Neural Networks | CNN | Unimodal | Deformable |
| Krebs et al. (2018) | Unsupervised probabilistic deformation modeling for robust diffeomorphic registration | SAEs | Unimodal | Deformable |
| Kearney et al. (2018) | An unsupervised convolutional neural network-based algorithm for deformable image registration | CNN | Unimodal | Deformable |
| Sheikhjafari et al. (2018) | Unsupervised deformable image registration with fully connected generative neural network | SAEs | Unimodal | Deformable |
| Ferrante et al. (2018) | On the adaptability of unsupervised CNN-based deformable image registration to unseen image domains | CNN | Multimodal | Deformable |
| Ito et al. (2018) | An Automated Method for Generating Training Sets for Deep Learning based Image Registration. | CNN | Unimodal | Deformable |
| Balakrishnan et al. (2019) | VoxelMorph: a learning framework for deformable medical image registration | CNN | Unimodal | Deformable |
| Yu et al. (2019) | Learning 3D non-rigid deformation based on an unsupervised deep learning for PET/CT image registration | CNN | Multimodal | Deformable |
| Che et al. (2019) | DGR-Net: Deep Groupwise Registration of Multispectral Images | CNN | Multimodal | Deformable |
| Van Kranen et al. (2019) | Unsupervised deep learning for fast and accurate CBCT to CT deformable image registration | CNN | Unimodal | Deformable |
| Che et al. (2019) | Deep Group-Wise Registration for Multi-Spectral Images From Fundus Images | CNN | Multimodal | Deformable |
| Hering et al. (2019) | Unsupervised learning for large motion thoracic CT follow-up registration | CNN | Unimodal | Deformable |
| Duan et al. (2019) | Adversarial learning for deformable registration of brain MR image using a multi-scale fully convolutional network | CNN | Unimodal | Deformable |
| de Vos et al. (2019) | A Deep Learning Framework for Unsupervised Affine and Deformable Image Registration | CNN | Unimodal | Deformable |
| Krebs et al. (2019) | Learning a probabilistic model for diffeomorphic registration | SAEs | Unimodal | Deformable |

Finally, since both supervised and unsupervised end-to-end image registration have their own drawbacks, the fifth generation belongs to the weakly/semi-supervised approaches. There are two different key paradigms in this category. Some approaches are based on the fully annotated grand-truth data with as many landmarks as possible. Routinely, these landmarks are contours, legions, corners, lines, turning points and so on each of which gets a distinct class label. The network trained on these fully labeled data; however, to be a few. Besides its main duty i.e. image registration, it learns to detect landmarks in any pair of inputted images. Detecting such kinds of landmarks are key to construct efficient models, and to enhance the accuracy of system. In addition, Target Registration Error (TRE) that is the most precious structural similarity measure can be used as the loss function to train the network, which is non-trivial. Hu *et al.* (2018) can be considered as the most complete representative for this paradigm where the framework is illustrated in Fig. 20. Another paradigm is based on the utilization of Generative Adversarial Network (GAN) (Goodfellow et al. 2014) (Fig. 12) where the generator takes the inputted fixed and moving images, and try to produce such transformation parameters so that the transformed moving image cannot be discriminated from the ground-truth by the discriminator, the situation expected from an expert registration agent. Based on the survival competition between the generator and the discriminator, just like the game theory, the network can be trained on a small set of data so that the generated samples cannot be discriminated, and the network goes for equilibrium. A list of works belong to this generation are presented in Table 6.

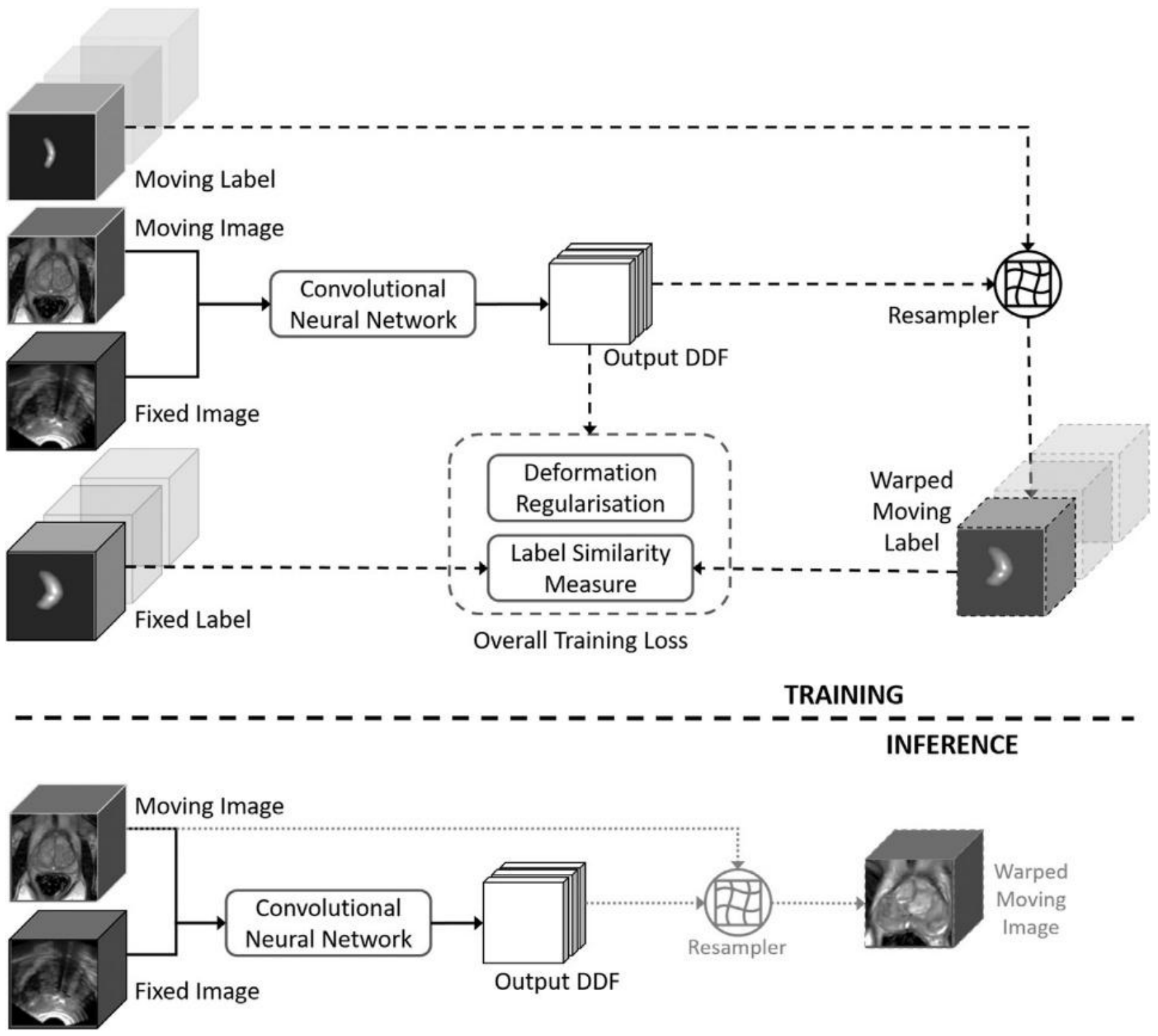

**Figure 20:** The main framework for weakly-supervised label-driven medical image registration (Hu et al. 2018)

**Table 6:** The Fourth Generation of Deep Learning Approaches for Medical Image Registration (Weakly/Semi-Supervised End-to-End Registration)

| Reference | Title | Technique | Modality | Transformation |
|---|---|---|---|---|
| Uzunova et al. (2017) | Training CNNs for image registration from few samples with model-based data augmentation | CNN | Unimodal | Deformable |
| Salehi et al. (2017) | Precise ultrasound bone registration with learning-based segmentation and speed of sound calibration | CNN | Multimodal | Deformable |
| Hu et al. (2018) | Weakly-supervised convolutional neural networks for multimodal image registration | CNN | Multimodal | Deformable |
| Fan et al. (2018) | Adversarial similarity network for evaluating image alignment in deep learning based registration | GAN | Unimodal | Deformable |
| Hu et al. (2018) | Label-driven weakly-supervised learning for multimodal deformarle image registration | CNN | Multimodal | Deformable |
| Hu et al. (2018) | Adversarial deformation regularization for training image registration neural networks | GAN | Multimodal | Deformable |
| Hering et al. (2019) | Enhancing Label-Driven Deep Deformable Image Registration with Local Distance Metrics for State-of-the-Art Cardiac Motion Tracking | CNN | Unimodal | Deformable |
| Fan et al. (2019) | BIRNet: Brain image registration using dual-supervised fully convolutional networks | CNN | Unimodal | Deformable |

In the following, some breakthroughs, generations' turning points and representatives are reviewed in more detail, and their network structures, advantages, disadvantages, novel ideas and key contributions are presented.

(Wu et al. 2013) should be properly considered as the first try to apply deep neural networks to medical image registration. The authors used a combination of CNN and Independent Subset Analysis (ISA) to extract proper features from inputted images. As illustrated in Fig. 21, the underlying architecture was a 2-layer network which takes input via both the layers. Also, a hierarchical training mechanism was used where small-sized patches in the size of 13×13×13 voxels were feeded to the first layer, and the first layer was trained accordingly. Afterwards, bigger-sized patches in the size of 21×21×21 voxels using sliding window with overlap were feeded to the second layer to train the second layer. The output of the network for each inputted patch was a 150-feature vector that was provided for two conventional registration algorithms namely HAMMER (Vercauteren, et al. 2009) and Diffeomorphic Demons (Shen 2007), and the final registration was achieved. The experimental study was conducted on two different datasets of IXI and ANDI containing MR images of human brain considering Dice coefficient as the comparison metric. Results from both the registration algorithms on both the aforementioned datasets showed tangible improvement considering significance test with 95% confidence interval.

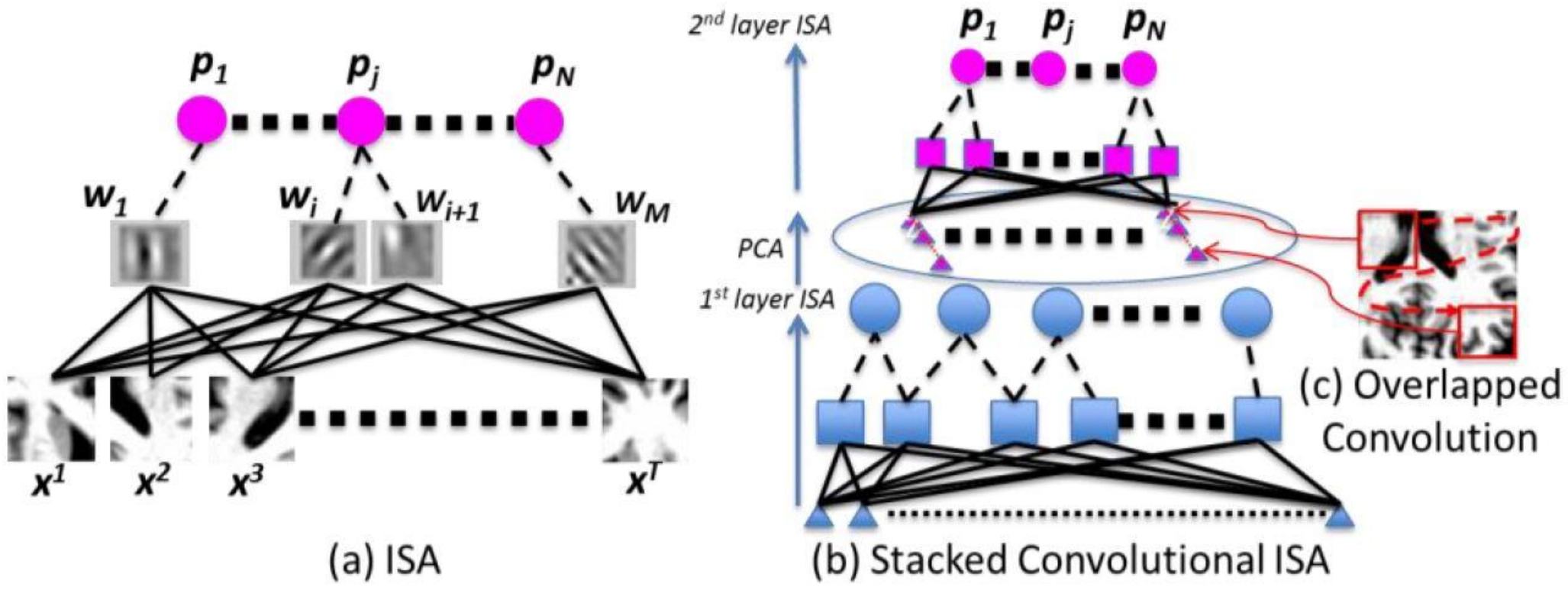

**Figure 21:** The architecture used in (Wu et al. 2013)

Cheng et al. (2016) trained a 5-layer SDAEs to approximate the amount of similarity between the couple of inputted CT and MR images of brain. Actually, the conventional similarity measure was substituted by the proposed network. The underlying reason was that multimodal similarity measures like Normalized Mutual Information (NMI) and Local Cross-Correlation (LCC) are far from the completeness, and have a lot of local minima in some applications. The input of the network, as illustrated in Fig. 22, is a couple of corresponding patches from inputted fixed and moving images, and the binary output indicates the correspondence/non-correspondence between these patches. In addition, the output before the Sigmoid activation function was extracted and used to compute the similarity measure. To train the binary network, the inputted CT and MR images were aligned rigidly by the authors. Then, the couple of inputted images were normalized to the zero mean and unit variance. To confine training data, patches were extracted from the central parts and around the scull considering the fact that edges and corners have the most informative and discriminative data. 2000 corresponding patches and 2000 non-corresponding ones in the size of 17×17 voxels were extracted to train the system. The proposed network was evaluated versus LCC and NMI on 300 patches as test data using cumulative sum of prediction error as the performance metric, and it showed significantly better performance.

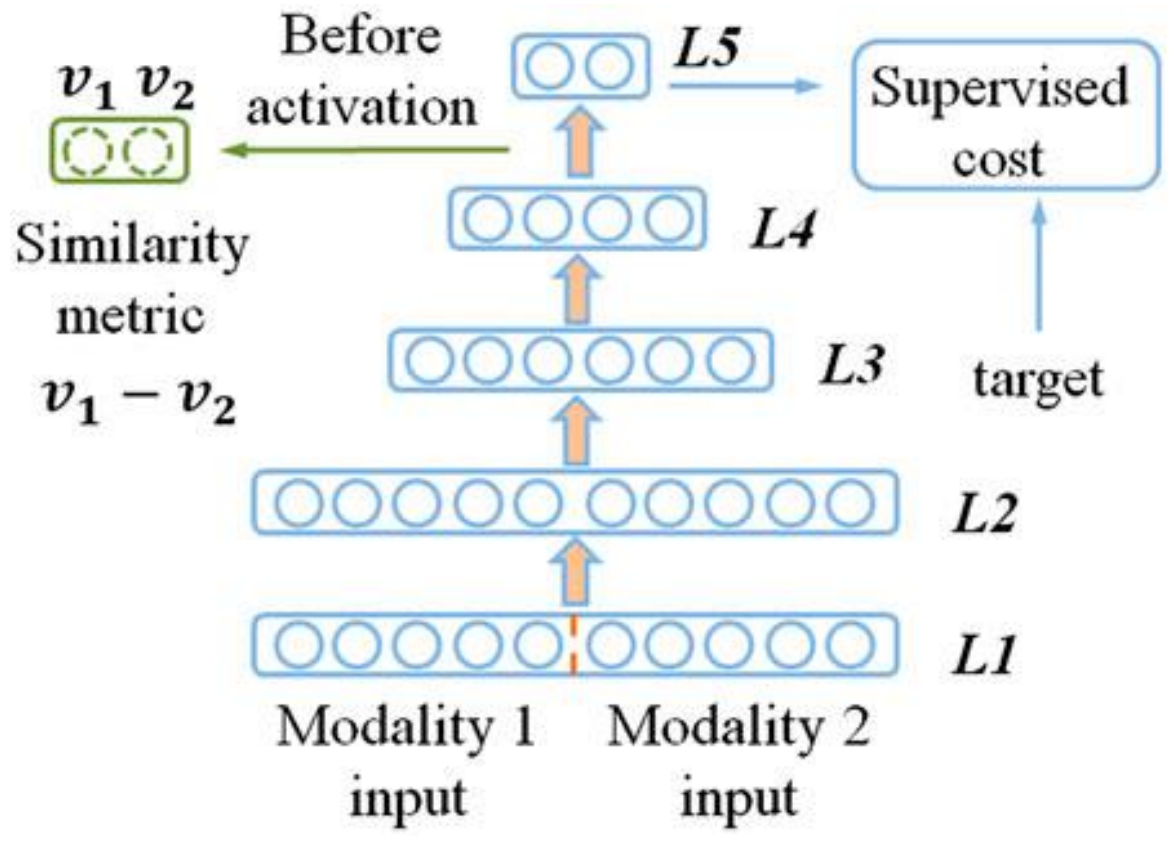

**Figure 22:** The architecture used in (Cheng et al. 2016)

Simonovsky et al. (2016) also proposed a deep learning model to use as a multimodal similarity measure, but despite the (Cheng et al. 2016), a CNN was exploited. Again, the reason was the same i.e. deficiency of the conventional multimodal similarity measures for all the modalities and all the organs of interest. At first, a rigid alignment was applied to the inputted images manually. Then, a large number of corresponding and non-corresponding patches in the size of 17×17×17 voxels were extracted from inputted images, and the network was trained accordingly. The network was a 5-layer 2-channel CNN with about two million weights. It was detected using stride in the pooling layers as well as Hinge loss-function (instead of cross-entropy) can contribute to the training convergence and the overall performance. Since the system was feeded by the couple of patches from fixed and moving images as an integration, it could not exploit the fact that the patch from fixed image does not need any manipulation. This fact was exploited in the next works by other authors using late fusion in the network. Using late fusion, the processed data over the channels/pipeline do not fused until the late layers of network i.e. we postpone sharing the channels' weights until the last layers. ALBERTs dataset was used for the evaluation study containing aligned and labeled T1 and T2-weighted MR images from the brain of 20 infants where there were 50 segmented anatomical regions labeled in each image. To worsen the situation, the system was trained based on the IXI dataset that contains 600 aligned T1 and T2-weighted MR images of adults' brain. The images of infants and adults have some differences in anatomy helping us better estimating the generalization potentiality of the proposed system. The proposed approach was compared versus Mutual Information (MI), and the results showed its superiority with 99% confidence interval. In addition, the running time used to response the network was about 2 times slower than MI which still indicates acceptable time for clinical use.

In all the aforementioned works, deep learning models were used to work as a similarity measure besides a conventional iterative approach exploiting this measure. For the first time, in a seminal work, Miao et al. (2016) used a CNN to directly regress the restricted Affine transformation parameters to register 2D X-ray images on 3D CT ones; however, their approach was based on the particles implanted in the patients' bodies, as illustrated in Fig. 5. At first, they considered an attention map from the 3D CT image, and extracted a Digitally Reconstructed Radiograph (DDR) from it. The duty of the proposed approach was to register the extracted DDR on the operational X-ray images. To enhance the system, they used the following three strategies: First of all, a novel similarity measure named Local Image Residual (LIR) was used, which could effectively keep its correlation with the difference caused by changes in transformation parameters. In other words, it was a better estimation of different alignments of the inputted images. Secondly, to decrease the complexity of registration, the images were segmented and decomposed to a predefined number of regions where each region had its own CNN; however, this strategy significantly increases the computational burden. Thirdly, parameters regression was made hierarchically to increase the precision. That is, the 6 transformation parameters were divided to 3 different sets of easy, moderate and hard, and would be calculated from easy to hard. As illustrated in Fig. 23, at first, the proposed approach extracted *N* patches around the implanted particle. There was a CNN for each patch, composed of 2 convolutional layers, 2 pooling layers, and a fully-connected layer with a 100-bit output. The outputs of all the CNNs were gathered, concatenated and feeded to a final 2-layer fully-connected network that results in the final 6 transformation parameters, 2 for translation, 1 for scaling, and 3 for rotation around the coordination axes. Three datasets named Total Knee Arthroplasty (TKA), Visual Implant Planning System (VIPS), and X-ray Echo Fusion (XEF) were used for comparison study. Also, a novel evaluation metric named mTREproj denoting mean target registration error for image projection from 8 corners of implanted particle was considered for evaluating the approaches. The proposed approach was compared versus different variants of the conventional Powell method, and it showed sensible superiority over all the three datasets. The running time was about 100ms on a typical system showing high potentially for real-time clinical use.

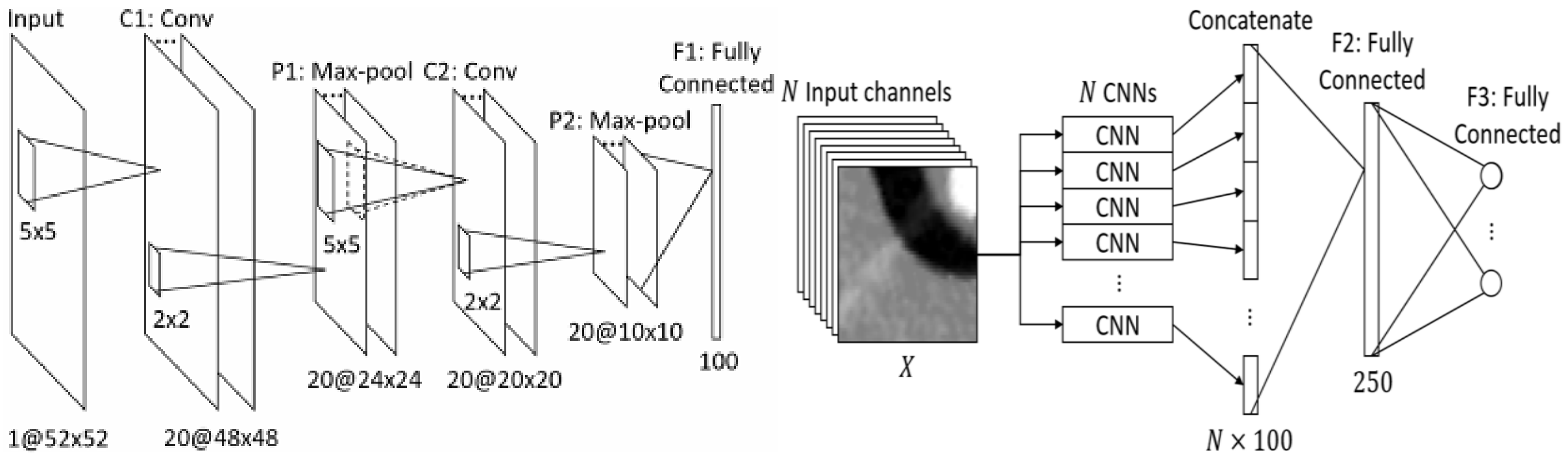

**Figure 23:** The architecture used in (Miao et al. 2016)

While the previous work considered only Affine transformation, in another seminal work, Sokooti et al. (2017) were successful to train a CNN to regress a Displacement Vector Field (DVF) capable of deformablely registering the inputted images in one shot. An overview on the mechanism of DVF is illustrated in Fig. 24. They extracted a large number of corresponding patches (about 2100 patches each image) based on the semi-automatically detected landmarks from the couple of inputted images, and feeded them to the network (Fig. 25). The patches were in the size of 29×29×29 voxels, while some other cases were extracted in the size of 54×54×54 voxels to keep the receptive field. These bigger patches were compressed to the half (to decrease the computational intensity), and then were feeded to the network via a different pipeline. For the first time, late fusion was used so that the different-sized patches used different pipelines in the network until the last layers, which could have a good contribution to the applicability and performance. The first and second pipelines (for 29×29×29 and 54×54×54 voxels patches, respectively) were composed of 3 sequential convolutional layers with 3×3×3 kernels followed by 2×2×2 pooling layers. Afterwards, the first and seconds pipelines would be subjected to 6 and 2 1×1×1 convolutional layers, respectively, to become the same-sized. Concatenation was done at this point, and after 4 other convolutional layers, there was a 2-layer fully-connected network to produce the final 3 translational parameters for each patches. The SPREAD dataset containing 19 couples of 3D CT images from chest was used to evaluate the proposed approach where 10, 2, and 7 of them were used for training, validation and testing, respectively. The Mean Absolute Error (MAE) metric was used for system training while Target Registration Error (TRE) was the evaluator of the approaches in the comparison study. The proposed approach was evaluated versus Affine, regular B-Spline, and 3-resolution B-Spline, and it outperformed the Affine and regular B-Spline methods, and was competitive with the 3-resulation B-Spline, one of the strongest methods in the literature.

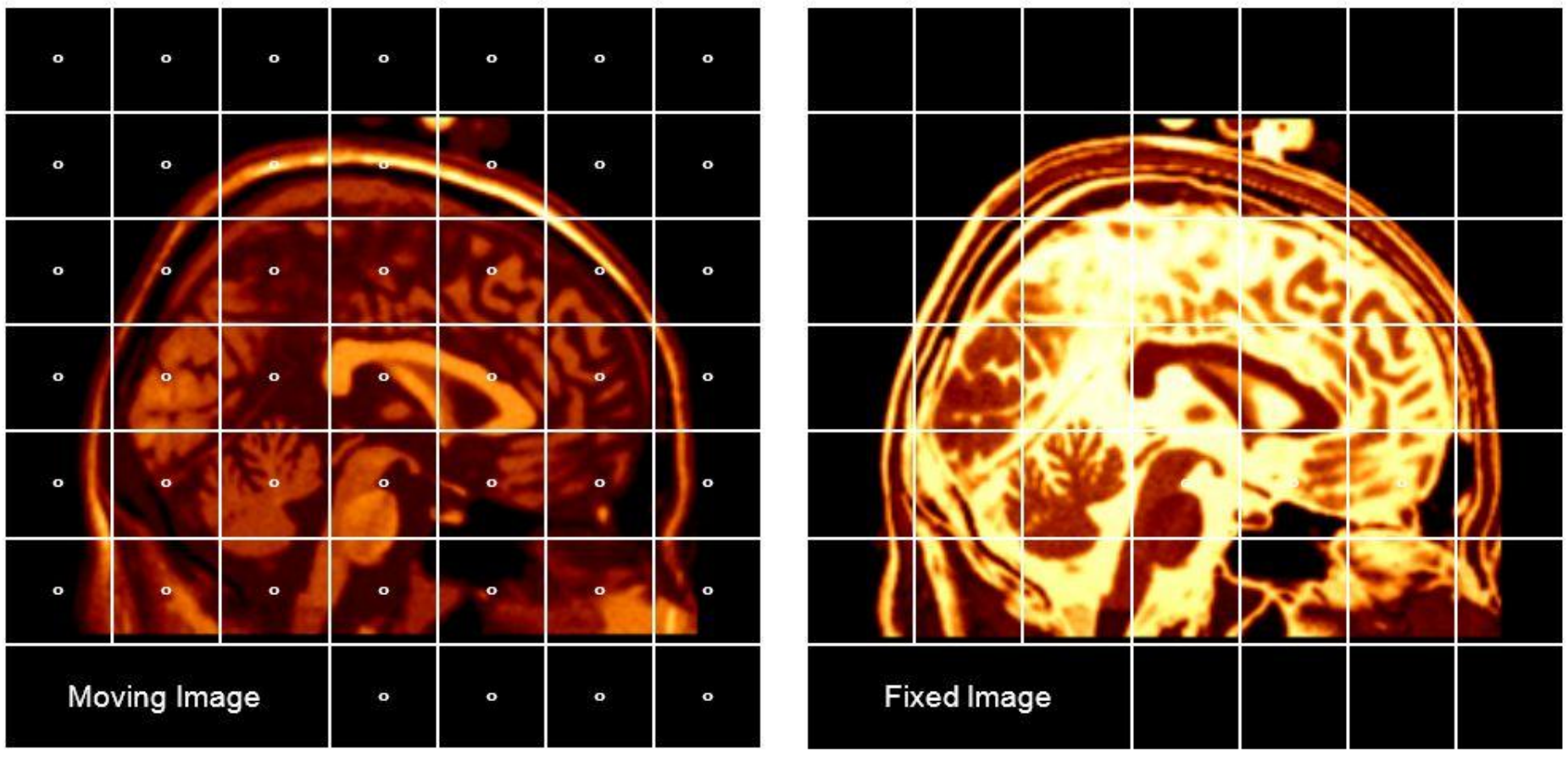

$$DVF = \begin{bmatrix} \langle dx_{11}|dy_{11}|dz_{11}\rangle & \langle dx_{12}|dy_{12}|dz_{12}\rangle & \cdots & \langle dx_{1m}|dy_{1m}|dz_{1m}\rangle \\ \langle dx_{21}|dy_{21}|dz_{21}\rangle & \langle dx_{22}|dy_{22}|dz_{22}\rangle & \cdots & \langle dx_{2m}|dy_{2m}|dz_{2m}\rangle \\ \langle dx_{31}|dy_{31}|dz_{31}\rangle & \langle dx_{32}|dy_{32}|dz_{32}\rangle & \cdots & \langle dx_{3m}|dy_{3m}|dz_{3m}\rangle \\ \vdots & \vdots & \ddots & \vdots \\ \langle dx_{n1}|dy_{n1}|dz_{n1}\rangle & \langle dx_{n2}|dy_{n2}|dz_{n2}\rangle & \cdots & \langle dx_{nm}|dy_{nm}|dz_{nm}\rangle \end{bmatrix}_{N\times M}$$

**Figure 24:** An overview on the mechanism of DVF

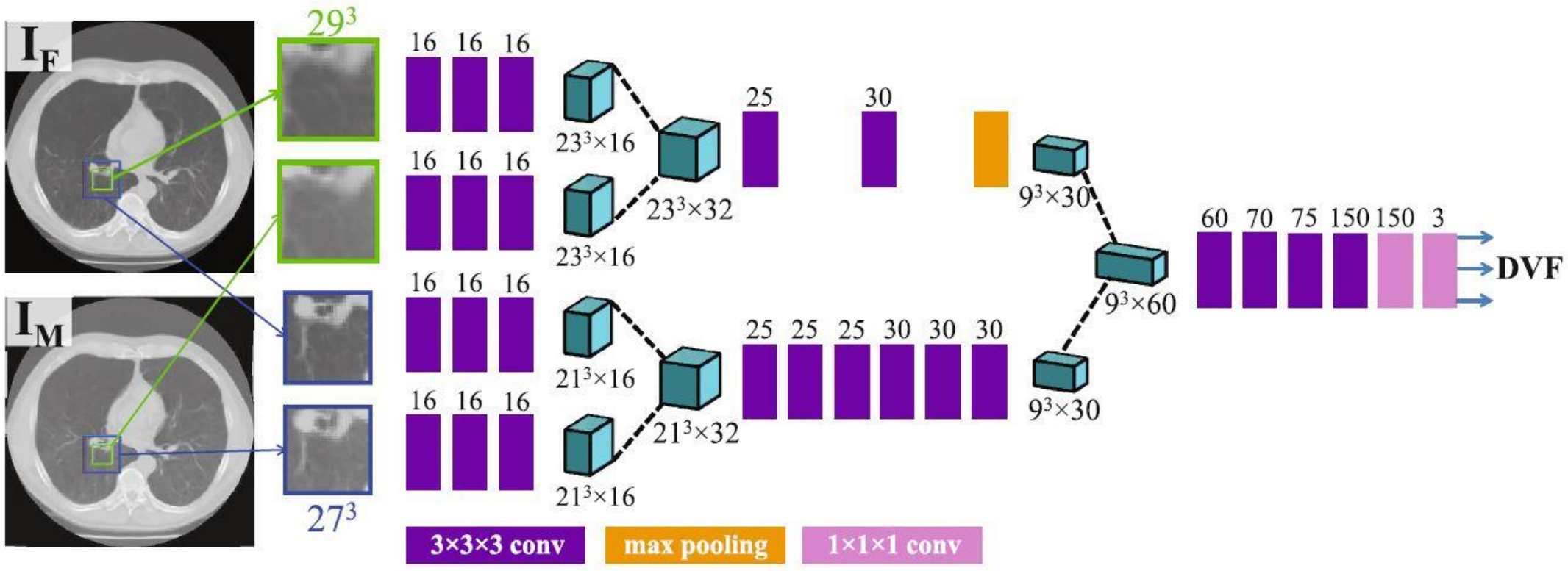

**Figure 25:** The architecture used in (Sokooti et al. 2017)

Hu et al. (2018) trained a CNN for deformablely registering pre-interventional multi-parametric MR images to operational Transrectal Ultrasound (TRUS) of the prostate gland in order to decrease the risk of image-guided intervention. Since the captured TRUS images are in low-quality with a high level of noise, fusing them with the high-quality 3D MR images can cause synergic effect to desirably guide and support the intervention process. The problem was that MRI and TRUS do not have representational abonnement except in the limited cases, which challenges the registration process. Accordingly, utilization of conventional approaches was not possible, and it justified the exploitation of machine learning approaches. Since there was not any comprehensive labeled dataset in this regard, and utilization of intensity-based similarity measures causing mechanical registrations has ambiguity aspects for the physicians and experts, a dataset that was partially manually labeled by the experts was used to train the network. Finally, corresponding labeled structures were exploited by the network to produce an exact voxel-vise registration. The promising point is that the system implicitly learned to detect the structures-in-interest after training; hence, the need to manual operation in the utilization phase is obviated. The similarity measure used by the proposed approach was multiscale and based on the registration distance of the corresponding labeled landmarks and structures in the inputted images. As illustrated in Fig. 26, the proposed 3D CNN was composed of 4 downsampling blocks followed by 4 upsampling ones just like typical U-Nets, but with far more connections to keep the impact of back-propagation over a large number of layers. The network output was a Dense Displacement Field (DDF) for registering and fusing whole the MR image on the continues TRUS ones. The SmartTarget dataset containing 108 couple of T2-weighted MRI and TRUS was used in which every patient has three kinds of images based on his/her treatment plan. Researcher fellowships and students consumed 200 hours in total to detect and label 834 couple of anatomical landmarks. 12-fold cross-validation as well as Dice and TRE measures were used to test the system, where the results, supported by statistical Wilcoxon rank-sum test with 95% confidence, certified the significant superiority of the proposed approach.

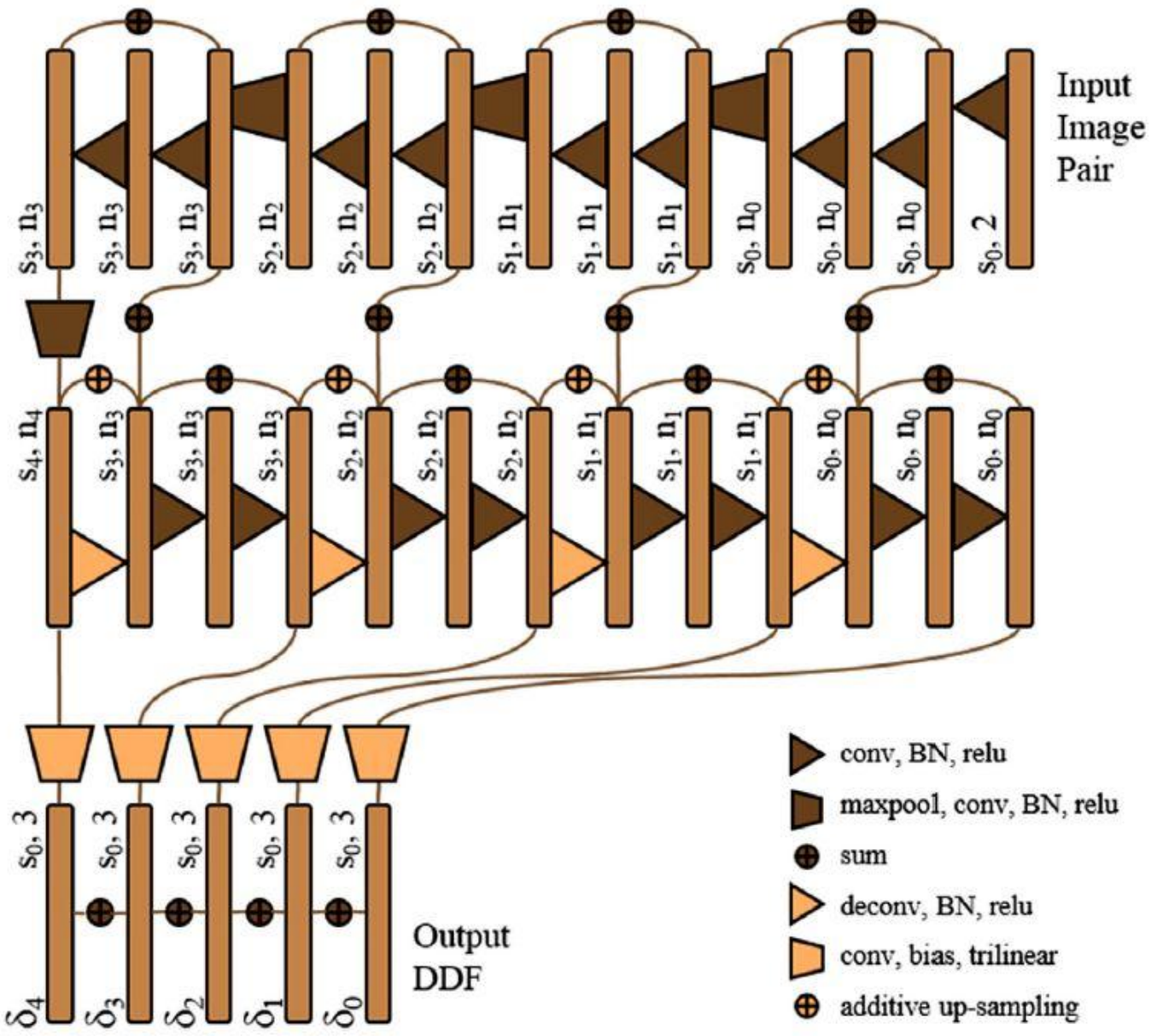

**Figure 26:** The architecture used in (Hu et al. 2018)

While the previous work is highly appreciated and valuable since the authors entered the time as the 4$^{th}$ dimension of TRUS in order to made the problem of registration far more complicated, the work of de Vos et al. (2019) should be considered as the most comprehensive deep learning framework based on the CNN to directly regress the deformable transformation parameters in one shot. In addition to regress the transformation parameters, their network was able to learn a predefined similarity measure so that the necessity to utilization of synthesized and labeled dataset is obviated, which is a big progress in applying DNNs to the field of medical image analysis where we are facing small-sized annotated datasets as regular. As illustrated in Fig. 27, the proposed method was a multi-resolution multi-stage 2-channel CNN-based approach. In the first stage, a 2-channel CNN with 5 layers of 3×3×3 kernels followed by 2×2×2 average pooling layers took the inputted images. The weights were shared between layers to decrease the number of the networks' weights. At the end, a concatenating layer followed by a 2-layer fully-connected networks produced the 12 parameters of Affine transformation. In the second stage, another CNN was used to made the final deformable registration based on the B-Spline. At first, a number of patches based on the specified landmarks were extracted and feeded to the network. The segmentation was based on the (Long et al. 2015) providing image analysis with arbitrary size. The network was composed of interleaving 3×3×3 convolutional layers followed by 2×2×2 pooling ones. The number of layers was variable based on the number of control points in the deformable mesh (i.e. based on the space among control points). After the final pooling layer, there were 2 layers of convolution to increase the receptive field, which is novel contribution to set the CNN architecture. Finally, a 2-layer fully-connected network regressed the B-Spline control point locations that was used as a reference to produce the correspondence DVF. Two different datasets were used to evaluate the proposed approach. The first one was Sunnybrook Cardiac Data (SCD) containing 900 MR images of 45 patients with 4 different states of heart attack. For each patient, 20 images covered a full cycle of heartbeat. The second dataset was composed of 2060 CT images randomly selected from the National Lung Screening Trial (NLST) with different size and quality. The proposed approach was evaluated versus the well-known Elastix (Marstal et al. 2016) using Dice coefficient, Hausdorff distance and Average Surface Distance (ASD) as the comparison metrics, where the proposed approach outperformed the Elastix in many cases while 350 time faster. A full deformable registration on the 4-core Xeon E5-1620 processors and NVIDIA Titan-X GPU was reported to take about 39 millisecond (ms), which is highly appreciated for real-time clinical use.

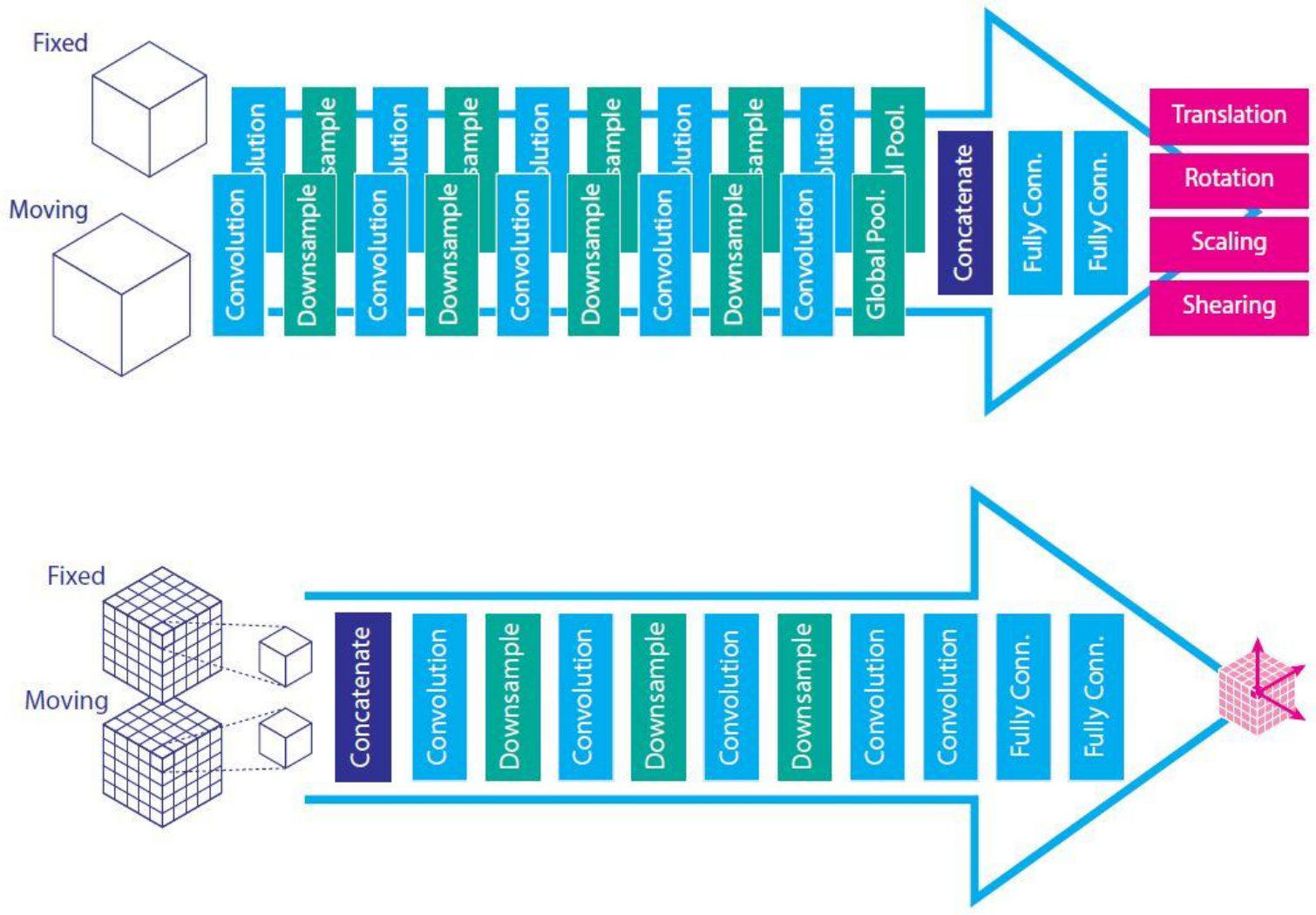

**Figure 27:** The architecture used in (de Vos et al. 2019)

### 6. 1. The Comparative Analysis, Advantages, Disadvantages, Main Contributions and Novelties

As stated in Section 6, we have faced five different generations or categories for medical image registration using deep neural networks based one the achievements and breakthroughs faced by the community. These categories can be enumerated as Deep Similarity Metrics (DSM), Supervised End-to-End Registration (SE2ER), Deep Reinforcement Learning (or Agent-Based Registration) (DRL), Unsupervised End-to-End Registration (UE2ER), Weakly/Semi-Supervised End-to-End Registration (WSE2ER), each of which with different paradigm to encounter the registration problem. The number of works in each category and a timeline from start to the end are depicted in the Fig. 28 and Fig. 29, respectively.

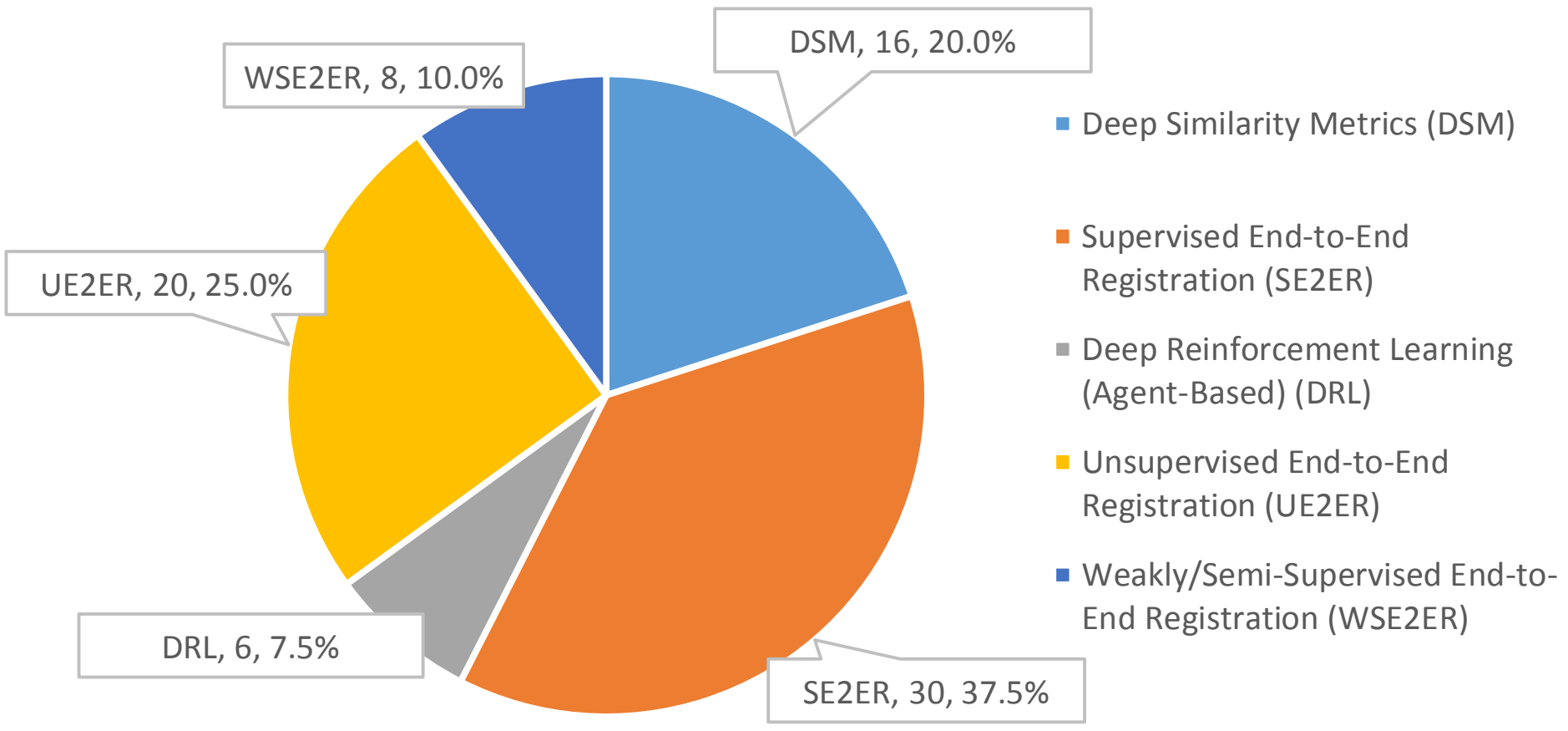

**Figure 28:** Different generations of deep learning approaches for medical image registration with their frequencies

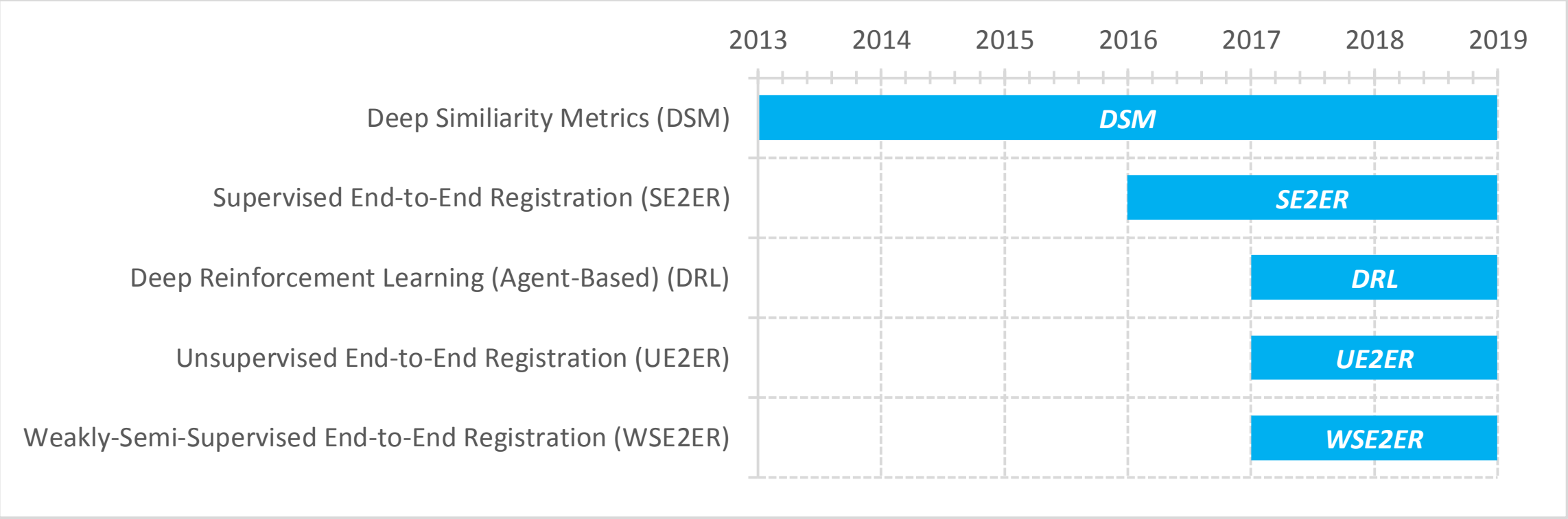

**Figure 29:** Different generations in timeline

At first, the topic started with Deep Similarity Metric (DSM) approaches in 2013 where different kinds of DNNs trained to learn visual similarity metrics from a large set of paired annotated ground-truths. The learned model after the train was able to precisely and meaningfully model the structural difference between the inputted pair of images/patches specially for deformable transformation with different modalities, where conventional similarity metrics (supposedly with the exception of MI) had a lot of difficulties. Two main drawbacks to this paradigm are that

- It is dependent on a large set of paired annotated ground-truths to train the network, which rarely is the case for medical applications.
- It is still dependent to conventional iterative optimization-based approaches that are very slow and impractical for clinical use.

Supervised End-to-End Registration (SE2ER), started in 2016, was a significant milestone that the community faced since it obviates the computational burden and time inconveniency of conventional iterative registration approaches, and by conducting the registration process in one-shot, practically makes the real-time clinical use possible. It started in 2016 by Miao et al (2016) for rigid registration, extended in 2017 by Sokooti et al. (2017) for deformable registration, and is the mainstream category that is remained active so far. Again, the main problem to this paradigm is that

- It is dependent on a large set of paired annotated ground-truths to train the network, which is a severe hindrance to develop any approach in this category.

Actually, a large number of authors abandoned this category, and went to try their chance on other different paradigms where the data annotation is out of the case. Deep Reinforcement Learning (DRL) (or Agent-Based Registration) is one of such paradigms, where a deep agent (or multiple agents) learns to produce the final transformation step-by-step so that the positive feedback from the environment (here from a similarity measure) can be maximized. Instead of the first deep similarity measure approaches, the similarity measures are routinely provided in a conventional way like NMI or LCC. The most confining factor to develop the generation of this paradigm is

- The inability of the agents to interact with the huge state-space introduced by deformable registration field where, as it can be seen in Table 4, all the approaches were proposed for rigid registration.

While they were still dependent to ground-truth to train the agents. To circumvent this problem, Unsupervised End-to-End Registration (UE2ER) paradigm was introduced, where different kinds of DNNs are trained without any ground-truth to construct regression models to produce the transformation parameters in one-shot. Instead of using enormous grand-truth set, they use data augmentation techniques on a few number of inputted samples as seeds where a traditional similarity measure (or a combination of them) is used as loss function to guide the learning process. Most of the approaches in this generation have been successful on unimodal registration, while multimodal registration is far more complicated, and can be regarded as the main challenge of this category, because

- Multimodal similarity measures that are used as loss function to conduct the network's learning are still inefficient, and a network trained on them, accordingly, inherits this inefficiency.

On this basis, this category needs to be looking forward to the introduction of much efficient and powerful novel similarity measures in the near future. Where both supervised and unsupervised manner have their own drawbacks, Weakly/Semi-Supervised End-to-End Registration (WSE2ER) found its way from 2017. It obviates the shortcomings associated to the two aforementioned paradigms while inherits the strengths of both. Some approaches are label-driven, i.e. based on a few fully-annotated ground-truth samples, they can implicitly learn to detect many paired landmarks in the inputted image, and conduct the registration process, accordingly. Some other approaches are dual-supervised based on the similarity measures (just like unsupervised approaches) and a few ground-truth samples to fine-tune the network. In addition, it has been demonstrated that having a few ground-truth samples, transfer learning from other body organs or modalities, is fully practical for medical image registration (Cao et al. 2017) and (Ferrante et al. 2018). Finally, another approaches use GANs as underlying technique, where the competitive interaction between the generator and discriminator needs a few ground-truth samples to construct and mature the model. Actually, weakly/semi supervision can be considered as the best practical paradigm so far, where we expect the significant research focus in the near future. The progress of this paradigm is heavily dependent of the theoretical progress and breakthroughs in the broader fields of machine vision, image processing, machine learning and pattern recognition, from which most of the progresses in the field of medical image registration have been inspired. Table 7 is a big picture on these five different categories as the taxonomy conducted on the literature.

**Table 7:** Deep Learning Approaches for Medical Image Registration: Paradigms, Frameworks, References, Advantages and Disadvantages

| Paradigm | Framework | List of References | Advantages | Disadvantage |
|---|---|---|---|---|
| Deep Similarity Metrics (DSM) | Fig. 16 | Table 2 | The constructed model after the train on ground-truth, whenever sufficient, were able to outperform the traditional similarity metrics specially for multimodal registration. | 1. The approaches are dependent on a large set of paired annotated ground-truths to train the network, which rarely is the case for medical applications. 2. The paradigm is still dependent to conventional iterative optimization-based approaches, which were very slow and impractical for clinical use. |
| Supervised End-to-End Registration (SE2ER) | Fig. 17 | Table 3 | It obviates the computational burden of conventional iterative registration approaches, and by conducting the registration process in one-shot, practically makes the real-time clinical use possible. | The paradigm is still dependent on a large set of paired annotated ground-truths to train the network, which is a severe hindrance to develop the approaches in this category. |
| Deep Reinforcement Learning (Agent-Based Registration) (DRL) | Fig. 19. | Table 4 | Instead of DSM approaches, which are dependent to conventional iterative optimization-based approaches to construct the final transformation, this step-by-step registration was still very faster for clinical use. | The inability of the agents to interact with the huge state space introduced by deformable registration field where, as it can be seen in Table 4, all the approaches were proposed for rigid registration. |
| Unsupervised End-to-End Registration (UE2ER) | Fig. 20 | Table 5 | Instead of using enormous grand-truth set, they use data augmentation techniques on a few number of inputted samples as seeds where a traditional similarity measure (or a combination of them) is used as loss function to guide the learning process. | Multimodal similarity measures that are used as loss function to conduct the network learning are still inefficient, and a network trained on them, accordingly, inherits this inefficiency. |
| Weakly-Semi-Supervised End-to-End Registration (WSE2ER) | Fig. 21 | Table 6 | It obviates the shortcomings associated to the SE2ER and UE2ER paradigms while inherits the strengths of both. The approaches are label-driven, dual-supervised, or based on adverbial learning (GAN). | The best practical paradigm so far, where we expect the significant research focus in near future. |

On the other hand, from technical point of view, we can have a different analysis on the introduced deep learning approaches for medical image registration. As stated, some authors use deep learning techniques to elicit the most influential and discriminative features to be feeded to the conventional optimization-based medical registration approaches in order to maximize the performance, while the others use deep neural networks as the regressor to directly estimate the transformation parameters in one-shot in order to maximize the runtime speed. Since the techniques are about same in nature, it can be stated that all the aforementioned works on medical image registration exploit the advantages of utilization of deep learning; the main contributing advantages are as follows:

- Deep learning obviates the burden of choosing, reducing, selecting, and normalizing handcrafted features that are the most important factor to achieve high performance in registration.
- Deep Learning techniques do not stalk in premature convergence or stagnation which are two prevalent confining dilemmas in the conventional optimization-based approaches specially where dealing with images from different modalities (multimodal image registration).
- Deep learning response time is effectively low (below a second in most the cases) instead of conventional iterative manner where runtimes in the tens of minutes are norm for common deformable image registration techniques. This is an actually decisive factor where practical use in clinical operations is real-time and such a prolonged wasting time is not appreciated.

On the other hand, based on the no-free-lunch theorem (Wolpert and Macready 1997), we are losing some parameters while getting some others, and deep learning cannot be an exception. Accordingly, utilization of deep neural networks imposes some disadvantages frequently reported by the authors as follows:

- Deep learning needs large-scale training data to elevate its performance and avoid over-fitting phenomena, while medical image datasets are inherently small-sized. Currently, endeavors from the both perspectives of expanding datasets, and devising novel ideas to circumvent the problem are in agenda; nevertheless, the problem is still reported as irritating by most the authors (Litjens et al. 2017).
- Computational burden of training deep neural networks is really high, and cannot be affordable but with the multiple contemporary GPUs (like a NVIDIA TitanX); actually, the authors need to restrict their set of experiments which is an influential obstacle to investigate novel ideas. Currently, commercial cloud computing environments have a good potential to contribute the issue; however, it is not available or affordable by the all, and the problem is still open (Agrawal et al. 2015).

Indeed, another informative knowledge for the readers is the main contribution and novelty proposed by each aforementioned seminal work; Table 8 is a collection of this information in an overview, which can be used as reference to compare the contribution of each work.

**Table 8:** The main contribution and novelty proposed by each aforementioned seminal work in an overview

| Author & Year | Deep Learning Technique | Main Contribution and Novelty |
|---|---|---|
| Wu et al. (2013) | CNN | The first utilization of DNNs on medical image registration. The CNN with ISA (Independent Subset Analysis) exploited to extract features from multimodal images to feed to the conventional HAMMER and Demons approaches. They used patches in the sizes of 13×13×13 as well as 21×21×21 voxels to broaden the network's receptive field. |
| Cheng et al. (2016) | SAEs | Utilization of SAEs to extract features and to work as a similarity metric for multimodal image registration, where the proposed approach outperformed conventional similarity metrics like NMI (Normalized Mutual Information) and LCC (Local Cross-Correlation). The patches were selected arbitrary from the center and corners of skull in the size of 17×17 voxels. |
| Simonovsky et al. (2016) | CNN | Utilization of CNN to extract features and to work as a similarity metric for multimodal image registration, where the proposed approach compared with MI (Mutual Information) and outperformed it with 99% confidence. The patches were selected arbitrary from the center and corners of skull in the size of 17×17×17 voxels. It was detected using stride in the pooling layers as well as Hinge loss-function instead of cross-entropy can contribute to the training convergence and the overall performance. |
| Miao et al. (2016) | CNN | For the first time, a CNN was used to directly regress the restricted Affine transformation parameters to register 2D X-ray images on 3D CT ones based on the particles implanted in the patients' bodies. The patches were selected arbitrary from the center and corners of implant in the size of 52×52 voxels. The runtime was reported as 100ms which is appreciable for clinical use. |
| Sokooti et al. (2017) | CNN | While the previous work considered only Affine transformation, this work was successful to train a CNN to regress a DVF (Displacement Vector Field) capable of deformablely registering inputted images in one shot; however, the registration was unimodal. The patches were in the size of 29×29×29 voxels, while some other cases were extracted in the size of 54×54×54 voxels to keep the receptive field. |
| Hu et al. (2018) | CNN | A CNN was trained for deformablely registering and fusing pre-interventional multi-parametric MR images to the operational Transrectal Ultrasound (TRUS) of the prostate gland in order to decrease the risk of image-guided intervention. The work is important considering it is multimodal, deformable, and one shot. In addition, for the first time, TRUS was used where the time as an extra dimension exacerbate the situation, and extra considerations should be taken into account. |
| de Vos et al. (2019) | CNN | This work should indeed be considered as the most comprehensive deep learning framework based on the CNN to directly regress the deformable transformation parameters in one shot. In addition to regress the transformation parameters, their multistage multiresolution approach was able to learn a predefined similarity measure so that the necessity to utilization of synthesized and labeled dataset is obviated, which is a big progress in applying DNNs to the field of medical image analysis where we are facing small-sized annotated datasets as regular. |

## 7. Literature Review Analysis

Fig. 30 presents the distribution of publications over the years from 2013 when the first related paper was published in the conference of Medical Image Computing and Computer-Assisted Intervention (MICCAI-2013) up to June 2019 (the submission time of the 2$^{nd}$ revise of this review). There are a 1-year gap in 2014 with no publication, and a pick of publications in 2018. Although this topic is in its infancy, and any conclusion may be premature at this time, based on a simple regression on the number of works published in each year, we expect this topic to ultimately finds its way, and we continue witnessing more publications and works in this challenging area.

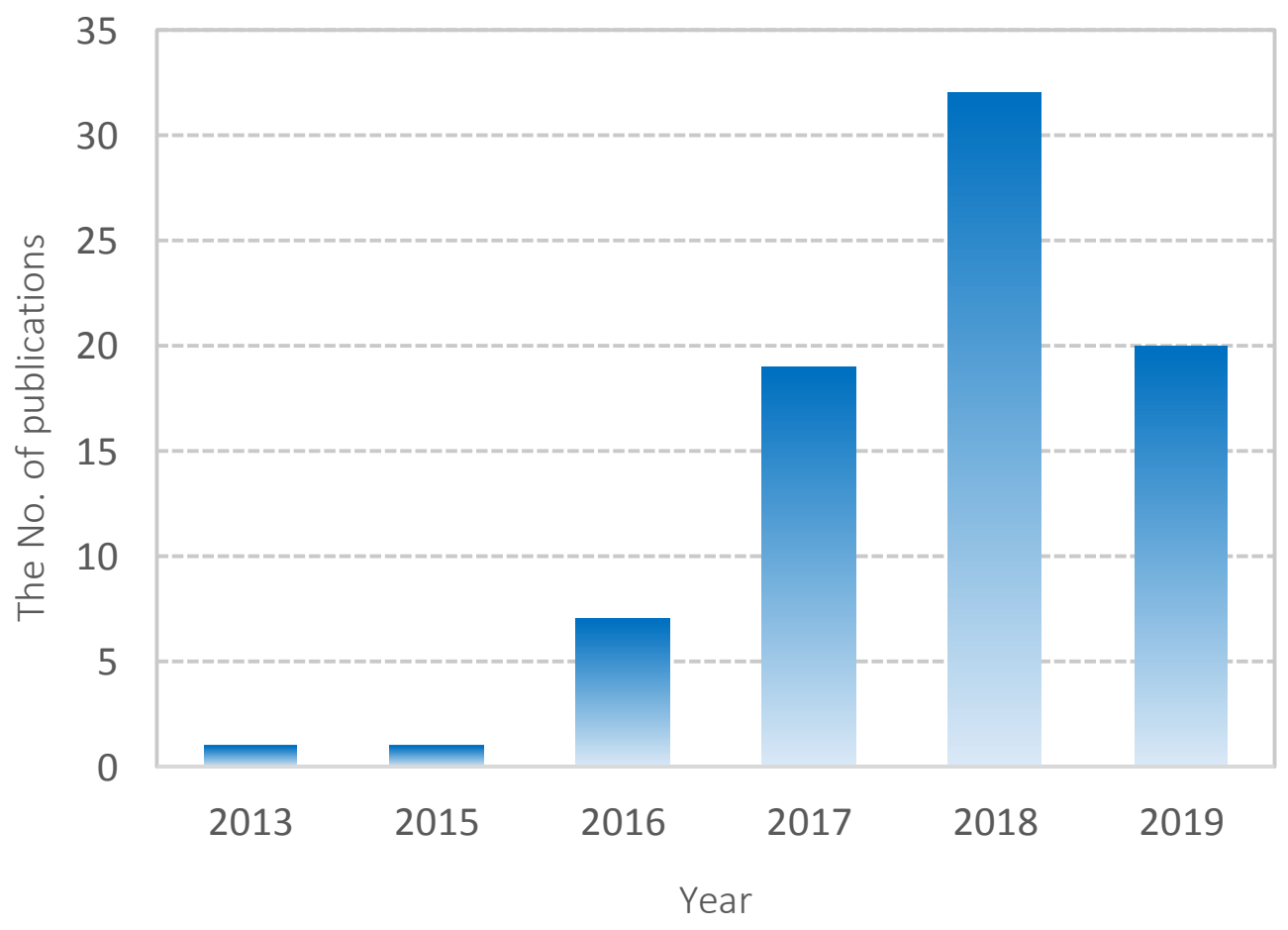

**Figure 30:** The number of publications based on the year

The number of publications based on the publication type is presented in Fig. 31. More than 60% of publications are from conference proceedings where journal articles and book chapters are in the next ranks. Accordingly, Table 9 lists top journals, conferences, and books in the field based on the number of related publications. As the first paper of the literature was published in, the conference of Medical Image Computing and Computer-Assisted Intervention (MICCAI) has been a valuable forum for the related researchers and their contributions. IEEE International Symposium on Biomedical Imaging, IEEE Transactions on Medical Imaging, International Workshop on Deep Learning in Medical Image Analysis, and Medical Imaging: Image Processing are in the second and third ranks. Accordingly, Fig. 32 shows top publication titles based the number of the published works where Springer Nature, IEEE, Elsevier BV, and SPIE are considered as top publications published over 85% of the works in this area.

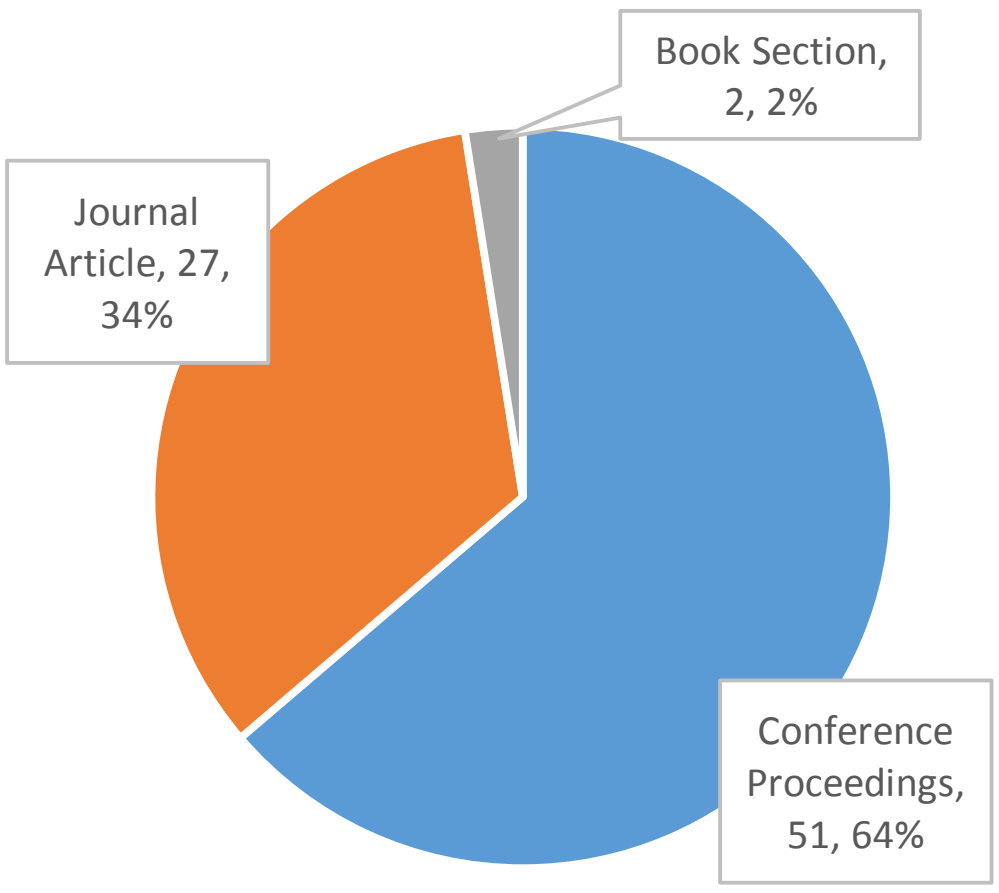

**Figure 31:** The number of publications based on the publication type

**Table 9:** Top journals, conferences, and books in the field based on the number of related publications

| | | | |
|---|---|---|---|
| International Conference on Medical Image Computing and Computer-Assisted Intervention | 14 | Biomedical Signal Processing and Control | 1 |
| IEEE International Symposium on Biomedical Imaging (ISBI) | 5 | Computational and Mathematical Methods in Medicine | 1 |
| IEEE Transactions on Medical Imaging | 4 | Computer Methods in Biomechanics and Biomedical Engineering: Imaging & Visualization | 1 |
| Medical Imaging: Image Processing | 4 | Conference on Medical Imaging with Deep Learning | 1 |
| International Workshop on Machine Learning in Medical Imaging | 4 | Simulation, Image Processing, and Ultrasound Systems for Assisted Diagnosis and Navigation | 1 |
| Deep Learning in Medical Image Analysis and Multimodal Learning for Clinical Decision Support | 3 | Sensors | 1 |
| International Journal of Computer Assisted Radiology and Surgery | 3 | Electronics Letters | 1 |
| Medical Image Analysis | 3 | Radiotherapy and Oncology | 1 |
| Bildverarbeitung fur die Medizin | 2 | International Workshop on Simulation and Synthesis in Medical Imaging | 1 |
| Deep Learning for Medical Image Analysis | 2 | Pattern Recognition Letters | 1 |
| IEEE Transactions on Biomedical Engineering | 2 | IEEE Workshop on Advances in Information, Electronic and Electrical Engineering | 1 |
| Image Analysis for Moving Organ, Breast, and Thoracic Images | 2 | IEEE/CVF conference on computer vision and pattern recognition | 1 |
| AAAI Conference on Artificial Intelligence | 2 | NeuroImage | 1 |
| International Joint Conference on Biomedical Engineering Systems and Technologies | 2 | Medical physics | 1 |
| International Workshop on Deep Learning in Medical Image Analysis | 2 | Medical Imaging: Digital Pathology | 1 |
| International Conference on Information Processing in Medical Imaging | 2 | Understanding and Interpreting Machine Learning in Medical Image Computing Applications | 1 |
| Physics in Medicine & Biology | 1 | Medical Imaging: Biomedical Applications in Molecular, Structural, and Functional Imaging | 1 |
| Annual Conference on Medical Image Understanding and Analysis | 1 | Medical & Biological Engineering & Computing | 1 |
| Asian Conference on Computer Vision | 1 | Journal of Medical Imaging | 1 |
| Technology in cancer research & treatment | 1 | IEEE Access | 1 |

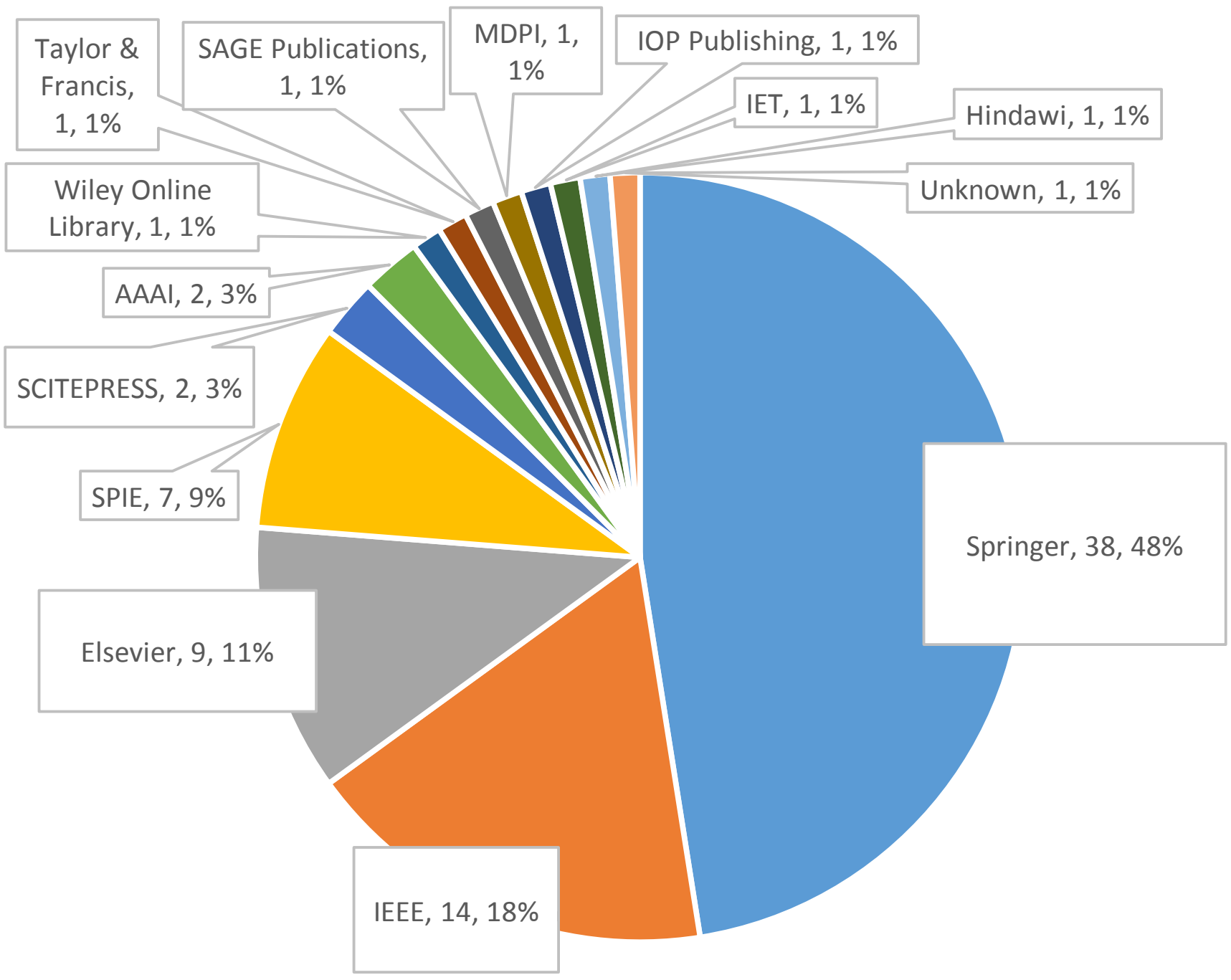

**Figure 32:** Top publication titles based the number of the published works

Table 10 shows a list of top 100 authors active in the field based on the number of their publications as well as total number of citations to their works. Totally 264 authors were detected in the field where other 164 authors accommodated in the Appendix 2 because of the space inconvenience. The number of citations have been gathered from "Scholar Google." Rui Liao and Shun Miao from Medical Imaging Technologies, Siemens Healthcare, USA can be considered as the top authors both from the number of publications and the total number of citations perspectives. Also, Dinggang Shen at the Department of Radiology, University of North Carolina at Chapel Hill is working in the next rank.

**Table 10:** Top 100 authors active in this field

| Author | Pub.s | Cite.s | Author | Pub.s | Cite.s | Author | Pub.s | Cite.s | Author | Pub.s | Cite.s |
|---|---|---|---|---|---|---|---|---|---|---|---|
| Liao, Rui | 9 | 269 | Ghavami, Nooshin | 3 | 34 | Sedai, Suman | 2 | 12 | Comaniciu, Dorin | 1 | 48 |
| Miao, Shun | 9 | 269 | Gibson, Eli | 3 | 34 | Mahapatra, D. | 2 | 12 | Grbic, Sasa | 1 | 48 |
| Shen, Dinggang | 8 | 205 | Hu, Yipeng | 3 | 34 | Ray, Nilanjan | 2 | 12 | de Tournemire, P. | 1 | 48 |
| Mansi, Tommaso | 6 | 113 | Yap, Pew-Thian | 3 | 32 | Zheng, Jiannan | 2 | 6 | Pennec, Xavier | 1 | 47 |
| Wang, Qian | 5 | 178 | Heinrich, Mattias P | 3 | 7 | Wood, Brad J | 2 | 5 | Sermesant, Maxime | 1 | 47 |
| Cao, Xiaohuan | 5 | 56 | Wu, Guorong | 3 | 0 | Xu, Sheng | 2 | 5 | Heimann, Tobias | 1 | 47 |
| Wang, Z Jane | 4 | 170 | Styner, Martin | 2 | 97 | Yan, Pingkun | 2 | 5 | Datar, Manasi | 1 | 47 |
| Kim, Minjeong | 4 | 168 | Viergever, Max A | 2 | 85 | Heldmann, Stefan | 2 | 4 | Rohe, Marc-Michel | 1 | 47 |
| Niethammer, Marc | 4 | 158 | Kamen, Ali | 2 | 83 | Hering, Alessa | 2 | 4 | Cheng, Xi | 1 | 45 |
| Kwitt, Roland | 4 | 158 | Navab, Nassir | 2 | 82 | Blendowski, Max | 2 | 4 | Maier, Andreas K | 1 | 35 |
| Yang, Xiao | 4 | 158 | Zhang, Li | 2 | 80 | Zhao, Bojun | 2 | 1 | Ghesu, Florin C | 1 | 35 |
| Staring, Marius | 3 | 143 | Zheng, Yefeng | 2 | 75 | Jiao, Wanzhen | 2 | 1 | Nie, Dong | 1 | 19 |
| Išgum, Ivana | 3 | 143 | Zhao, Amy | 2 | 59 | Cong, Jinyu | 2 | 1 | Fan, Yong | 1 | 15 |
| Berendsen, Floris | 3 | 143 | Delingette, Herve | 2 | 45 | Jiang, Yanyun | 2 | 1 | Li, Hongming | 1 | 15 |
| de Vos, Bob | 3 | 143 | Zhang, Jun | 2 | 30 | Zheng, Yuanjie | 2 | 1 | Bandula, Steven | 1 | 13 |
| Sabuncu, Mert R | 3 | 87 | Yang, Jianhua | 2 | 30 | Che, Tongtong | 2 | 1 | Wang, Guotai | 1 | 13 |
| Guttag, John | 3 | 87 | Modat, Marc | 2 | 30 | Yang, Xiaodong | 2 | 0 | Li, Wenqi | 1 | 13 |
| Balakrishnan, Guha | 3 | 87 | Fan, Jingfan | 2 | 21 | Munsell, Brent C | 1 | 81 | Ehrhardt, Jan | 1 | 12 |
| Dalca, Adrian V | 3 | 87 | Barratt, Dean C | 2 | 21 | Komodakis, Nikos | 1 | 72 | Handels, Heinz | 1 | 12 |
| Sokooti, Hessam | 3 | 69 | Noble, J Alison | 2 | 21 | Mateus, Diana | 1 | 72 | Wilms, Matthias | 1 | 12 |
| Ayache, Nicholas | 3 | 51 | Vercauteren, Tom | 2 | 21 | Gutierrez-Becker, B. | 1 | 72 | Uzunova, Hristina | 1 | 12 |
| Krebs, Julian | 3 | 51 | Pluim, Josien PW | 2 | 20 | Simonovsky, Martin | 1 | 72 | Veta, Mitko | 1 | 12 |
| Emberton, Mark | 3 | 34 | Eppenhof, Koen | 2 | 20 | Liao, Shu | 1 | 62 | Moeskops, Pim | 1 | 12 |
| Moore, Caroline M | 3 | 34 | Xue, Zhong | 2 | 17 | Gao, Yaozong | 1 | 62 | Lafarge, Maxime | 1 | 12 |
| Bonmati, Ester | 3 | 34 | Mailhe, Boris | 2 | 16 | Lelieveldt, Boudewijn | 1 | 58 | Jeong, Won-Ki | 1 | 11 |

Fig. 33 is a TagCloud visualization based on the frequencies of keywords used by the authors. The most frequent keywords are Deep Learning (22 times), Convolutional Neural Network (16 times), Image Registration (12 times), Deformable Image Registration (8 times), and Deformable Registration (6 times).

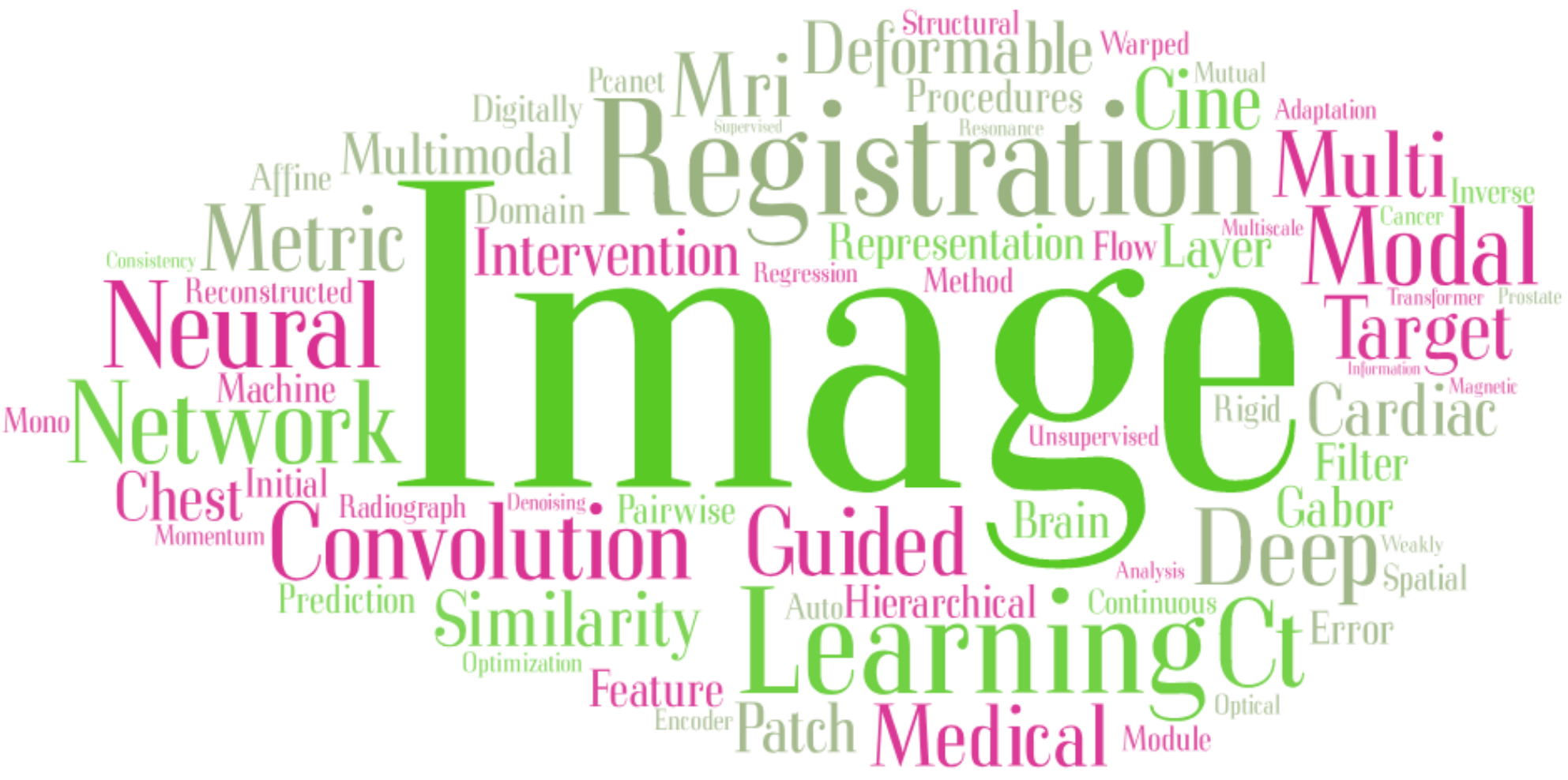

**Figure 33:** TagCloud keyword frequency diagram

Top metrics used to evaluate the approaches are illustrated in Fig. 34. They really need special considerations where Dice Coefficient (DSC) and Target Registration Error (TRE) are the most frequently used metrics. Dice coefficient is an accredited non-parametric measure to quantify the amount of overlapping regions in the inputted fixed and moving images. It can be calculated using (3).

$$DSC = \frac{2|A \cdot B|}{|A| \times |B|} \quad (3)$$

where *A* and *B* are in the intensity histogram of the inputted images. Its value is in range of [0, 1] where 0 indicates no overlap, while the value of 1 indicates perfect one. The most interesting point about DSC is its non-parametric nature and no need to any manual operation. On the other hand, there is Target Registration Error (TRE) for which to be computed, a number of corresponding landmarks should be specified in both the inputted images. The sum of distances between these points at the end of the registration is considered as TRE using (4) regularly uttered in mm. TRE is also of the most accredited performance metrics whose only drawback is its need to manual operations; however, the corresponding landmarks can be specified automatically to remove the burden.

$$TRE = \sum_{i=1}^{n} \left| l_i^A - l_i^B \right| \quad (4)$$

where *A* and *B* are the inputted images, $l_i^A$ and $l_i^B$ are the *i*-the corresponding landmarks in *A* and *B*, and *n* is the total number of landmarks in both images.

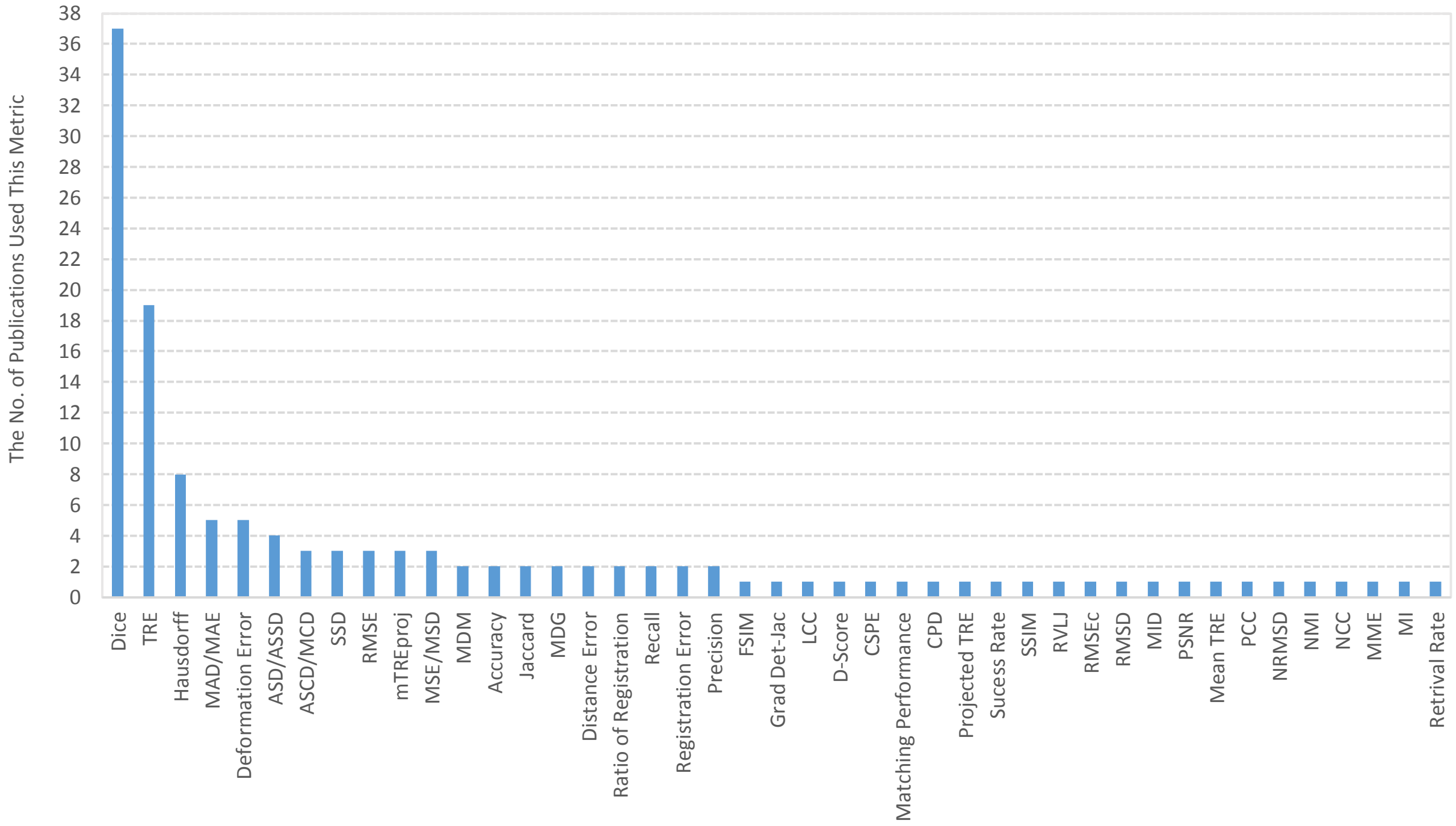

**Figure 34:** Top metrics used to evaluate the approaches

Top datasets used for the implementation and evaluation studies are illustrated in Fig. 35. The most frequently used datasets are the set of {Private, ANDI, LONI, IXI, OASIS, and DIRLAB}, each of which has been utilized in more than 5 papers. Private stands for the private datasets that were not publicly available at the time the paper had been published. Also, in the second rank is the set of { Sunnybrook, ACDC, MCIC, MGH10, XEF, Harvard GSP, HABS, PPMI, CUMC12, IBSR18, BrainWeb, SmartTarget, ADHD200,ABIDE, TKA, and VIPS}.

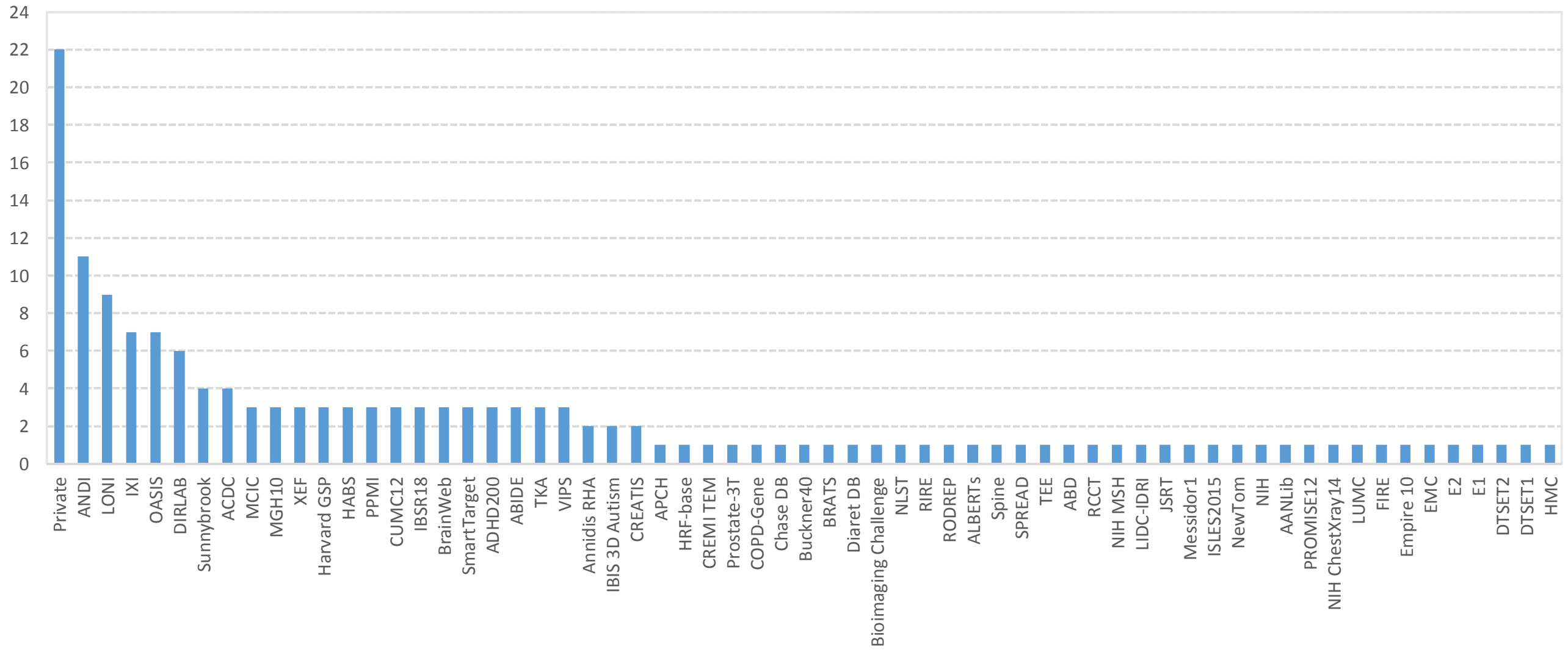

**Figure 35:** Top datasets used to implement the approaches based on the number of publications

The top organs of interest based on the number of publications are presented in the Fig. 36 where the brain is by far the most interesting organ to which 77 papers are belonged. It is the most likely because the brain is enclosed by the skull that is solid and makes the process of alignment faster and simpler. In addition, the structures of interest are mostly visible for brain in different modalities, which also makes the validation process more accurate.

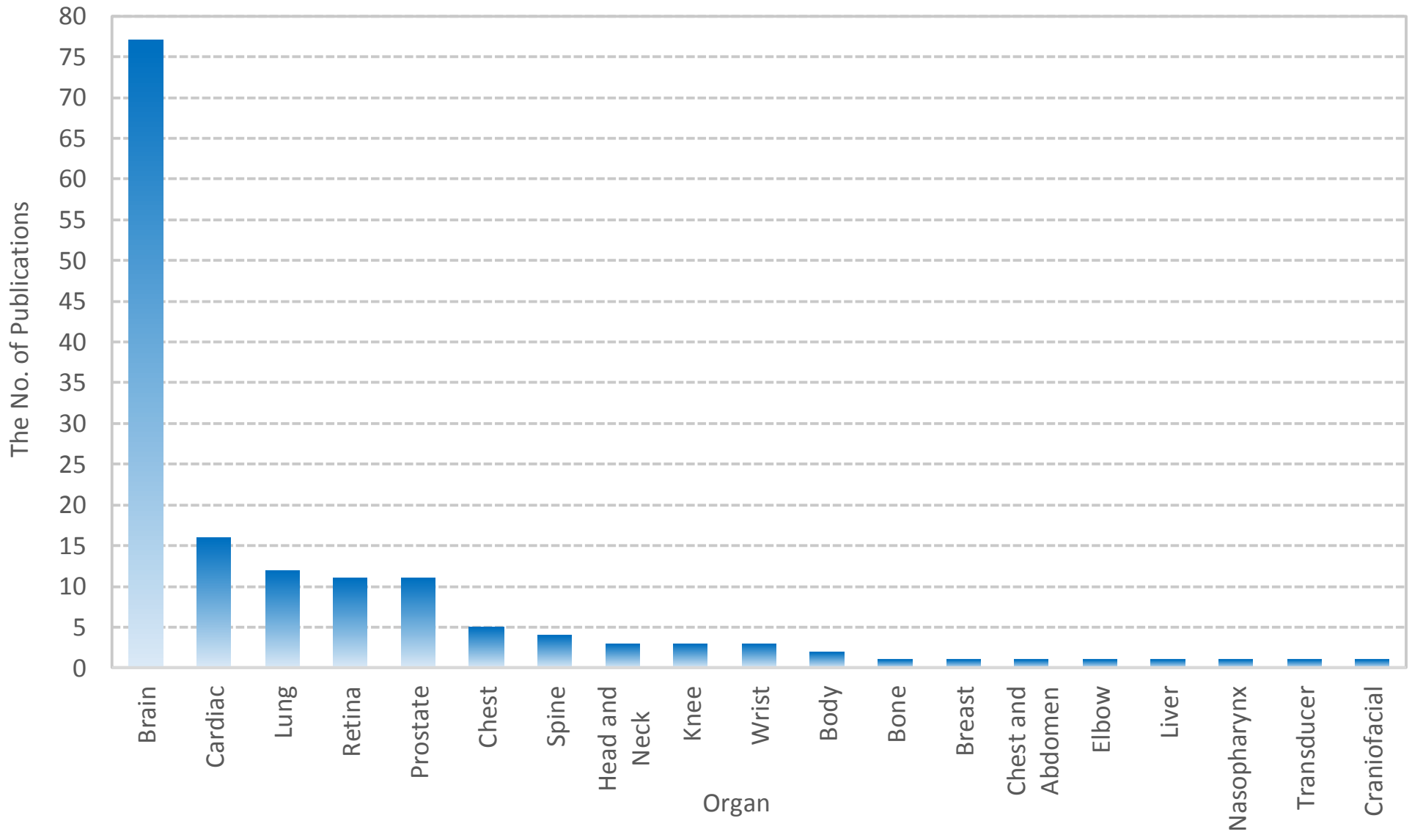

**Figure 36:** Top organs of interest

Fig. 37 shows the pie diagram of top modalities used in the literature. MR imaging was used for about 50%, CT and X-ray which are in the next ranks constitute about other 27%, and there are CFI and OCT for the retina imaging accumulately about 10%. Based on the Fig. 38, multimodal registration is the case for 29 publications (36%) and the remaining 51 works (64%) were on unimodal registration. Worse mentioning that registration of consubstantial modalities like CT and X-ray is considered as unimodal while the utilization of same modalities with different parameters like T1 and T2-weighted MRI is considered as multimodal.

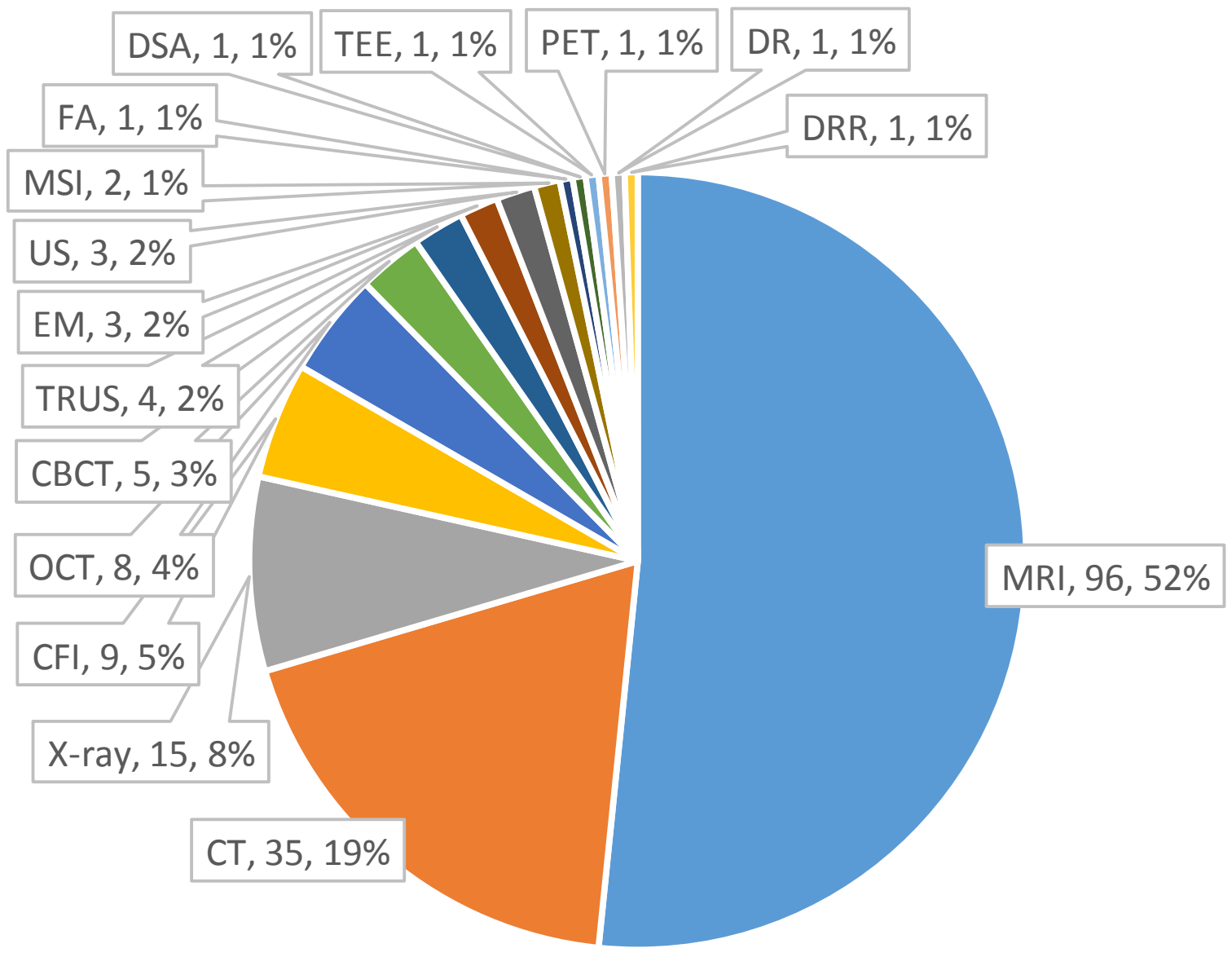

**Figure 37:** Top modalities used in the literature

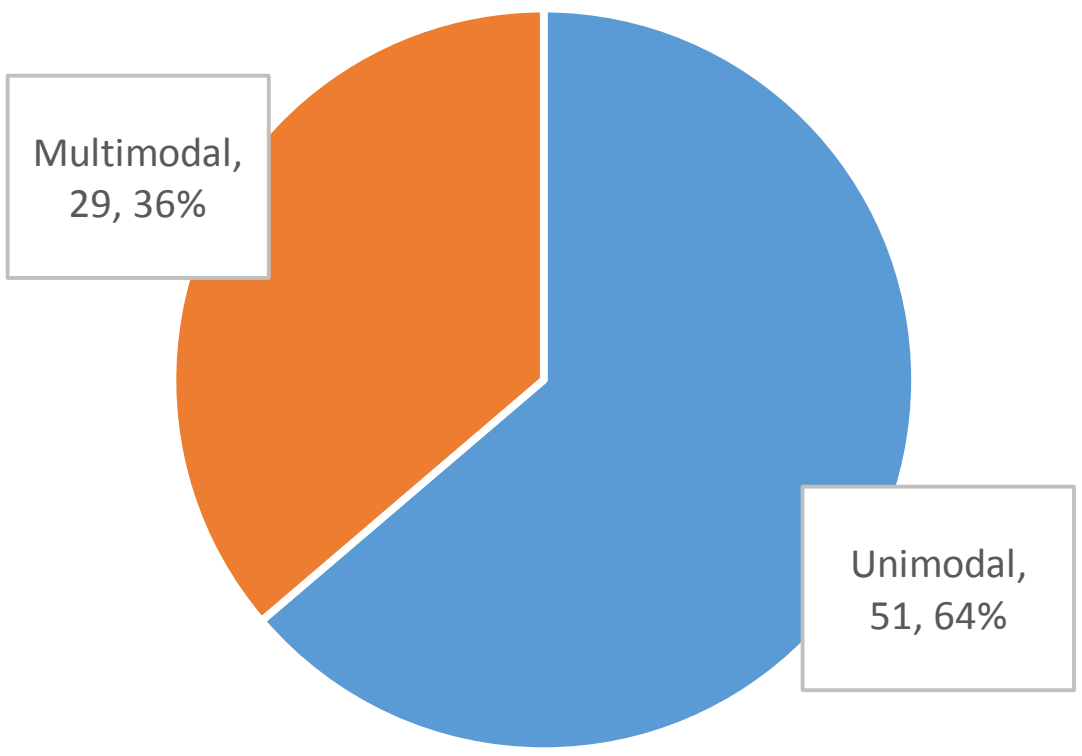

**Figure 38:** Multimodal versus unimodal registration

Finally, Fig. 39 shows the pie diagram of transformation models used in the literature. Accordingly, in 78% of cases (62 times), authors tried for deformable registration, while only 22% (18 cases) belongs to the rigid registration; where we see that mostly the organ-of-interest is the brain that has relatively a rigid form, it can better reveal us the importance of deformable registration for almost all the organs.

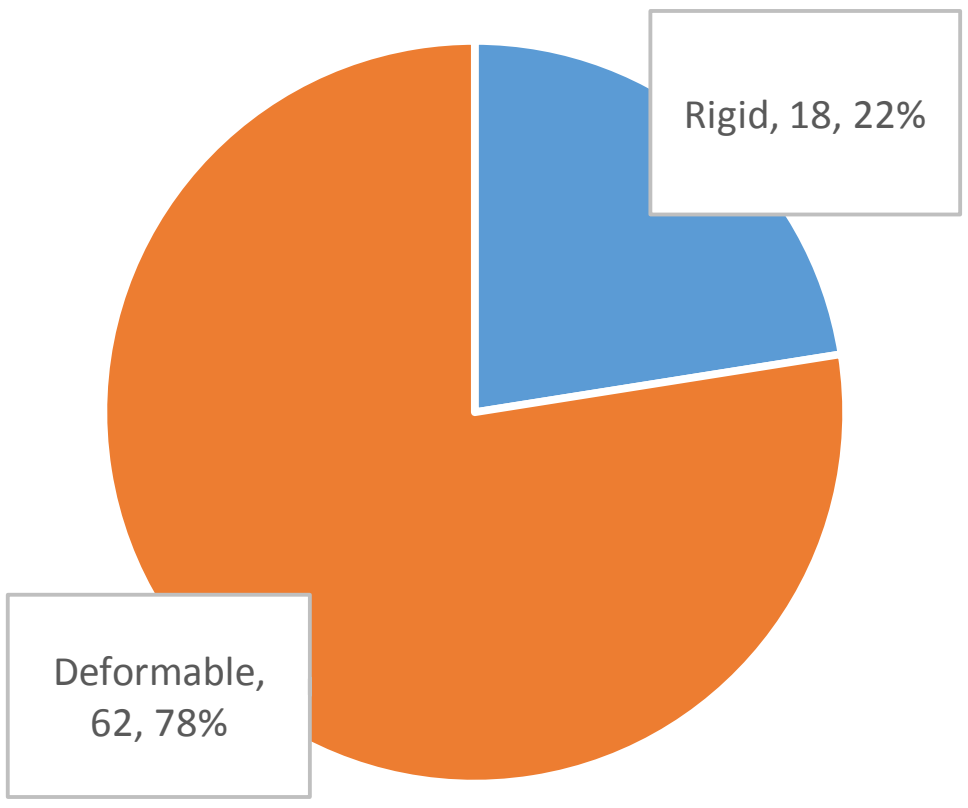

**Figure 39:** Transformation models used in the literature

## 8. Discussion on Confiding Challenges, Open Problems and Promising Directions

The study on the publications in the field of applying deep learning approaches on medical imaging reveals us some frequent challenges. On the other hand, the set of novel solutions proposed by the related authors, experts and researchers in their publications can be a rich reference and guideline to the prospective problems in the area of medical image registrations using deep neural networks. Some of these challenges and solutions are enumerated as follows:

- **Challenge:** Medical datasets are often small-sized.
  **Solution:** Utilization of augmentation techniques to artificially increase the number of samples. Utilization of transfer learning to train the network from other datasets and then fine-tune it with the dataset in consideration. Utilization of weakly-supervised learning to train the network from semi-annotated data. Finally, utilization of dropout which is a technique to stochastically drop some inputs out of each layer to decrease the overfitting effect.
- **Challenge:** Medical datasets' annotated labels are noisy to a large extent since the physicians and experts do not have a consent in a lots of cases.
  **Solution:** Modeling the noise distribution and feeding it to the network, or use e.g. fuzzy logic to tackle the issue.

- **Challenge:** In contrast to Decision Support Systems (DSSs), deep learning approaches do not present the rule-chain for their inferences where it is may be unacceptable for the physicians even though the system has had a high precision and accuracy.
  **Solution:** Some promising studies have already been conducted to represent the way of network's inference based on the visualization of network's internal (hidden) layers, but the overcomes are limited, and the problem is still open.

- **Challenge:** Background knowledge and context can be highly informative as the physicians ask the patients a number of related questions, and see their different records and experiments' results.
  **Solution:** Likewise, patients' clinical records, their genomics, biopsies and other experiments' results can also be gathered and feeded to the network via different channels to enhance the performance; while to investigate the impact, there are too few integrated datasets which worsens the situation.

- **Challenge:** Medical imaging is inherently 3D, but is processed 2D or 2.5D by most of the current deep neural networks. The reason is that 3D processing with 3D DNNs is computationally unaffordable in many cases.
  **Solution:** Nothing to do! We should seat and see whether the progress in the infrastructure will finally enable us to attract the computational intensity requisite to do that.

The review on the literature of medical image registration based on the deep neural networks reveals that the proposed approaches try to technically enhance the two following parameters:

1. **Registration Runtime:** The architectures of the proposed approaches are so that to decrease the registration run-time while keeping the performance; in other words, the authors did not want to over-engineer a network by increasing the number of layers, connections and parameters to drastically improve the performance. In (de Vos et al. 2019), it has been asserted that the proposed approach is able register a typical couple of inputted images in less than 50 ms, which is highly suitable and appreciated for real-time clinical use.
2. **Network Receptive Field:** The extracted patches from the couple of inputted images are typically selected using a small sliding window with some overlap. The patches are usually small in range of 13×13×13 up to 30×30×30 voxels for reducing the computational intensity to be tractable, and this confides the network receptive field while the background context may be fully informative. To issue the problem, in some studies, some bigger patches are extracted apart from the regular patches; these patches are compressed and shrunken (to reduce the computational burden), and feeded to the network via a different channel.

## 9. Conclusions and Future Trends

In this paper, a taxonomy was developed on deep learning based approaches for medical image registration with five categories named Deep Similarity Metrics (DSM), Supervised End-to-End Registration (SE2ER), Deep Reinforcement Learning (or Agent-Based Registration) (DRL), Unsupervised End-to-End Registration (UE2ER), Weakly/Semi-Supervised End-to-End Registration (WSE2ER). The approaches in each categories shares some identical specifications, underlying philosophies, paradigms, advantages and disadvantages. Generally, deep reinforcement learning proved not be optimistic as the huge state-space associated to deformable registration is out-of-tolerate for the learning agents avoid them to be properly converged. Among others, unsupervised and weakly-supervised approaches have less dependency to the ground-truth that is very costy-to-obtain for medical imaging applications; hence, beside the hindrance and confining challenges they are faced, we expect increasing attention and research focus on them. Unless, we can be encountered the publication of huge publicly-available annotated datasets for different organs-of-interest with different modalities. Transfer learning is another choice for supervised approaches, which was successfully applied for some organs and modalities, but its generalization needs much more evidences. Among weakly-supervised end-to-end registration, adverbial learning e.g. utilization of GANs has the major contribution; also, dual-supervision that is based on the learning from a similarity measure like MI (just like unsupervised approaches) and a few ground-truth samples to fine-tune the network is very promising, and needs a further consideration; however, as we go farer from the real-world ground-truth, the results get questionable for clinicians who want to know about how realistic, feasible and practical are them. We are seeing a gap here, and hope for further research on the registration validation to obviate this doubt. The following statements can also be concluded from this review:

- Multistage policy, where we define a rigid registration before goes for deformable one, has a positive impact as reported by many authors.

- Multiresolution policy, where we gradually conduct the registration process from low-resolution to the highest one, also has reported as a strong paradigm to increase the registration precision.
- As reported in the literature, having a few ground-truth samples, transfer learning from other body organs or modalities, is fully practical for medical image registration.
- Incorporating the theory of geometry to the approaches is a novel idea, and need future consideration.
- Spatial Transformer Network (STN) is one the key contributors, when coupled to a e.g. CNN, it can have a significant improvement on the performance.
- Over-engineering, i.e. increasing the number of layers, connections and parameters to drastically improve the performance, is of the case here since the application is real-time.
- Since the background context can be fully informative, increasing the network's receptive field is a positive contributing factor followed by many authors.

We believe that most of the future trends and contributions cannot be intrinsic in the field of medical image registration, but they will come from other medical imaging problems, or even a bit farer i.e. computer vision and machine learning fields. From the application prospective, deep learning techniques applied to the medical image registration are restricted to the CNNs, SAEs, GANs, DRL, and deep RNN while other models have a high potential for contribution. To exemplify, Gated Recurrent Units (GRD), which is a recurrent deep learning model, has a high potential to be applied where the time is the case as the 4th dimension e.g. in continues US images or discrete fluoroscopy.

On the other hand, from the technique point of view, the fields of machine learning and computer vision are in a continuous progress where promising techniques are introducing constantly. For example, Spiking Neural Networks (SNNs), as the 3rd generation of neural networks, aim at bridging the gap between neuroscience and machine learning via utilization of biologically-realistic models of neurons for computation. A SNN is basically different from the conventional neural networks knowing by the community of machine learning. SNNs operate using spikes, which are discrete events occurring at time instances, rather than to be continuous values. Taking place of a spike is determined by differential equations that represent various biological processes where the most important is the membrane potential. In general, once a neuron reaches a certain potential, it spikes, and the potential of that neuron is reset. The most common model for this is the Leaky Integrate-and-Fire (LIF) model. Moreover, SNNs are often sparsely connected and take advantage of sparse network topologies.

**Appendix 1:** Acronyms

| | | | |
|---|---|---|---|
| CNN | Convolutional Neural Networks | PCC | Pearson's Correlation Coefficient |
| DNN | Deep Neural Networks | RMSD | Root Mean Squared Error |
| SNN | Spiking Neural Network | NRMSD | Normalized RMSD |
| AEs | Auto-Encoders | MDM | Mean Deformation Magnitude |
| DAEs | Denoising AEs | MDG | Mean Deformation Gradient |
| SAEs | Staked AEs | MID | Mean Intensity Difference |
| GAN | Generative Adversarial Network | CSPE | Cumulative Sum of Prediction Error |
| DRL | Deep Reinforcement Learning | TRE | Target Registration Error |
| LIR | Local Image Residual | D-Scores | Discriminator Scores |
| LSTM | Long Short-Term Memory | SSIM | Structural Similarity Index |
| ISA | Independent Subset Analysis | PSNR | Peak Signal To Noise Ratio |
| LDDMM | Large Deformation Diffeomorphic Metric Mapping | DSC | Dice Coefficient |
| SVF | Stationary Velocity Field | ASD | Average Surface Distance |
| TPS | Thin-Plate Spline | ASSD | Average Symmetric Surface Distance |
| LIF | Leaky Integrate-and-Fire | ASCD -or- MCD | Average Symmetric Contour Distance -or- Mean Contour Distance |
| DR | Digital Radiograph | RVLJ | Relative Variance Log-Jacobian |
| DDR | Digitally Reconstructed Radiograph | Grad Det-Jac | Mean magnitude of the Gradients of the Determinant of the Jacobian |
| DVF | Displacement Vector Field | LLC | Local Correlation Coefficient |
| IGRT | Image-Guided Radio-Therapy | MI | Mutual Information |
| RoI | Region of Interest | NMI | Normalized Mutual Information |
| OoI | Organ of Interest | FSIM | Feature Similarity Index Metric |
| Linac | Linear Accelerator Machine | RMSEc | Root Mean Squared Error of the 3D Canny Edge |
| DSS | Decision Support System | LCC | Local Cross-Correlation |
| TPS | Thin-Plate Spline | NCC | Normalized Cross-Correlation |
| CT | Computed Tomography | RPD | Re-Projection Distance |
| MRI | Magnetic Resonance Imaging | CPD | Closest Point Distance (CPD) |
| PET | Positron Emission Tomography | DSM | Deep Similarity Metrics |
| TEE | Transesophageal Echocardiography | SE2ER | Supervised End-to-End Registration |
| EM | Electron Microscopy | DRL | Deep Reinforcement Learning |
| FA | Fluorescein Angiography | UE2ER | Unsupervised End-to-End Registration |
| US | Ultrasound | WSE2ER | Weakly/Semi-Supervised End-to-End Registration |
| MSI | Multi-Spectral Imaging | TRUS | Trans-Rectal Ultra-Sound |
| DSA | Digital Subtraction Angiography | SSD | Sum of Squared Differences |
| CFI | Color Fundus Images | MSE -or- MSD | Mean Squared Error -or- Distance |
| OCT | Optical Coherence Tomography | MAE -or- MAD | Mean Absolute Error -or- Distance |
| DoF | Degree of Freedom | | |

**Appendix 2:** Other 164 authors active in this field

| Author | Pub.s | Cite.s | Author | Pub.s | Cite.s | Author | Pub.s | Cite.s | Author | Pub.s | Cite.s |
|---|---|---|---|---|---|---|---|---|---|---|---|
| Lee, Wei-Chung A. | 1 | 11 | Zha, Hongbin | 1 | 5 | Ito, Masato | 1 | 1 | Rozendaal, R | 1 | 0 |
| Tobin, Willie F | 1 | 11 | Xu, Tianmin | 1 | 5 | Werner, Rene | 1 | 1 | Kanehira, T | 1 | 0 |
| Hildebrand, David | 1 | 11 | Guo, Yuke | 1 | 5 | Madesta, Frederic | 1 | 1 | Van Kranen, SR | 1 | 0 |
| Yoo, Inwan | 1 | 11 | Ma, Gengyu | 1 | 5 | Sentker, Thilo | 1 | 1 | Sui, Xiaodan | 1 | 0 |
| Garnavi, Rahil | 1 | 10 | Qin, Haifang | 1 | 5 | Nikos, Paragios | 1 | 1 | Fujita, Hiroshi | 1 | 0 |
| Antony, Bhavna | 1 | 10 | Zhang, Yungeng | 1 | 5 | Stavroula, M. | 1 | 1 | Hara, Takeshi | 1 | 0 |
| Wein, Wolfgang | 1 | 10 | Pei, Yuru | 1 | 5 | Marie-Pierre, Revel | 1 | 1 | Wang, Zhiguo | 1 | 0 |
| Moctezuma, J-L | 1 | 10 | Rastinehad, Ardeshir R | 1 | 4 | Guillaume, C. | 1 | 1 | Kang, Hongjian | 1 | 0 |
| Prevost, Raphael | 1 | 10 | Kuckertz, Sven | 1 | 3 | Maria, Vakalopoulou | 1 | 1 | Jiang, Huiyan | 1 | 0 |
| Salehi, Mehrdad | 1 | 10 | Gholipour, Ali | 1 | 3 | Mihir, Sahasrabudhe | 1 | 1 | Zhou, Xiangrong | 1 | 0 |
| Solberg, Timothy D | 1 | 9 | Erdogmus, Deniz | 1 | 3 | Stergios, C. | 1 | 1 | Yu, Hengjian | 1 | 0 |
| Valdes, Gilmer | 1 | 9 | Khan, Shadab | 1 | 3 | Han, Hua | 1 | 1 | Hoogeman, Mischa | 1 | 0 |
| Sudhyadhom, Atchar | 1 | 9 | Salehi, Seyed Sadegh | 1 | 3 | Xie, Qiwei | 1 | 1 | Marijnen, CAM | 1 | 0 |
| Haaf, Samuel | 1 | 9 | Punithakumar, K. | 1 | 3 | Chen, Xi | 1 | 1 | Incrocci, Luca | 1 | 0 |
| Kearney, Vasant | 1 | 9 | Noga, Michelle | 1 | 3 | Shu, Chang | 1 | 1 | Yousefi, Sahar | 1 | 0 |
| Mewes, Philip | 1 | 9 | Sheikhjafari, Ameneh | 1 | 3 | Zhang, Xuming | 1 | 1 | Shahzad, Rahil | 1 | 0 |
| Tuysuzoglu, Ahmet | 1 | 9 | Milone, Diego H | 1 | 2 | Jin, Xiaomeng | 1 | 1 | Qiao, Yuchuan | 1 | 0 |
| Fischer, Peter | 1 | 9 | Glocker, Ben | 1 | 2 | Huang, Tao | 1 | 1 | Zinkstok, R Th | 1 | 0 |
| Piat, Sebastien | 1 | 9 | Oktay, Ozan | 1 | 2 | Ding, Mingyue | 1 | 1 | Jagt, Thyrza | 1 | 0 |
| Ghosal, Sayan | 1 | 9 | Ferrante, Enzo | 1 | 2 | Zhu, Xingxing | 1 | 1 | Elmahdy, Mohamed S | 1 | 0 |
| Jia, Kebin | 1 | 8 | Zhang, Songtao | 1 | 2 | Hristov, Dimitre | 1 | 0 | Wu, Xi | 1 | 0 |
| Zhao, Liya | 1 | 8 | Sun, Li | 1 | 2 | Hancock, Steven | 1 | 0 | Song, Qi | 1 | 0 |
| Aylward, Stephen | 1 | 7 | Chakravorty, Rajib | 1 | 2 | Han, Bin | 1 | 0 | Hu, Jinrong | 1 | 0 |
| Park, Eunbyung | 1 | 7 | Ge, Zongyuan | 1 | 2 | Najafi, Mohammad | 1 | 0 | Yao, Mingqing | 1 | 0 |
| Han, Xu | 1 | 7 | Siebert, J Paul | 1 | 2 | Zhu, Ning | 1 | 0 | Hu, Jing | 1 | 0 |
| e Delingette, Herve | 1 | 6 | Goatman, Keith A | 1 | 2 | Zheng, Jian | 1 | 0 | Sun, Shanhui | 1 | 0 |
| Chen, Terrence | 1 | 6 | Sloan, James M | 1 | 2 | Chen, Xinjian | 1 | 0 | Song, Zhijian | 1 | 0 |
| Wimmer, Andreas | 1 | 6 | Ding, Yanhui | 1 | 1 | Fu, Tianxiao | 1 | 0 | Wang, Manning | 1 | 0 |
| Chang, Yao-Jen | 1 | 6 | Niu, Yi | 1 | 1 | Gong, Lun | 1 | 0 | Jiang, Dongsheng | 1 | 0 |
| Tamersoy, Birgi | 1 | 6 | van Walsum, Theo | 1 | 1 | Yuan, Gang | 1 | 0 | Liu, Xueli | 1 | 0 |
| Singh, Vivek | 1 | 6 | Niessen, Wiro J | 1 | 1 | Duan, Luwen | 1 | 0 | Estepar, Raul San J. | 1 | 0 |
| Wang, Jiangping | 1 | 6 | Moelker, Adriaan | 1 | 1 | Joshi, Sarang C | 1 | 0 | de la Puente, Maria P. | 1 | 0 |
| Ma, Kai | 1 | 6 | Sun, Yuanyuan | 1 | 1 | Sawant, Amit | 1 | 0 | Marti-Fuster, Berta | 1 | 0 |
| Wang, Shaoyu | 1 | 6 | Pinto, Peter A | 1 | 1 | Zimmerman, Blake E | 1 | 0 | Onieva, Jorge Onieva | 1 | 0 |
| Wang, Li | 1 | 5 | Kruger, Uwe | 1 | 1 | Foote, Markus D | 1 | 0 | Rajpoot, Nasir | 1 | 0 |
| Yang, Jianhuan | 1 | 5 | Kruecker, Jochen | 1 | 1 | Xu, Jingyun | 1 | 0 | Awan, Ruqayya | 1 | 0 |
| Mountney, Peter | 1 | 5 | Haskins, Grant | 1 | 1 | Jin, Xiance | 1 | 0 | Zhou, Xiang Sean | 1 | 0 |
| Rhode, Kawal | 1 | 5 | Matuzevieius, Dalius | 1 | 1 | Lu, Zheming | 1 | 0 | Zhan, Yiqiang | 1 | 0 |
| Rinaldi, Christopher | 1 | 5 | Stankevieius, G. | 1 | 1 | Ma, Longhua | 1 | 0 | Harder, Martin | 1 | 0 |
| Kurzendorfer, Tanja | 1 | 5 | Abanovie, Eldar | 1 | 1 | Liu, Cong | 1 | 0 | Reda, Fitsum | 1 | 0 |
| Toth, Daniel | 1 | 5 | Ino, Fumihiko | 1 | 1 | Sonke, J | 1 | 0 | Bhatia, Parmeet S | 1 | 0 |